# Rapid protein assignments and structures from raw NMR spectra with the deep learning technique ARTINA

Piotr Klukowski[1]*, Roland Riek[1]*, Peter Güntert[1,2,3]*

[1] Laboratory of Physical Chemistry, ETH Zurich, Vladimir-Prelog-Weg 2, 8093 Zurich, Switzerland

[2] Institute of Biophysical Chemistry, Goethe University Frankfurt, Max-von-Laue-Str. 9, 60438 Frankfurt am Main, Germany

[3] Department of Chemistry, Tokyo Metropolitan University, 1-1 Minami-Osawa, Hachioji, 192-0397 Tokyo, Japan

* e-mail: piotr.klukowski@phys.chem.ethz.ch; roland.riek@phys.chem.ethz.ch; peter.guentert@phys.chem.ethz.ch

## Abstract

Nuclear Magnetic Resonance (NMR) spectroscopy is one of the major techniques in structural biology with over 11,800 protein structures deposited in the Protein Data Bank. NMR can elucidate structures and dynamics of small and medium size proteins in solution, living cells, and solids, but has been limited by the tedious data analysis process. It typically requires weeks or months of manual work of a trained expert to turn NMR measurements into a protein structure. Automation of this process is an open problem, formulated in the field over 30 years ago. Here, we present a solution to this challenge that enables the completely automated analysis of protein NMR data within hours after completing the measurements. Using only NMR spectra and the protein sequence as input, our machine learning-based method, ARTINA, delivers signal positions, resonance assignments, and structures strictly without any human intervention. Tested on a 100-protein benchmark comprising 1329 multidimensional NMR spectra, ARTINA demonstrated its ability to solve structures with 1.44 Å median RMSD to the PDB reference and to identify 91.36% correct NMR resonance assignments. ARTINA can be used by non-experts, reducing the effort for a protein assignment or structure determination by NMR essentially to the preparation of the sample and the spectra measurements.

Studying structures of proteins and ligand-protein complexes is one of the most influential endeavors in molecular biology and rational drug design. All key structure determination techniques, X-ray crystallography, electron microscopy, and NMR spectroscopy, have led to remarkable discoveries, but suffer from their respective experimental limitations. NMR can elucidate structures and dynamics of small and medium size proteins in solution[1] and even in living cells[2]. However, the analysis of NMR spectra and the resonance assignment, which are indispensable for NMR studies, remain time-consuming even for a skilled and experienced spectroscopist. Attributed to this, the percentage of NMR protein structures in the Protein Data Bank (PDB) has decreased from a maximum of 14.6% in 2007 to 7.3% in 2021 (https://www.rcsb.org/stats). The problem has sparked research towards automating different tasks in NMR structure determination[3,4], including peak picking[5-9], resonance assignment[10-12], and the identification of distance restraints[13,14]. Several of these methods are available as webservers[15,16]. This enabled semi-automatic[17,18] but not yet unsupervised automation of the entire NMR structure determination process, except for a very small number of favorable proteins[7,19].

The advance of machine learning techniques[20] now offers unprecedented possibilities for reliably replacing decisions of human experts by efficient computational tools. Here, we present a method that achieves this goal for NMR assignment and structure determination. We show for a diverse set of 100 proteins that NMR resonance assignments and protein structures can be determined within hours after completing the NMR measurements. Our method, *Art*ificial *I*ntelligence for *N*MR *A*pplications, ARTINA (Fig. 1), combines machine learning for tasks that are difficult to model otherwise with existing algorithms—evolutionary optimization for resonance assignment with FLYA[12], chemical shift database searches for torsion angle restraint generation with TALOS-N[21], ambiguous distance restraints, network-anchoring and constraint combination for NOESY assignment[14,22] and simulated annealing by torsion angle dynamics for structure calculation with CYANA[23]. Machine learning is used in multiple flavors—deep residual neural networks[24] for visual spectrum analysis to identify peak positions (pp-ResNet) and to deconvolve overlapping signals (deconv-ResNet) in 25 different types of spectra (Supplementary Table 1), kernel density estimation (KDE) to reconstruct original peak positions in folded spectra, a deep graph neural network[25,26] (GNN) for chemical shift estimation within the refinement of chemical shift assignments, and a gradient boosted trees[27] (GBT) model for the selection of structure proposals.

A major challenge in developing ARTINA was the collection and preparation of a large training data set that is required for machine learning, because, in contrast to assignments and structures, NMR spectra are generally not archived in public data repositories. Instead, we were obliged to collect from different sources and standardize complete sets of multidimensional NMR spectra for the assignment and structure determination of 100 proteins.

In the following, we describe the algorithm, training and test data, and results of ARTINA automated structure determination, which are on par with those achieved in weeks or months of human experts' labor.

## Results

**Benchmark dataset**. One of the major obstacles for developing deep learning solutions for protein NMR spectroscopy is the lack of a large-scale standardized benchmark dataset of protein NMR spectra. To date, published manuscripts presenting the most notable methods for computational NMR, typically refer to less than 50 2D/3D/4D NMR spectra in their experimental sections. Even the well-recognized CASD-NMR competition cannot serve as a major source of training data for deep learning, since only the NOESY spectra of 10 proteins were used in the last round of the event[28].

To make our study possible, we established a standardized benchmark of 1329 2D/3D/4D NMR spectra, which allows 100 proteins to be recalculated using their original spectral data (Fig. 2, Supplementary Table 2). Each protein record in our dataset contains 5–20 spectra together with manually identified chemical shifts (usually depositions at the Biological Magnetic Resonance Data Bank, BMRB) and the previously determined ("ground truth") protein structure (PDB record; Supplementary Table 3). The benchmark covers protein sizes typically studied by NMR spectroscopy with sequence lengths between 35 and 175 residues (molecular mass 4–20 kDa).

**Automated protein structure determination.** The accuracy of protein structure determination with ARTINA was evaluated in a 5-fold cross-validation experiment with the aforementioned benchmark dataset. Five instances of pp-ResNet and GBT were trained, each one using data from about 80% of the proteins for training and the remaining ones for testing. Since each protein was present exactly once in the test set, reported quality metrics were obtained directly in the cross-validation experiment, and no averaging between data splits was required. To deploy pp-ResNet and GBT models in our online system, we constructed an ensemble by averaging predictions of all 5 cross-validation models. The other models were trained only once using either generated data (deconv-ResNet, Supplementary Fig. 1) or BMRB depositions excluding all benchmark proteins (GNN, KDE).

In this experiment, we reproduced 100 structures in fully automated manner using only NMR spectra and the protein sequences as input. Since ARTINA has no tunable parameters and does not require any manual curation of data, each structure was calculated by a single execution of the ARTINA workflow. All benchmark datasets were analyzed by ARTINA in parallel with execution times of 4–20 h per protein.

All automatically determined structures, overlaid with the corresponding reference structures from the PDB, are visualized in Fig. 3, Supplementary Fig. 2, and Supplementary Movie 1. ARTINA was able to reproduce the reference structures with a median backbone root-mean-square deviation (RMSD) of 1.44 Å between the mean coordinates of the ARTINA structure bundle and the mean coordinates of the corresponding reference PDB structure bundle for the backbone atoms N, $C^{\alpha}$, C' in the residue ranges determined by CYRANGE[29] (Fig. 4a, Supplementary Table 4). ARTINA automatically identified between 459 and 4678 distance restraints (2198 on average over 100 proteins), which corresponds to 4.25–33.20 restraints per residue (Fig. 4b). This number is mainly influenced by the extent of unstructured regions and the quality of the NOESY spectra. In agreement with earlier findings[30], it correlates only weakly with the backbone RMSD to reference (linear correlation coefficient –0.38). As a more expressive validation measure for the structures from ARTINA, we computed a predicted RMSD to the PDB reference structure on the basis of the RMSDs between the 10 candidate structure bundles calculated in ARTINA (see Methods, Fig. 5, Supplementary Table 5). The average deviation between actual and predicted RMSDs for the 100 proteins in this study is 0.35 Å, and their linear correlation coefficient is 0.77 (Fig. 5). In no case, the true RMSD exceeds the predicted one by more than 1 Å.

Additional structure validation scores obtained from ANSSUR[31] (Supplementary Table 6), RPF[32] (Supplementary Table 7), and consensus structure bundles[33] (Supplementary Table 8) confirm that overall the ARTINA structures and the corresponding reference PDB structures are of equivalent quality. Energy refinement of the ARTINA structures in explicit water using OPALp[34] (not part of the standard ARTINA workflow) does not significantly alter the agreement with the PDB reference structures (Supplementary Table 9). The benchmark data set comprises 78 protein structures determined by the Northeast Structural Genomics Consortium (NESG). On average, ARTINA yielded structures of the same accuracy for NESG targets (median RMSD to reference 1.44 Å) than for proteins from other sources (1.42 Å). Likewise, no remarkable difference was observed for proteins measured with different NMR spectrometer brands (data not shown).

On average, ARTINA correctly assigned 90.39% of the chemical shifts (Fig. 4c), as compared to the manually prepared assignments, including both "strong" (high-reliability) and "weak" (tentative) FLYA assignments[12]. Backbone chemical shifts were assigned more accurately (96.03%) than side-chain ones (86.50%), which is mainly due to difficulties in assigning lysine/arginine (79.97%) and aromatic (76.87%) side-chains. Further details on the assignment accuracy for individual amino acid types in the protein cores (residues with less than 20% solvent accessibility) are given in Supplementary Table 10. Assignments for core residues, which are important for the protein structure, are generally more accurate than for the entire protein, in particular for core Ala, Cys and Asp residues, which show a median assignment accuracy of 100% over the 100 proteins. The lowest accuracies are observed for core His (83.3%), Phe (83.3%), and Arg (87.5%) residues. The three proteins with highest RMSD to reference, 2KCD, 2L82, and 2M47 (see below), show 68.2, 83.8, and 75.7% correct aromatic assignments, respectively, well below the corresponding median of 85.5%. On the other hand, the assignment accuracies for the methyl-containing residues Ala, Ile, Val are above average and reach a median of 100, 97.6, 98.6%, respectively.

The quality of automated structure determination and chemical shift assignment reflects the performance of deep learning-based visual spectrum analysis, presented qualitatively in Figs. 6–7, Supplementary Fig. 3, and Supplementary Movies 2–4. In this experiment, our models (pp-ResNet, deconv-ResNet) automatically identified 1,168,739 cross-peaks with high confidence ($\geq 0.50$) in the benchmark spectra. All 1329 peak lists, together with automatically determined protein structures and chemical shift lists, are available for download.

**Error analysis.** The largest deviations from the PDB reference structure were observed for the proteins 2KCD, 2L82, and 2M47, for which the pRMSD consistently indicated low accuracy (Fig. 5). Significant deviations are mainly due to displacements of terminal secondary structure elements (e.g., a tilted α-helix near a chain terminus), or inaccurate loop conformations (e.g., more flexible than in the PDB deposition). We investigated the origin of these discrepancies.

2KCD is a 120-residue (14.4 kDa) protein from *Staphylococcus saprophyticus* with an α-β roll architecture. Its dataset comprises 19 spectra (8 backbone, 6 side-chain, and 5 NOESY). The ARTINA structure has a backbone RMSD to PDB reference of 3.13 Å, which is caused by the displacement of the C-terminal α-helix (residues 105–109; Supplementary Fig. 4a). Excluding this 5-residue fragment decreases the RMSD to 2.40 Å (Supplementary Table 11). The positioning of this helix appears to be uncertain, since an ARTINA calculation without the 4D CC-NOESY spectrum yields a significantly lower RMSD of 1.77 Å (Supplementary Table 12).

2L82 is a *de novo* designed protein of 162 residues (19.7 kDa) with an αβ 3-layer (αβα) sandwich architecture. Although only 9 spectra (4 backbone, 2 side-chain and 3 NOESY) are available, ARTINA correctly assigned 97.87% backbone and 81.05% side-chain chemical shifts. The primary reason for the high RMSD value of 3.55 Å is again a displacement of the C-terminal α-helix (residues 138–153). The remainder of the protein matches closely the PDB deposition (1.04 Å RMSD, Supplementary Fig. 4b).

The protein with highest RMSD to reference (4.72 Å) in our benchmark dataset is 2M47, a 163-residue (18.8 kDa) protein from *Corynebacterium glutamicum* with an α-β 2-layer sandwich architecture, for which 17 spectra (7 backbone, 7 side chain and 3 NOESY) are available. The main source of discrepancy are two α-helices spanning residues 111–157 near the C-terminus. Nevertheless, the residues contributing to the high RMSD value are distributed more extensively than in 2L82 and 2KCD just discussed. Interestingly, 2 of the 10 structure proposals calculated by ARTINA have an RMSD to reference below 2 Å (1.66 Å and 1.97 Å). In the final structure selection step, our GBT model selected the 4.72 Å RMSD structure as the first choice and 1.66 Å as the second one (Supplementary Fig. 4c). Such results imply that the automated structure determination of this protein is unstable. Since ARTINA returns the two structures selected by GBT with the highest confidence, the user can, in principle, choose the better structure based on contextual information.

In addition to these three case studies, we performed a quantitative analysis of all regular secondary structure elements and flexible loops present in our 100-protein benchmark in order to assess their impact on the backbone RMSD to reference (Supplementary Table 11). All residues in the structurally well-defined regions determined by CYRANGE[29] were assigned to 6 partially overlapping sets: (a) first secondary structure element, (b) last secondary structure element, (c) α-helices, (d) β-sheets, (e) α-helices and β-sheets, and (f) loops. Then, the RMSD to reference was calculated 6 times, each time with one set excluded. In total, for 66 of the 100 proteins the lowest RMSD was obtained if set (f) was excluded from RMSD calculation, and 13% benefited most from removal of the first or last secondary structure element (a or b). Moreover, for 18 out of the 19 proteins with more than 0.5 Å RMSD decrease compared to the RMSD for all well-defined residues, (a), (b) or (f) was the primary source of discrepancy. These results are consistent with our earlier statement that deviations in automatically determined protein structures are mainly caused by terminal secondary structure elements or inaccurate loop conformations.

**Ablation studies.** During the experiment, we captured the state of each structure determination at 9 time-points, 3 per structure determination cycle: (a) after the initial FLYA shift assignment, (b) after GNN shift refinement, and (c) after structure calculation (Fig. 1). Comparative analysis of these states allowed us to quantify the contribution of different ARTINA components to the structure determination process (Table 1).

The results show a strong benefit of the refinement cycles, as quantities reported in Table 1 consistently improve from cycle 1 to 3. The majority of benchmark proteins converge to the correct fold after the first cycle (1.56 Å median backbone RMSD to reference), which is further refined to 1.52 Å in cycle 2 and 1.44 Å in cycle 3. Additionally, within each chemical shift refinement cycle, improvements in assignment accuracy resulting from the GNN predictions are observed. This quantity also increases consistently across all refinement cycles, in particular for side-chains. Refinement cycles are particularly advantageous for large and challenging systems, such as 2LF2, 2M7U or 2B3W, which benefit substantially in cycles 2 and 3 from the presence of approximate protein fold in the chemical shift assignment step.

**Impact of 4D NOESY experiments.** As presented in Fig. 2, 26 out of 100 benchmark datasets contain 4D CC-NOESY spectra, which require long measurement times and were used in the manual structure determination. To quantify their impact, we performed automated structure determinations of these 26 proteins with and without the 4D CC-NOESY spectra (Supplementary Table 12).

On average, the presence of 4D CC-NOESY improves the backbone RMSD to reference by 0.15 Å (decrease from 1.88 Å to 1.73 Å) and has less than 1% impact on chemical shift assignment accuracy. However, the impact is non-uniform. For three proteins, 2KIW, 2L8V, and 2LF2, use of the 4D CC-NOESY decreased the RMSD by more than 1 Å. On the other hand, there is also one protein, 2KCD, for which the RMSD decreased by more than 1 Å by *excluding* the 4D CC-NOESY.

These results suggest that overall the amount of information stored in 2D/3D experiments is sufficient for ARTINA to reach close to optimal performance, and only modest improvement can be achieved by introducing additional information redundancy from 4D CC-NOESY spectra.

**Automated chemical shift assignment.** Apart from structure determination, our data analysis pipeline for protein NMR spectroscopy can address an array of problems that are nowadays approached manually or semi-manually. For instance, ARTINA can be stopped after visual spectrum analysis, returning positions and intensities of cross-peaks that can be utilized for any downstream task, not necessarily related to protein structure determination.

Alternatively, a single chemical shift refinement cycle can be performed to get automatically assigned cross-peaks from spectra and sequence. We evaluated this approach with three sets of spectra: (i) Exclusively backbone assignment spectra were used to assign N, $C^\alpha$, $C^\beta$, C' and $H^N$ shifts. With this input, ARTINA assigned 92.40% (median value) of the backbone shifts correctly. (ii) All through-bond but no NOESY spectra were used to assign the backbone and side-chain shifts. This raised the percentage of correct backbone assignments to 94.20%. (iii) The full data set including NOESY yielded 96.60% correct assignments of the backbone shifts. These three experiments were performed for the 45 benchmark proteins, for which CBCANH and CBCAcoNH, as well as either HNCA and HNcoCA or HNCO and HNcaCO experiments were available. The availability of NOESY spectra had a large impact on the side-chain assignments: 86.00% were correct for the full spectra set iii, compared to 73.70% in the absence of NOESY spectra (spectra set ii). The presence of NOESY spectra consistently improved the chemical shift assignment accuracy of all amino acid types (Supplementary Tables 13 and 14). The improvement is particularly strong for aromatic residues (Phe, 61.6 to 76.5%, Trp 52.5 to 80%, and Tyr 71.4 to 89.7%), but not limited to this group.

## Discussion

The results obtained with ARTINA differ in several aspects substantially from previous approaches towards automating protein NMR analysis[3,4,7,12,17-19,35]. First, ARTINA comprehends the entire workflow from spectra to structures rather than individual steps in it, and there are strictly no manual interventions or protein-specific parameters to be adapted. Second, the quality of the results regarding peak identification, resonance assignments, and structures have been assessed on a large and diverse set of 100 proteins; for the vast majority of which they are on par with what can be achieved by human experts. Third, the method provides a two-orders-of-magnitude leap in efficiency by providing assignments and a structure within hours of computation time rather than weeks or months of human work. This reduces the effort for a protein structure determination by NMR essentially to the preparation of the sample and the measurement of the spectra. Its implementation in the https://nmrtist.org

webserver (Supplementary Movie 5) encapsulates its complexity, eliminates any intermediate data and format conversions by the user, and enables the use of different types of high-performance hardware as appropriate for each of the subtasks. ARTINA is not limited to structure determination but can be used equally well for peak picking and resonance assignment in NMR studies that do not aim at a structure, such as investigations of ligand binding or dynamics.

Although ARTINA has no parameters to be optimized by the user, care should be given to the preparation of the input data, i.e., the choice, measurement, processing, and specification of the spectra. Spectrum type, axes and isotope labelling declarations must be correct, and chemical shift referencing consistent over the entire set of spectra. Slight variations of corresponding chemical shifts within the tolerances of 0.03 ppm for $^1$H and 0.4 ppm for $^{13}$C/$^{15}$N can be accommodated, but larger deviations, resulting, for instance, from the use of multiple samples, pH changes, protein degradation, or inaccurate referencing, can be detrimental. Where appropriate, ARTINA proposes corrections of chemical shift referencing[36]. Furthermore, based on the large training data set, which comprises a large variety of spectral artifacts, ARTINA largely avoids misinterpreting artifacts as signals. However, with decreasing spectral quality, ARTINA, like a human expert, will progressively miss real signals.

Regarding protein size and spectrum quality, limitations of ARTINA are similar to those encountered by a trained spectroscopist. Machine-learning based visual analysis of spectra requires signals to be present and distinguishable in the spectra. ARTINA does not suffer from accidental oversight that may affect human spectra analysis. On the other hand, human experts may exploit contextual information to which the automated system currently has no access because it identifies individual signals by looking at relatively small, local excerpts of spectra.

In this paper, we used all spectra that are available from the earlier manual structure determination. For most of the 100 proteins, the spectra data set has significant redundancy regarding information for the resonance assignment. Our results indicate that one can expect to obtain good assignments and structures also from smaller sets of spectra[37], with concomitant savings of NMR measurement time. We plan to investigate this in a future study.

The present version of ARTINA can be enhanced in several directions. Besides improving individual models and algorithms, it is conceivable to integrate the so far independently trained collection of machine learning models, plus additional models that replace conventional algorithms, into a coherent system that is trained as a whole. Furthermore, the reliability of machine learning approaches depends strongly on the quantity and quality of training data available. While the collection of the present training data set for ARTINA was cumbersome, from now on it can be expected to expand continuously through the use of the https://nmrtist.org website, both quantitatively and qualitatively with regard to greater variability in terms of protein types. spectral quality, source laboratory, data processing (including non-linear sampling), etc., which can be exploited in retraining the models.

ARTINA can also be extended to use additional experimental input data, e.g., known partial assignments, stereospecific assignments, $^3J$ couplings, residual dipolar couplings, paramagnetic data, H-bonds. Structural information, e.g., from AlphaFold[38], can be used in combination with reduced sets of NMR spectra for rapid structure-based assignment. Finally, the range of application of ARTINA can be generalized to small molecule-protein complexes relevant for structure-activity relationship studies in drug research, protein-protein complexes, RNA, solid state and in-cell NMR.

Overall, ARTINA stands for a paradigm change in biomolecular NMR from a time-consuming technique for specialists to a fast method open to researchers in molecular biology and medicinal chemistry. At the same time, in a larger perspective, the appearance of generally highly accurate structure predictions by AlphaFold[38] is revolutionizing structural biology. Nevertheless, there remains space for the experimental methods, for instance, to elucidate various states of proteins under different conditions or in dynamic exchange, or for studying protein-ligand interaction. Regarding ARTINA, one should keep in mind that its applications extend far beyond structure determination. It will accelerate virtually any biological NMR studies that require the analysis of multidimensional NMR spectra and chemical shift assignments. Protein structure determination is just one possible ARTINA application, which is both demanding in terms of the amount and quality of required experimental data and amenable to quantitative evaluation.

## Methods

**Spectrum benchmark collection**. To collect the benchmark of NMR spectra (Fig. 2, Supplementary Table 2), we implemented a crawler software, which systematically scanned the FTP server of the BMRB data bank[39], identifying data files relevant to our study. Additional datasets were obtained by setting up a website for the deposition of published data (https://nmrdb.ethz.ch), from our collaboration network, or had been acquired internally in our laboratory. NMR data was collected from these channels either in the form of processed spectra (Sparky[40], NMRpipe[41], XEASY[42], Bruker formats), or in the form of time-domain data accompanied by depositor-supplied NMRpipe processing scripts. No additional spectra processing (e.g., baseline correction) was performed as part of this study.

The most challenging aspects of the benchmark collection process were: scarcity of data – only a small fraction of all BMRB depositions are accompanied by uploaded spectra (or time-domain data), lack of standards for NMR data depositions – each protein data set had to be prepared manually, as the original data was stored in different formats (spectra name conventions, axis label standards, spectra data format), and difficulties in correlating data files deposited in the BMRB FTP site with contextual information about the spectrum and the sample (e.g., sample characteristics, measurement conditions, instrument used).

Different approaches to 3D $^{13}$C-NOESY spectra measurement had to be taken into account: (i) Two separate $^{13}$C NOESY for aliphatic and aromatic signals. These were analyzed by ARTINA without any special treatment. We used `ALI`, `ARO` tags (Supplementary Movie S5) to provide the information that only either aliphatic or aromatics shifts are expected in a given spectrum. (ii) Simultaneous NC-NOESY. These spectra were processed twice to have proper scaling of the $^{13}$C and $^{15}$N axes in ppm units, and cropped to extract $^{15}$N-NOESY and $^{13}$C-NOESY spectra. If nitrogen and carbon cross-peaks have different signs of signal amplitude, we used `POS`, `NEG` tags to provide the information that only either positive or negative signals should be analyzed. (iii) Aliphatic and aromatic signals in a single $^{13}$C-NOESY spectrum. These measurements do not require any special treatment, but proper cross-peak unfolding plays a vital role in aromatic signals analysis.

**Overview of the ARTINA algorithm.** ARTINA uses as input only the protein sequence and a set of NMR spectra, which may contain any combination of 25 experiments currently supported by the method (Supplementary Table 1). Within 4–20 hours of computation time (depending on protein size, number of spectra, and computing hardware load), ARTINA determines: (a) cross-peak positions for each spectrum, (b) chemical shift assignments, (c) distance restraints from NOESY spectra, and (d) the protein structure. The whole process does not require any human involvement, allowing rapid protein NMR assignment and structure determination by non-experts.

The ARTINA workflow starts with *visual spectrum analysis* (Fig. 1), wherein cross-peak positions are identified in frequency-domain NMR spectra using deep residual neural networks (ResNet)[24]. Coordinates of signals in the spectra are passed as input to the FLYA automated assignment algorithm[12], yielding initial chemical *shift assignments*. In the subsequent chemical *shift refinement* step, we bring to the workflow contextual information about thousands of protein structures solved by NMR in the past using a deep Graph Neural Network (GNN)[25] that was trained on BMRB/PDB depositions. Its goal is to predict expected values of yet missing chemical shifts, given the shifts that have already been confidently and unambiguously assigned by FLYA. With these GNN predictions as additional input, the cross-peak positions are reassessed in a second FLYA call, which completes the *chemical shift refinement cycle* (Fig. 1).

In the *structure refinement cycle*, 10 variants of NOESY peak lists are generated, which differ in the number of cross-peaks selected from the output of the visual spectrum analysis by varying the confidence threshold of a signal selected by ResNet between 0.05 and 0.5. Each set of NOESY peak lists is used in an independent CYANA structure calculation[22,23], yielding *10 intermediate structure proposals* (Fig. 1). The structure proposals are ranked in the *intermediate structure selection* step based on 96 features with a dedicated Gradient Boosted Trees (GBT) model. The selected best structure proposal is used as contextual information in a consecutive FLYA run, which closes the *structure refinement cycle*.

After the two initial steps of visual spectrum analysis and initial chemical shift assignment, ARTINA interchangeably executes refinement cycles. The chemical shift refinement cycle provides

FLYA with tighter restraints on expected chemical shifts, which helps to assign ambiguous cross-peaks. The structure refinement cycle provides information about possible through-space contacts, allowing identified cross-peaks (especially in NOESY) to be reassigned. The high-level concept behind the interchangeable execution of refinement cycles is to iteratively update the protein structure given fixed chemical shifts, and update chemical shifts given the fixed protein structure. Both refinement cycles are executed three times.

**Automated visual analysis of the spectrum**. We established two machine learning models for the visual analysis of multidimensional NMR spectra (see downloads in the Code availability section). In their design, we made no assumptions about the downstream task and the 2D/3D/4D experiment type. Therefore, the proposed models can be used as the starting point of our automated structure determination procedure, as well as for any other task that requires cross-peak coordinates.

The automated visual analysis starts by selecting all extrema $x = \{x_1, x_2, ..., x_N\}$, $x_n \in \mathbb{N}^D$ in the NMR spectrum, which is represented as a $D$-dimensional regular grid storing signal intensities at discrete frequencies. We formulated the peak picking task as an object detection problem, where possible object positions are confined to $x$. This task was addressed by training a deep residual neural network[24], in the following denoted as peak picking ResNet (pp-ResNet), which learns a mapping $x_n \rightarrow [0, 1]$ that assigns to each signal extremum a real-valued score, which resembles its probability of being a true signal rather than an artefact.

Our network architecture is strongly linked to ResNet-18[24]. It contains 8 residual blocks, followed by a single fully connected layer with sigmoidal activation. After weight initialization with Glorot Uniform[43], the architecture was trained by optimizing a binary cross-entropy loss using Adam[44] with learning rate $10^{-4}$ and gradient clipping of 0.5.

To establish an experimental training dataset for pp-ResNet, we normalized the 1329 spectra in our benchmark with respect to resolution (adjusting the number of data grid points per unit chemical shift (ppm) using linear interpolation) and signal amplitude (scaling the spectrum by a constant). Subsequently, 675,423 diverse 2D fragments of size $256 \times 32 \times 1$ were extracted from the normalized spectra and manually annotated, yielding 98,730 positive and 576,693 negative class training examples. During the training process, we additionally augmented this dataset by flipping spectrum fragments along the second dimension (32 pixels), stretching them by 0–30% in the first and second dimensions, and perturbing signal intensities with Gaussian noise addition.

The role of the pp-ResNet is to quickly iterate over signal extrema in the spectrum, filtering out artefacts and selecting approximate cross-peak positions for the downstream task. The relatively small network architecture (8 residual blocks) and input size of 2D $256 \times 32$ image patches make it possible to analyze large 3D $^{13}$C-resolved NOESY spectra in less than 5 minutes on a high-end desktop computer. Simultaneously, the first dimension of the image patch (256 pixels) provides long-range contextual information on the possible presence of signals aligned with the current extremum (e.g., $C^\alpha$, $C^\beta$ cross-peaks in an HNCACB spectrum).

Extrema classified with high confidence as true signals by pp-ResNet undergo subsequent analysis with a second deep residual neural network (deconv-ResNet). Its objective is to perform signal deconvolution, based on a 3D spectrum fragment ($64 \times 32 \times 5$ voxels) that is cropped around a signal extremum selected by pp-ResNet. This task is defined as a regression problem, where deconv-ResNet outputs a $3 \times 3$ matrix storing 3D coordinates of up to 3 deconvolved peak components, relative to the center of the input image. To ensure permutation invariance with respect to the ordering of components in the output coordinate matrix, and to allow for a variable number of 1–3 peak components, the architecture was trained with a Chamfer distance loss[45].

Since deconv-ResNet deals only with true signals and their local neighborhood, its training dataset can be conveniently generated. We established a spectrum fragment generator, based on rules reflecting the physics of NMR, which produced 110,000 synthetic training examples (Extended Data Fig. 1) having variable (a) numbers of components to deconvolve (1–3), (b) signal-to-noise ratio, (c) component shapes (Gaussian, Lorentzian, and mixed), (d) component amplitude ratios, (e) component separation, and (f) component neighborhood type (i.e., NOESY-like signal strips or HSQC-like 2D signal clusters). The deconv-ResNet model was thus trained on fully synthetic data.

**Signal unaliasing**. To use ResNet predictions in automated chemical shift assignment and structure calculation, detected cross-peak coordinates must be transformed from the spectrum coordinate system to their true resonance frequencies. We addressed the problem of automated signal unfolding with the classical machine learning approach to density estimation.

At first, we generated $10^5$ cross-peaks associated with each experiment type supported by ARTINA (Supplementary Table 1). In this process, we used randomly selected chemical shift lists deposited in the BMRB database, excluding depositions associated with our benchmark proteins. Subsequently, we trained a Kernel Density Estimator (KDE):

$$p_e(\boldsymbol{x}) = \frac{1}{N_e} \sum_{i=1}^{N_e} \kappa \left( \boldsymbol{x} - \boldsymbol{x}_i^{(e)} \right)$$

which captures distribution $p_e(\boldsymbol{x})$ of true peaks being present at position $\boldsymbol{x}$ in spectrum type $e$, based on $N_e = 10^5$ cross-peaks coordinates $\boldsymbol{x}_i^{(e)}$ generated with BMRB data, and $\kappa$ being the Gaussian kernel.

Unfolding a $k$-dimensional spectrum is defined as a discrete optimization problem, solved independently for each cross-peak position $\boldsymbol{x}_j^{(e)}$ observed in a spectrum of type $e$:

$$\boldsymbol{s}^* = \arg \max_{\boldsymbol{s}} p_e(\boldsymbol{x}_j^{(e)} + \boldsymbol{w} \circ \boldsymbol{s})$$

where $\boldsymbol{w} \in \mathbb{R}^k$ is a vector storing the spectral width in each dimension (ppm units), $\circ$ is element-wise multiplication, $\boldsymbol{s} \in \mathbb{Z}^k$ is a vector indicating how many times cross-peak is unfolded in each dimension, and $\boldsymbol{s}^* \in \mathbb{Z}^k$ is the optimal cross-peak unfolding.

As long as regular and folded signals do not overlap or have different signs in the spectrum, KDE can unfold the peak list regardless of spectrum dimensionality. The spectrum must not be cropped in the folded dimension, i.e., the folding sweep width must equal the width of the spectrum in the corresponding dimension.

All 2D/3D spectra in our benchmark were folded in at most one dimension and satisfy the aforementioned requirements. However, the 4D CC-NOESY spectra satisfy neither, as regular and folded peaks both overlap and have the same signal amplitude sign. This introduces ambiguity in the spectrum unfolding that prevents direct use of the KDE technique. To retrieve original signal positions, 4D CC-NOESY cross-peaks were unfolded to overlap with signals detected in 3D $^{13}$C- NOESY. In consequence, 4D CC-NOESY unfolding depended on other experiments, and individual 4D cross-peaks are retained only if they are confirmed in a 3D experiment.

**Chemical shift assignment**. Chemical shift assignment is performed with the existing FLYA algorithm[12] that uses a genetic algorithm combined with local optimization to find an optimal matching between expected and observed peaks. FLYA uses as input the protein sequence, lists of peak positions from the available spectra, chemical shift statistics, either from the BMRB[39] or the Graph Neural Network (GNN) described in the next section, and, if available, the structure from the previous refinement cycle. The tolerance for the matching of peak positions and chemical shifts was set to 0.03 ppm for $^1$H, and 0.4 ppm for $^{13}$C/$^{15}$N shifts. Each FLYA execution comprises 20 independent runs with identical input data that differ in the random numbers used in the optimization algorithm. Nuclei for which at least 80% of the 20 runs yield, within tolerance, the same chemical shift value are classified as reliably assigned[12] and used as input for the following chemical shift refinement step.

**Chemical shift refinement.** We used a graph data structure to combine FLYA-assigned shifts with information from previously assigned proteins (BMRB records) and possible spatial interactions. Each node corresponds to an atom in the protein sequence, and is represented by a feature vector composed of (a) a one-hot encoded atom type code (e.g., $C^\alpha$, $H^\beta$), (b) a one-hot encoded amino acid type, (c) the value of the chemical shift assigned by FLYA (only if a confident assignment is available, zero otherwise), (d) atom-specific BMRB shift statistics (mean and standard deviation), and (e) 30 chemical shift values obtained from BMRB database fragments. The latter feature is obtained by searching BMRB records for assigned 2–3-residue fragments that match the local protein sequence and have minimal mean-squared-error (MSE) to shifts confidently assigned by FLYA (non-zero values of feature (c) in the local neighborhood of the atom). The edges of the graph correspond to chemical bonds or skip connections. The latter connect the $C^\beta$ atom of a given residue with $C^\beta$ atoms 2, 3, and 5 residues apart in the amino acid sequence, and have the purpose to capture possible through-space influence on the chemical shift that is typically observed in secondary structure elements.

The chemical shift refinement task is defined as a node regression problem, where an expected value of the chemical shift is predicted for each atom that lacks a confident FLYA assignment. This task is addressed with a 26-layer DeepGCN model[25,26] that was trained on 28,400 graphs extracted from 2840 referenced BMRB records[39]. Each training example was created by building a fully assigned graph out of a single BMRB record, and dropping chemical shift values (feature (c) above) for randomly chosen atoms that FLYA typically assigns either with low confidence or inaccurately.

Our DeepGCN model is designed specifically for *de novo* structure determination, as it uses only the protein sequence and partial shift assignments to estimate values of missing chemical shifts. Its predictions are used to guide the FLYA genetic algorithm optimization[12] by reducing its search range for assignments. The precise final chemical shift value is always determined by the position of a signal in the spectrum, rather than the model prediction alone.

**Torsion angle restraints**. Before each structure calculation step, torsion angle restraints for the $\phi$ and $\psi$ angles of the polypeptide backbone were obtained from the current backbone chemical shifts using the program TALOS-N[21]. Restraints were only generated if TALOS-N classified the prediction as 'Good', 'Strong', or 'Generous'. Given a TALOS-N torsion angle prediction of $\phi \pm \Delta\phi$, the allowed range of the torsion angle was set to $\phi \pm \max(\Delta\phi, 10º)$ for 'Good' and 'Strong' predictions, and $\phi \pm 1.5 \max(\Delta\phi, 10º)$ for 'Generous' predictions.

**Structure calculation and selection**. Given the chemical shift assignments and NOESY cross-peak positions and intensities, the structure is calculated with CYANA[23] using the established method[22] that comprises 7 cycles of NOESY cross-peak assignment and structure calculation, followed by a final structure calculation. In total, $8 \times 100$ conformers are calculated for a given input data set using 30,000 torsion angle dynamics steps per conformer. The 20 conformers with the lowest final target function value are chosen to represent the solution structure proposal. The entire combined NOESY assignment and structure calculation procedure is executed independently 10 times based on 10 variants of NOESY peak lists, which differ in the number of cross-peaks selected from the output of the visual spectrum analysis. The first set generously includes all signals selected by ResNet with confidence $\geq 0.05$. The other variants of NOESY peak lists follow the same principle with increasingly restrictive confidence thresholds of 0.1, 0.15, …, 0.5.

The CYANA structures calculations are followed by a structure selection step, wherein the 10 intermediate structure proposals are compared pairwise by a Gradient Boosted Tree (GBT) model that uses 96 features from each structure proposal (including the CYANA target function value[23], number of long-range distance restraints, etc.; for details, see downloads in the Code availability section) to rank the structures by their expected accuracy. The best structure from the ranking is subsequently used as contextual information for the chemical shift refinement cycle (Fig. 1), or returned as the final outcome of ARTINA. The second-best final structure is also returned for comparison.

To train GBT, we collected a set of successful and unsuccessful structure calculations with CYANA. Each training example was a tuple $(s_i, r_i)$, where $s_i$ is the vector of features extracted from the CYANA structure calculation output, and $r_i$ is the RMSD of the output structure to the PDB reference. The GBT was trained to take the features $s_i$ and $s_j$ of two structure calculations with CYANA as input, and to predict a binary order variable $o_{ij}$, such that $o_{ij} = 1$ if $r_i < r_j$, and 0 otherwise. Importantly, the deposited PDB reference structures were not used directly in the GBT model training (they are used only to calculate the RMSDs). Consequently, the GBT model is unaffected by methodology and technicalities related to PDB deposition (e.g., the structure calculation software used to calculate the deposited reference structure).

**Structure accuracy estimate**. As an accuracy estimate for the final ARTINA structure, a predicted RMSD to reference (pRMSD) is calculated from the ARTINA results (without knowledge of the reference PDB structure). It aims at reproducing the actual RMSD to reference, which is the RMSD between the mean coordinates of the ARTINA structure bundle and the mean coordinates of the corresponding reference PDB structure bundle for the backbone atoms N, $C^{\alpha}$, C' in the residue ranges as given in Supplementary Table 4. The predicted RMSD is given by pRMSD $= (1 - t) \times 4$ Å, where, in analogy to the GDT_HA value[46], $t$ is the average fraction of the RMSDs $\leq 0.5, 1, 2, 4$ Å between the mean coordinates of the best ARTINA candidate structure bundle and the mean coordinates of the structure bundles of the 9 other structure proposals. Since $t \in [0, 1]$, the pRMSD is always in the range of 0–4 Å, grouping all "bad" structures with expected RMSD to reference $\geq 4$ Å at pRMSD $= 4$ Å.

**Reporting summary.** Further information on research design is available in the Nature Research Reporting Summary linked to this article.

## Data availability

References structures: PDB Protein Data Bank (https://www.rcsb.org/; accession codes in Fig. 2 and Supplementary Table 3)

Spectra and reference assignments: BMRB Biological Magnetic Resonance Data Bank (https://bmrb.io/; entry IDs in Supplementary Table 3)

Peak lists, assignments, and structures:
https://nmrtist.org/media/publications/ARTINA/ARTINA_results.zip

Source data for Figs. 2, 4, and 5 is available in Supplementary Tables 2, 4, and 5, respectively.

## Code availability

The ARTINA algorithm is available as a webserver at https://nmrtist.org. pp-ResNet, deconv-ResNet, GNN, and GBT are available for download in binary form, together with architecture schemes, example input data, model input description, and source code that allows to read model files and make predictions (https://github.com/PiotrKlukowski/ARTINA, https://nmrtist.org/static/public/publications/artina/models/{ARTINA_peak_picking.zip, ARTINA_peak_deconvolution.zip, ARTINA_shift_prediction.zip, ARTINA_structure_ranking.zip}). These files provide a full technical specification of the components developed within ARTINA, and allow for their independent use in Python.

Existing software used: Python (https://www.python.org/), CYANA (https://www.las.jp/), TALOS-N (https://spin.niddk.nih.gov/bax/software/TALOS-N).

## Acknowledgements

We thank Drs. Frédéric Allain, Fred Damberger, Hideo Iwai, Harindranath Kadavath, Julien Orts, and Dean Strotz for providing unpublished spectra. This project has received funding from the European Union's Horizon 2020 research and innovation programme under the Marie Sklodowska-Curie grant agreement No 891690 (P.K.), and a Grant-in-Aid for Scientific Research of the Japan Society for the Promotion of Science (P.G., 20 K06508).

## Author contributions

P.K. prepared training and test data sets, designed and trained machine learning models, performed experiments described in the manuscript, and implemented ARTINA within nmrtist.org web platform. P.K. and P.G. wrote software. All authors conceived the project, analyzed the results, and wrote the manuscript.

## Competing interests

The authors declare no competing interests.

## Additional information

**Supplementary information** The online version contains supplementary material available at https://doi.org/...

**Correspondence** and requests for materials should be addressed to P.K., R.R., or P.G.

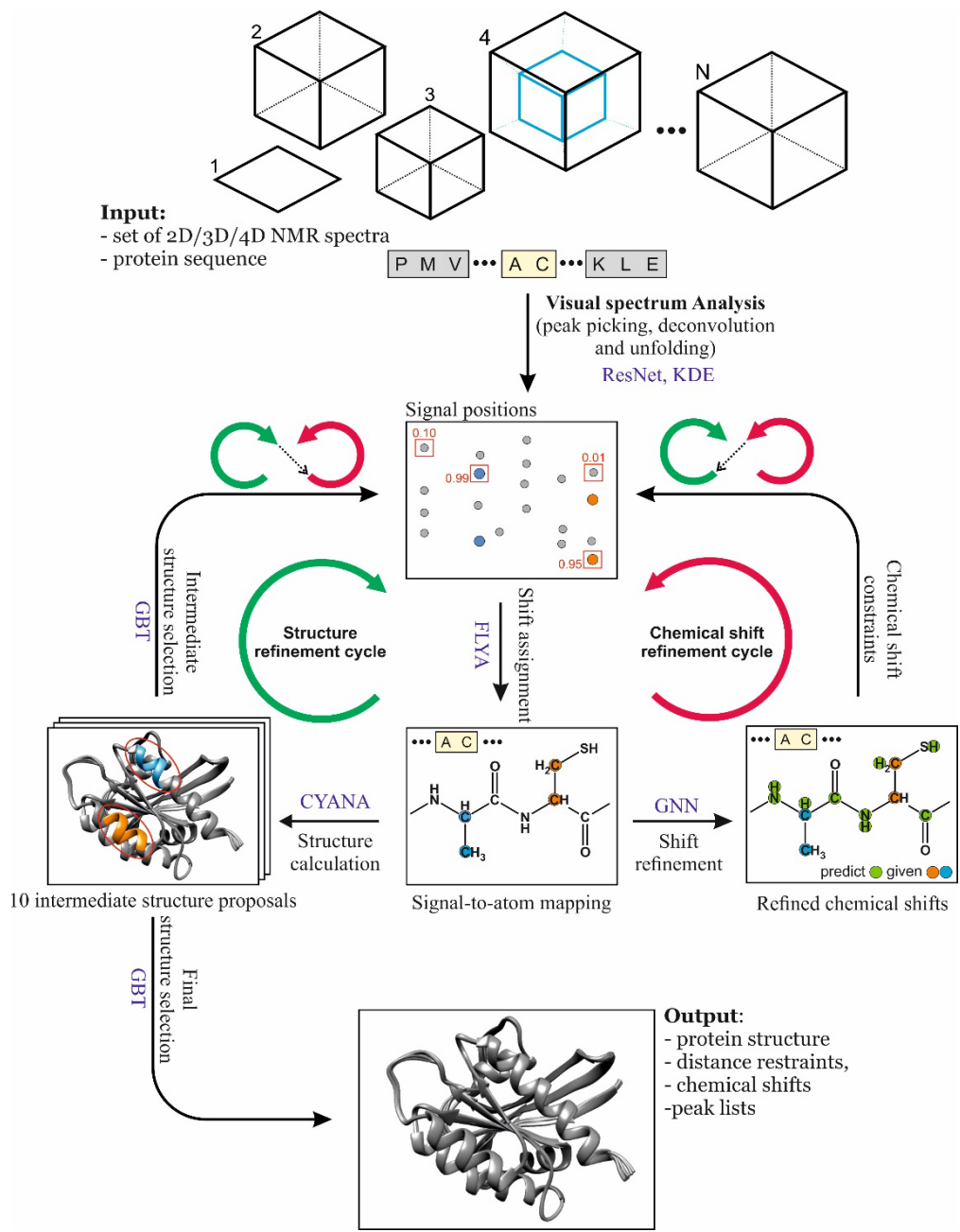

**Fig. 1 Graphical abstract of the ARTINA workflow for automated NMR protein structure determination.**

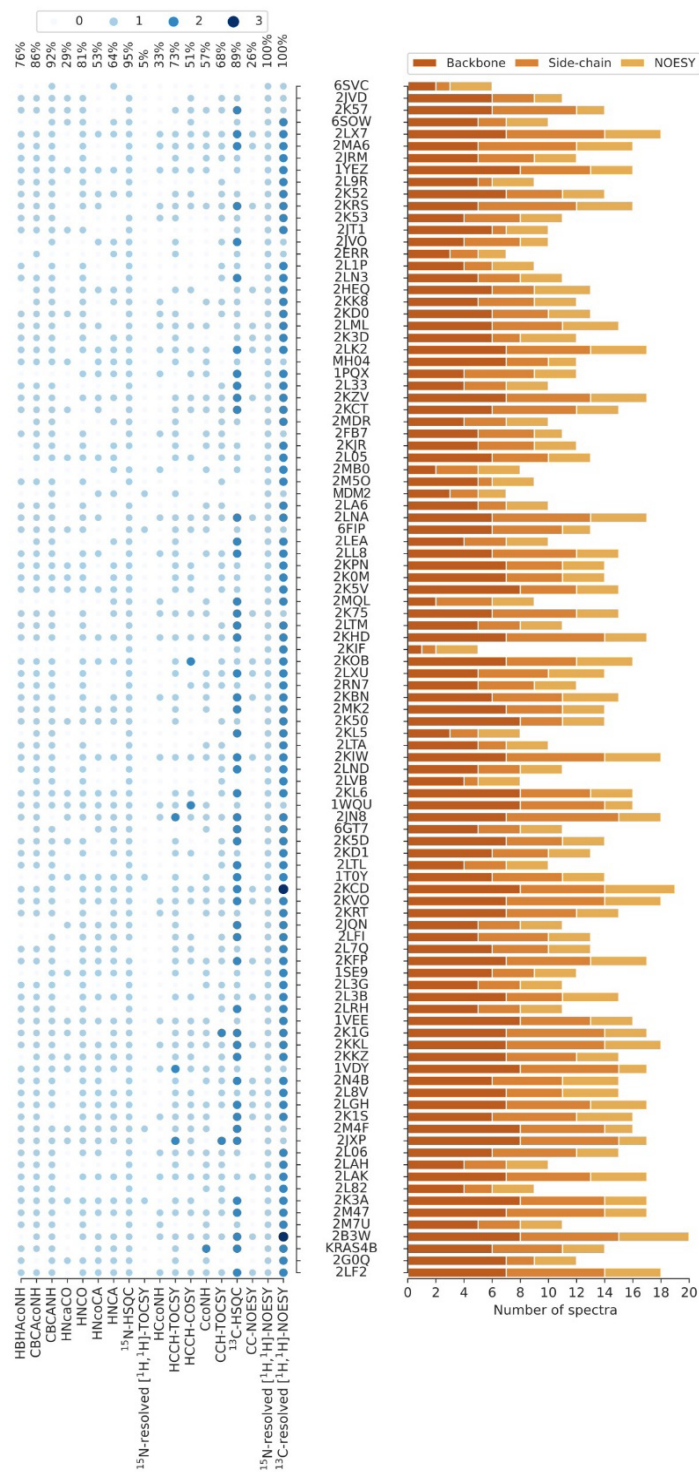

**Fig. 2 NMR benchmark dataset content.** PDB codes (or names, MH04, MDM2, KRAS4B, if PDB code unavailable) of the 100 benchmark proteins are ordered by the number of residues. The histogram shows the number of spectra for backbone assignment, side-chain assignment, and NOE measurement. Spectrum types in each data set are shown by light to dark blue circles indicating the number of individual spectra of the given type. The percentages of benchmark records that contain a given spectrum type are given at the top. Spectrum types present in less than 5% of the data sets have been omitted.

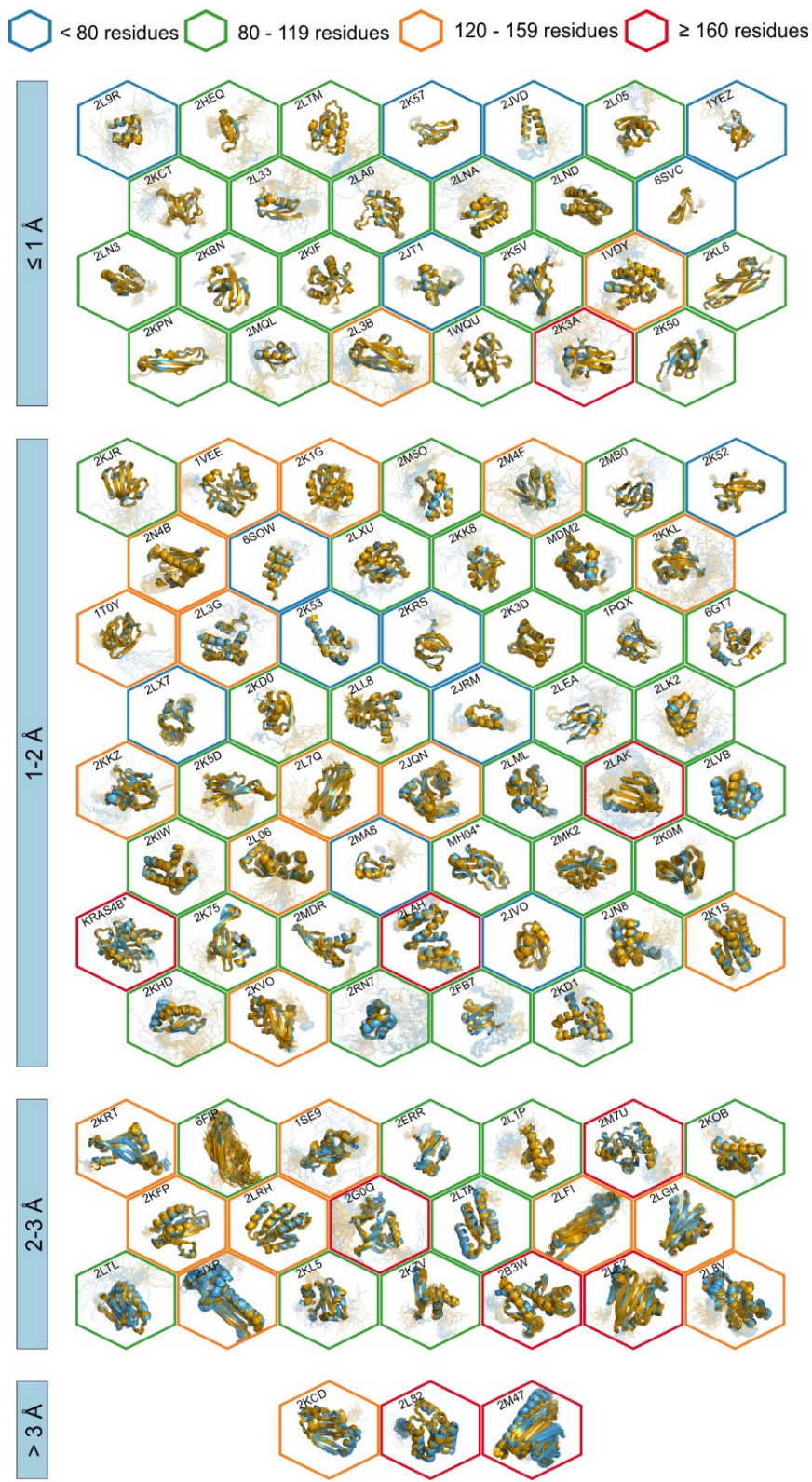

**Fig. 3 100 protein structures determined automatically by ARTINA (blue) overlaid with corresponding PDB depositions (orange).** The structures are aligned with the RMSD to reference range as indicated on the left and hexagonal frames color-coded by their size as indicated above. Structures with no corresponding PDB depositions are marked by an asterisk.

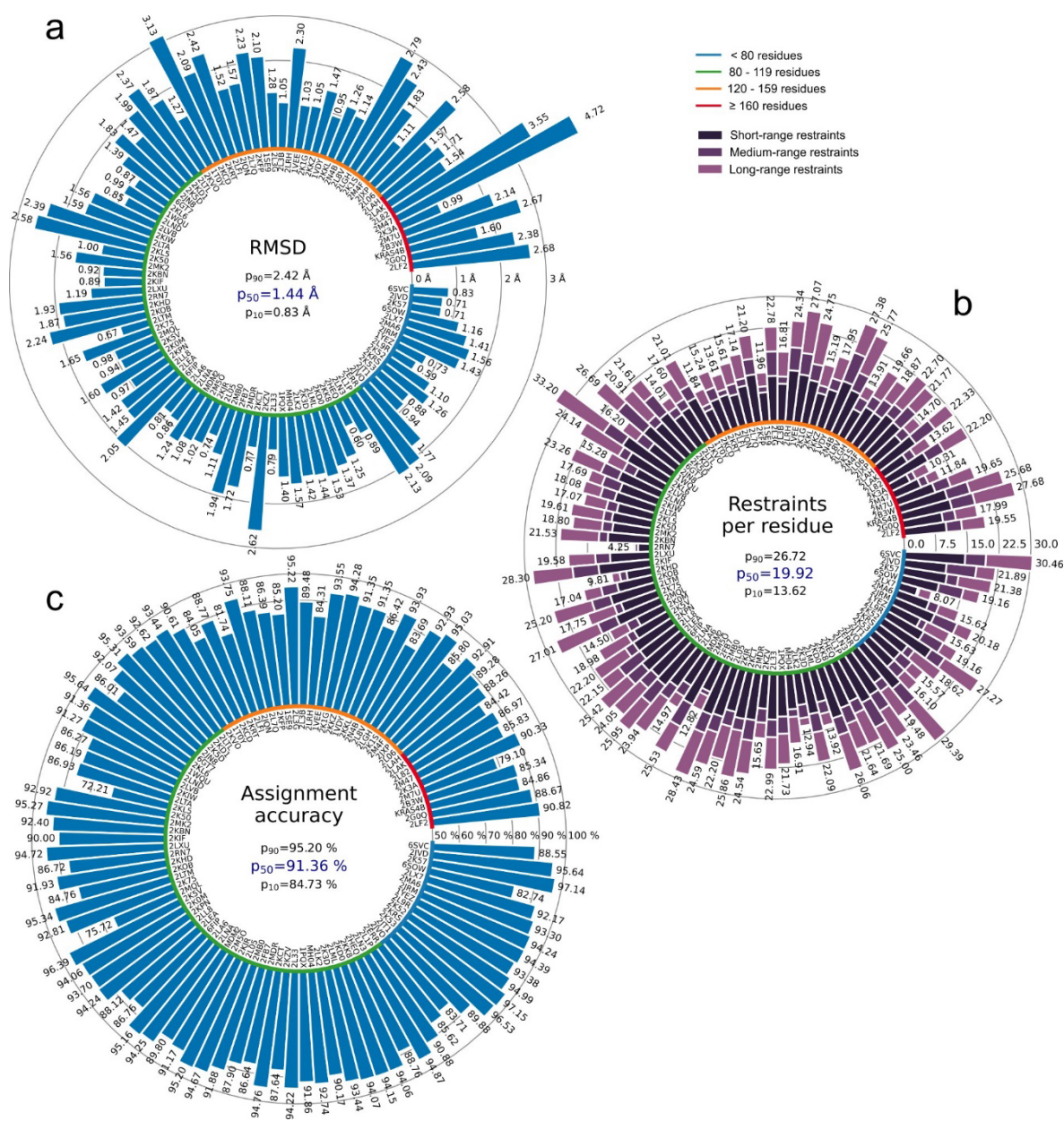

**Fig. 4 Results of the automated structure determination of 100 proteins. a** Backbone RMSD to reference. **b** Number of distance restraints per residue. **c** Chemical shift assignment accuracy. Bars represent quantity values for benchmark proteins, identified by PDB codes (or protein names). Proteins are ordered by size, which is indicated by a color-coded circle. Values in the center of each panel are $10^{th}$, $50^{th}$ and $90^{th}$ percentiles of values presented in the bar plot. Short/medium/long-range restraints are between residues $i$ and $j$ with $|i - j| \leq 1$, $2 \leq |i - j| \leq 4$, and $|i - j| \geq 5$, respectively.

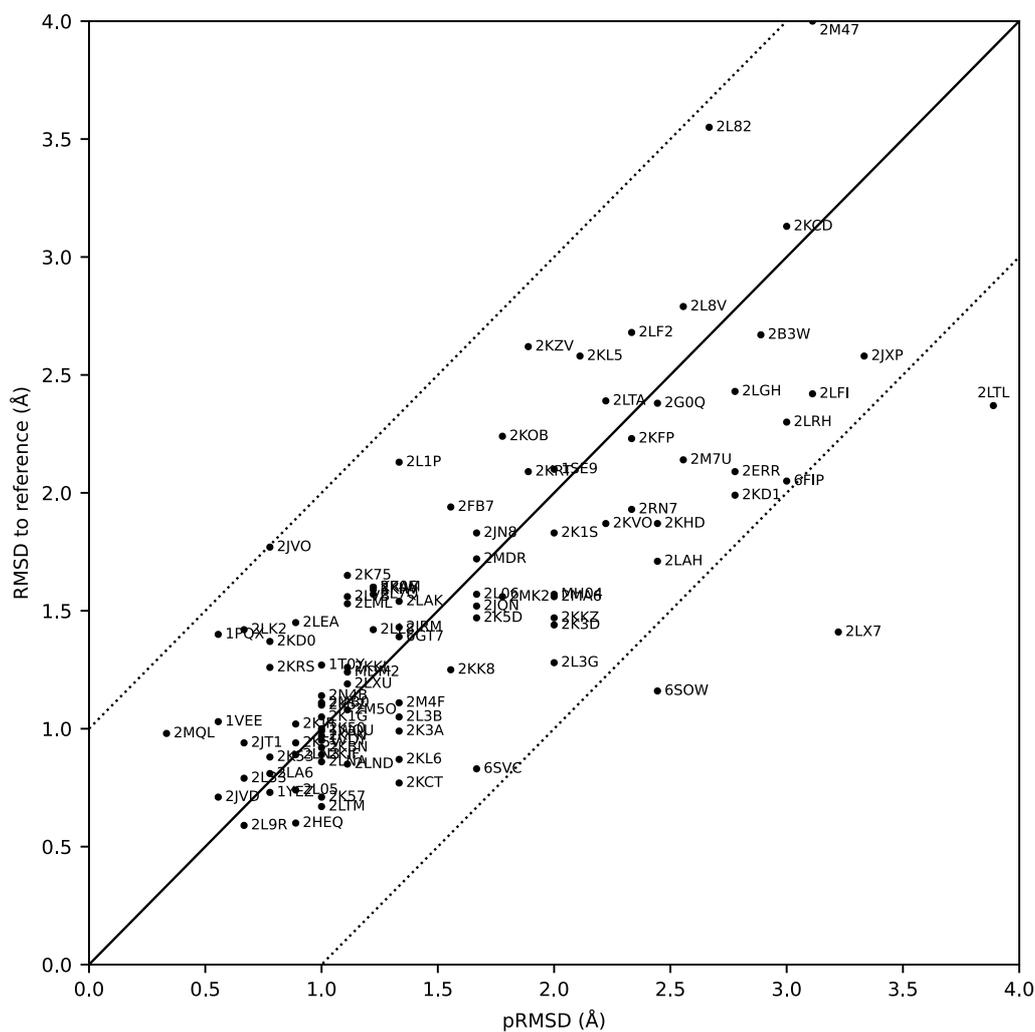

**Fig. 5** Actual and predicted RMSD between ARTINA and reference PDB structures. The predicted RMSD to reference (pRMSD) is calculated from the ARTINA results without knowledge of the reference PDB structure (see Methods) and, by definition, always in the range of 0–4 Å. For comparability, actual RMSD values to reference are also truncated at 4 Å (protein 2M47 with RMSD 4.47 Å). The dotted lines represent deviations of ±1 Å between the two RMSD quantities.

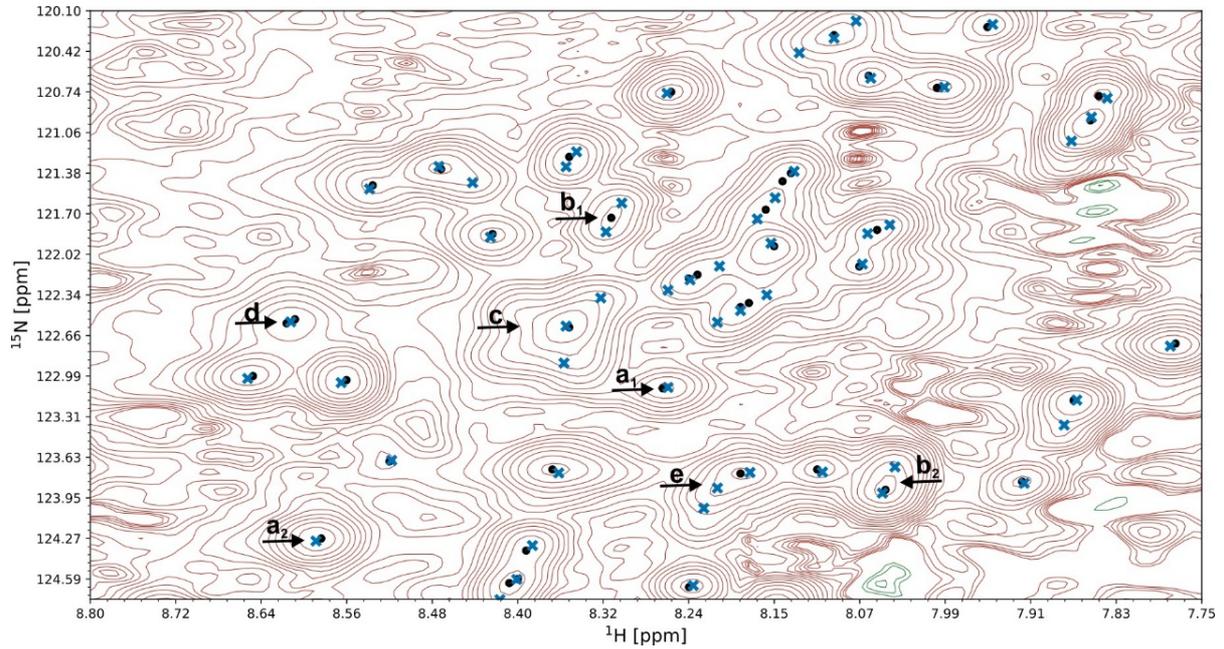

**Fig. 6 Commonly occurring challenges in visual spectrum analysis.** A fragment of a $^{15}$N-HSQC spectrum of the protein 1T0Y is shown. Initial signal positions identified by the peak picking model pp-ResNet (black dots) are deconvolved by deconv-ResNet, yielding the final coordinates used for automated assignment and structure determination (blue crosses). **$a_1$, $a_2$** Initial peak picking marker position is refined by the deconvolution model. **$b_1$, $b_2$** pp-ResNet output is deconvolved into two components. **c** The deconvolution model supports maximally 3 components per initial signal. **d** Two peak picking markers are merged by the deconvolution model. **e** Peak picking output deconvolved into 3 components.

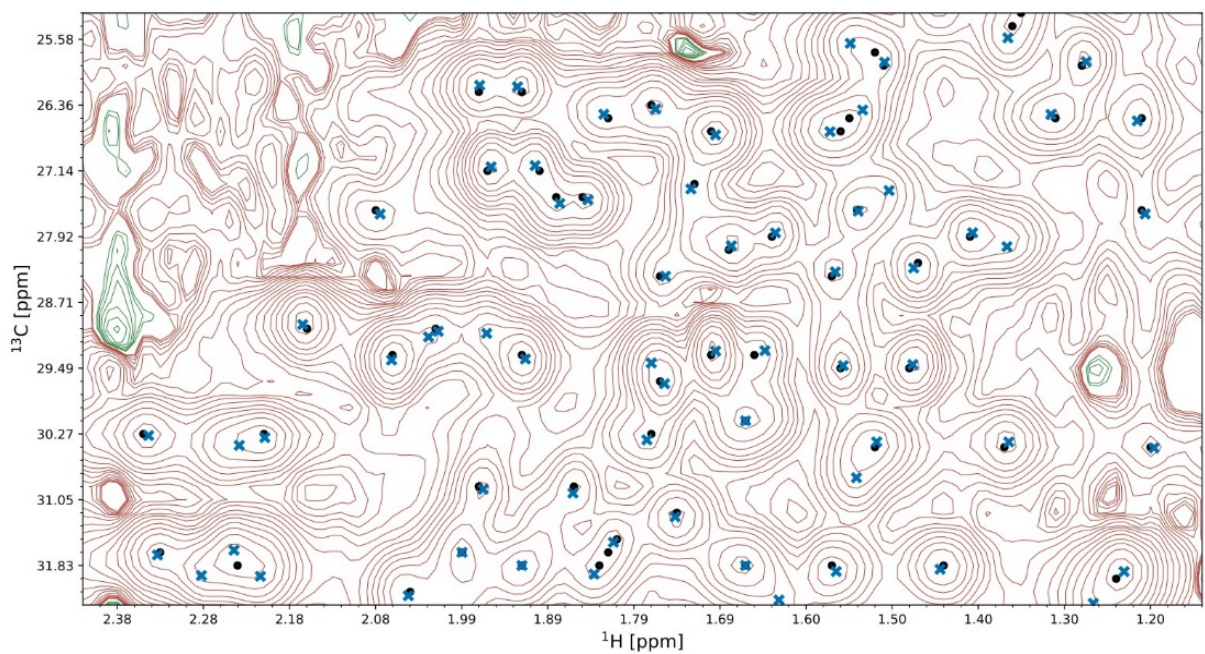

**Fig. 7 Performance of the peak picking model on a spectrum fragment with high peak overlap.** A fragment of the $^{13}$C-HSQC spectrum of protein 2K0M is shown. Initial signal positions identified by the peak picking model pp-ResNet (black dots) are deconvolved by deconv-ResNet, yielding the final coordinates used for automated assignment and structure determination (blue crosses).

**Table 1 Quality of assignments and structures in each refinement cycle.**

| | | Quantity | Refinement cycle | | |
|---|---|---|---|---|---|
| | | | 1 | 2 | 3 |
| **Chemical shift assignment** | **Initial** | Backbone assignment accuracy [%] | 96.12 | 96.19 | **96.34** |
| | | Side-chain assignment accuracy [%] | 84.90 | 86.83 | **86.95** |
| | | All-atom assignment accuracy [%] | 89.20 | 90.51 | **90.79** |
| | **Refined** | Backbone assignment accuracy [%] | 96.78 | 96.92 | **97.22** |
| | | Side-chain assignment accuracy [%] | 86.02 | 87.75 | **88.04** |
| | | All-atom assignment accuracy [%] | 90.17 | 91.31 | **91.36** |
| **Structure calculation** | | CYANA target function value [$Å^2$] | 4.53 | 4.61 | **4.03** |
| | | Backbone RMSD to reference [Å] | 1.56 | 1.52 | **1.44** |
| | | Heavy-atom RMSD to reference [Å] | 2.21 | 2.07 | **2.02** |
| | | Proteins with backbone RMSD to reference ≤ 1 Å [%] | 17 | 20 | 26 |
| | | Proteins with backbone RMSD to reference 1–2 Å [%] | 51 | 54 | 51 |
| | | Proteins with backbone RMSD to reference 2–3 Å [%] | 18 | 17 | 20 |
| | | Proteins with backbone RMSD to reference > 3 Å [%] | 14 | 9 | 3 |

Reported quantities (except for the RMSD distribution in the 4 bottom rows) are median values over the 100 proteins in the benchmark data set. The best metric value in each row is presented in bold. A refinement cycle is a single ARTINA iteration, composed of one execution of the chemical shift refinement cycle (comprising two FLYA[12] executions; rows initial and refined), and the structure refinement cycle (comprising 10 CYANA runs[22]).

**Supplementary information**

**Rapid protein assignments and structures from raw NMR spectra with the deep learning technique ARTINA**

Piotr Klukowski[1]*, Roland Riek[1]*, Peter Güntert[1,2,3]*

[1] Laboratory of Physical Chemistry, ETH Zurich, Vladimir-Prelog-Weg 2, 8093 Zurich, Switzerland

[2] Institute of Biophysical Chemistry, Goethe University Frankfurt, Max-von-Laue-Str. 9, 60438 Frankfurt am Main, Germany

[3] Department of Chemistry, Tokyo Metropolitan University, 1-1 Minami-Osawa, Hachioji, 192-0397 Tokyo, Japan

* e-mail: piotr.klukowski@phys.chem.ethz.ch; roland.riek@phys.chem.ethz.ch; peter.guentert@phys.chem.ethz.ch

## Contents

**Supplementary Fig. 1** Examples of generated spectrum fragments for deconvolution model training

**Supplementary Fig. 2** Visualization of 100 protein structures determined automatically with ARTINA overlaid with corresponding PDB depositions

**Supplementary Fig. 3** Robustness of peak picking to background artefacts

**Supplementary Fig. 4** The three proteins with backbone RMSD > 3 Å between the automatically determined structure and the PDB deposition

**Supplementary Table 1** Supported NMR spectrum types

**Supplementary Table 2** Proteins and spectra used for automated structure determination experiments

**Supplementary Table 3** Metadata for PDB reference structures

**Supplementary Table 4** Results of automated structure determination of 100 proteins

**Supplementary Table 5** Structure accuracy prediction

**Supplementary Table 6** ANSURR structure evaluation scores

**Supplementary Table 7** RPF structure evaluation scores

**Supplementary Table 8** Consensus structure bundles

**Supplementary Table 9** Results of restrained energy refinement

**Supplementary Table 10** Chemical shift assignment accuracy of protein core residues

**Supplementary Table 11** Quantitative analysis of sources of backbone RMSD in automatically determined structures

**Supplementary Table 12** Quality metrics of 26 protein structures determined without and with 4D CC-NOESY spectra

**Supplementary Table 13** Accuracy of automated chemical shift assignment using all input spectra

**Supplementary Table 14** Accuracy of automated chemical shift assignment using all except NOESY-type input spectra

**Supplementary Movie 1** Visualization of 100 automatically determined protein structures along with their PDB depositions

**Supplementary Movie 2** Visualization of automated visual spectrum analysis

**Supplementary Movie 3** Visualization of the ability of automated visual spectrum analysis to handle strong background artefacts

**Supplementary Movie 4** Visualization of automated visual spectrum analysis of an HCCH-TOCSY spectrum

**Supplementary Movie 5** Video tutorial presenting all steps of automated protein structure determination with ARTINA as available on the nmrtist.org website

**Supplementary Fig. 1 Examples of generated spectrum fragments for deconvolution model training.** Each row presents one generated spectrum fragment of size $64 \times 32 \times 5$ that was used for the deconvolution model training. Ground truth positions of signal components are marked by red crosses.

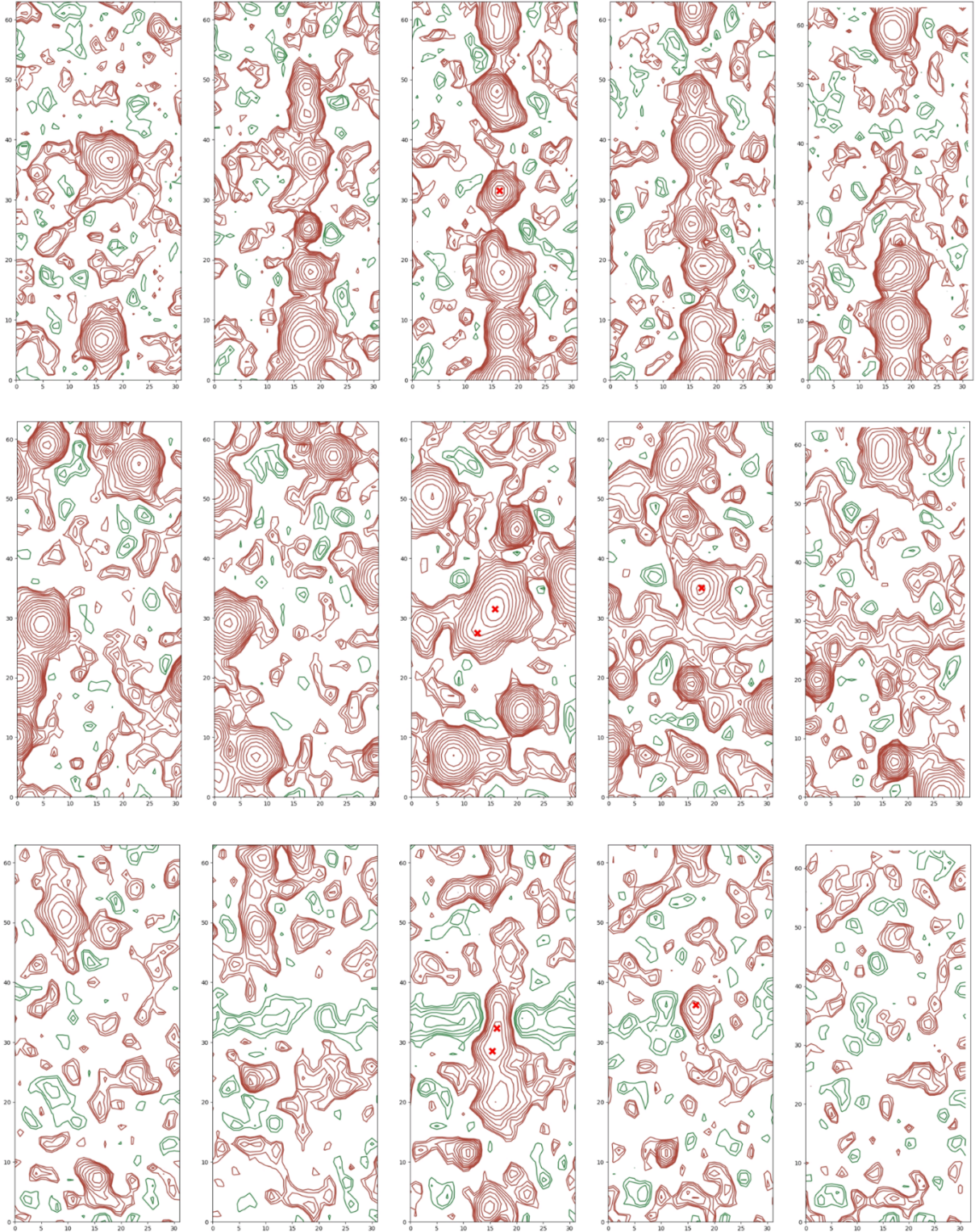

**Supplementary Fig. 2 Visualization of 100 protein structures determined with ARTINA (blue) overlaid with corresponding PDB depositions (orange).** Each panel contains information about PDB code, RMSD between ARTINA structure and PDB deposition, and backbone/side-chain chemical shift assignment accuracy. Well-defined and disordered regions (Supplementary Table 4) are shown in and strong and light colors, respectively.

2L9R (0.59Å; 97.93/90.44%)    2HEQ (0.60Å; 94.48/84.63%)    2LTM (0.67Å; 99.31/92.79%)    2JVD (0.71Å; 99.57/92.94%)

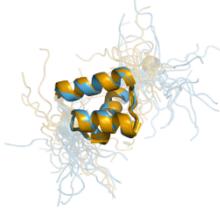 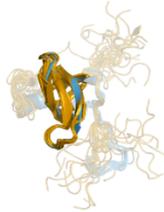 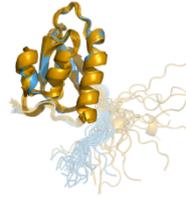 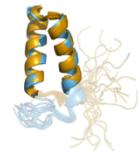

2K57 (0.71Å; 99.62/95.38%)    1YEZ (0.73Å; 99.06/90.95%)    2L05 (0.74Å; 99.50/91.56%)    2KCT (0.77Å; 96.59/93.42%)

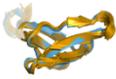 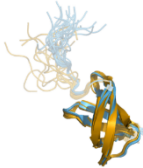 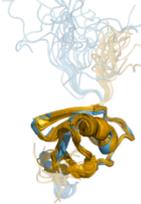 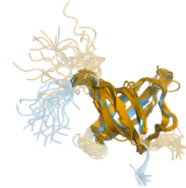

2L33 (0.79Å; 98.68/91.76%)    2LA6 (0.81Å; 98.78/92.60%)    6SVC (0.83Å; 97.52/83.97%)    2LND (0.85Å; 95.52/88.24%)

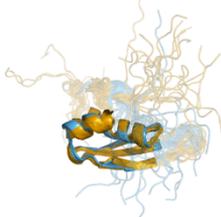 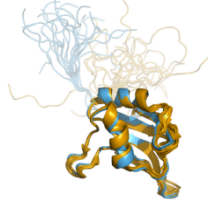 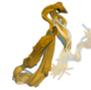 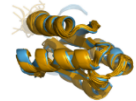

2LNA (0.86Å; 99.30/90.78%)    2KL6 (0.87Å; 98.64/93.52%)    2K53 (0.88Å; 98.53/95.30%)    2KIF (0.89Å; 94.43/87.78%)

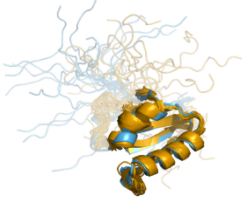 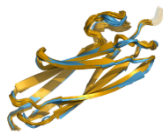 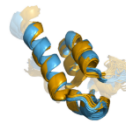 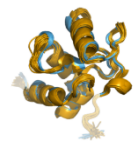

2LN3 (0.89Å; 98.37/92.52%)    2KBN (0.92Å; 97.98/88.49%)    2JT1 (0.94Å; 88.66/90.75%)    2K5V (0.94Å; 99.37/94.25%)

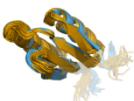 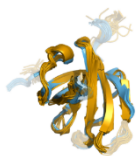 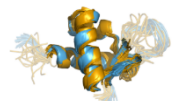 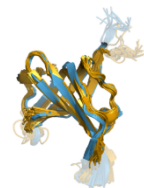

**Supplementary Fig. 2 (continued)**

1VDY (0.95Å; 95.48/88.37%)

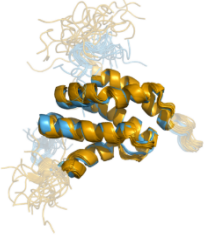

2KPN (0.97Å; 96.95/91.16%)

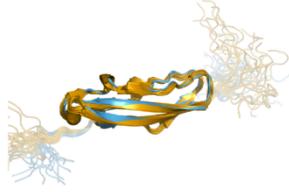

2MQL (0.98Å; 83.59/70.55%)

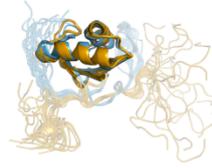

2K3A (0.99Å; 96.79/84.40%)

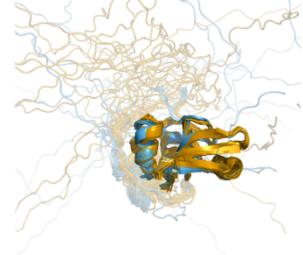

1WQU (0.99Å; 98.05/86.80%)

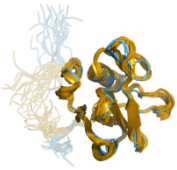

2K50 (1.00Å; 97.39/89.74%)

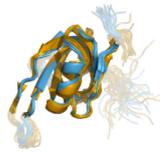

2KJR (1.02Å; 99.76/91.62%)

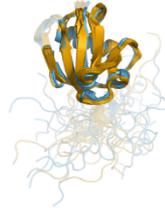

1VEE (1.03Å; 97.81/90.30%)

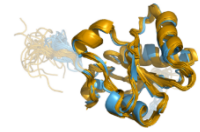

2K1G (1.05Å; 97.22/92.18%)

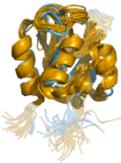

2L3B (1.05Å; 93.97/86.19%)

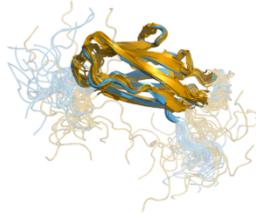

2M5O (1.08Å; 92.48/90.20%)

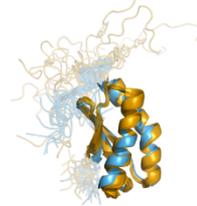

2K52 (1.10Å; 97.17/93.54%)

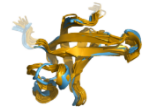

2M4F (1.11Å; 90.80/81.83%)

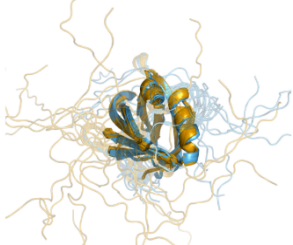

2MB0 (1.11Å; 97.69/88.46%)

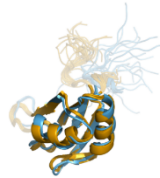

2N4B (1.14Å; 98.94/90.25%)

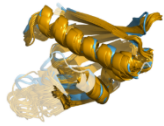

6SOW (1.16Å; 88.97/78.57%)

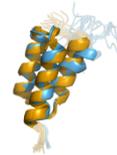

2LXU (1.19Å; 99.40/91.33%)

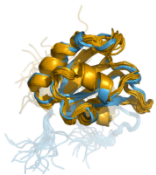

MDM2 (1.24Å; 98.08/84.88%)

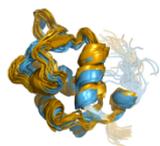

2KK8 (1.25Å; 99.18/90.26%)

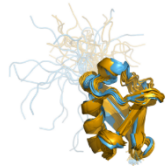

2KKL (1.26Å; 94.06/80.60%)

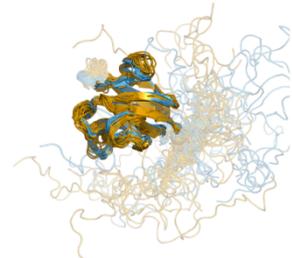

**Supplementary Fig. 2 (continued)**

2KRS (1.26Å; 98.47/96.18%)      1T0Y (1.27Å; 97.27/85.04%)      2L3G (1.28Å; 98.04/93.19%)      2KD0 (1.37Å; 97.24/91.96%)

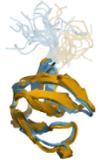 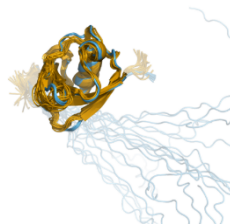 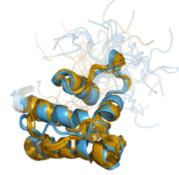 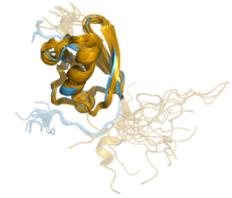

6GT7 (1.39Å; 96.93/79.31%)      1PQX (1.40Å; 99.03/86.51%)      2LX7 (1.41Å; 98.54/87.84%)      2LK2 (1.42Å; 96.08/86.08%)

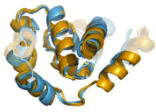 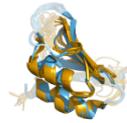 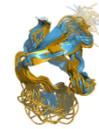 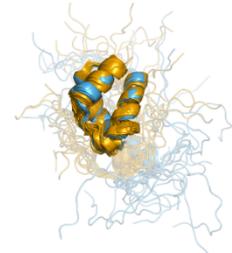

2LL8 (1.42Å; 98.04/91.42%)      2JRM (1.43Å; 97.21/92.14%)      2K3D (1.44Å; 98.96/89.76%)      2LEA (1.45Å; 96.17/83.33%)

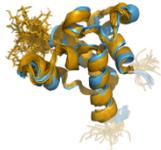 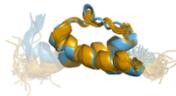 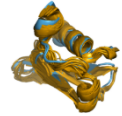 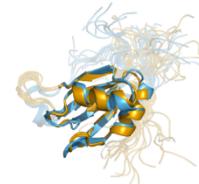

2K5D (1.47Å; 99.25/92.40%)      2KKZ (1.47Å; 96.50/87.83%)      2JQN (1.52Å; 97.81/90.68%)      2LML (1.53Å; 97.44/91.71%)

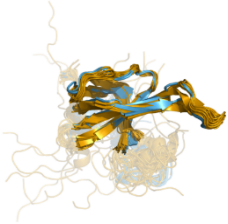 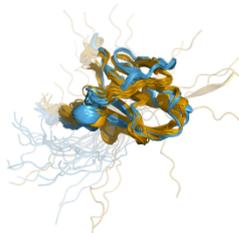 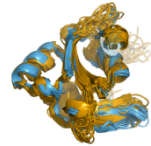 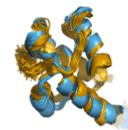

2LAK (1.54Å; 92.15/78.69%)      2LVB (1.56Å; 89.27/84.11%)      2MA6 (1.56Å; 98.94/89.36%)      2MK2 (1.56Å; 99.58/91.97%)

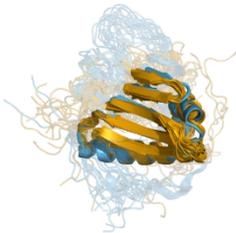 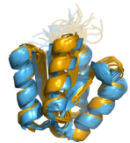 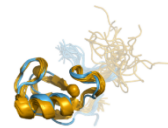 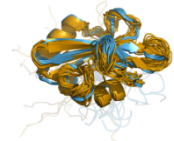

**Supplementary Fig. 2 (continued)**

2L06 (1.57Å; 96.46/84.28%)    2L7Q (1.57Å; 96.11/82.62%)    MH04 (1.57Å; 98.58/88.57%)    2KIW (1.59Å; 93.99/80.94%)

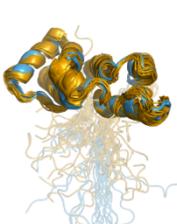 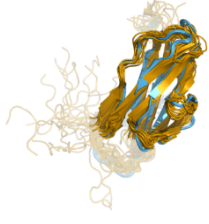 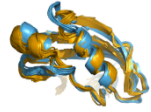 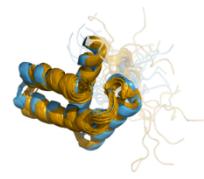

2K0M (1.60Å; 96.87/92.08%)    KRAS4B (1.60Å; 98.00/76.81%)    2K75 (1.65Å; 98.77/88.36%)    2LAH (1.71Å; 93.23/85.67%)

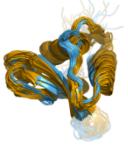 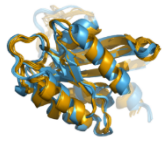 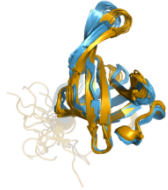 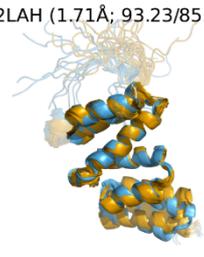

2MDR (1.72Å; 93.17/82.89%)    2JVO (1.77Å; 96.64/76.42%)    2JN8 (1.83Å; 95.45/89.68%)    2K1S (1.83Å; 98.83/92.02%)

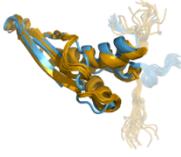 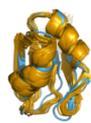 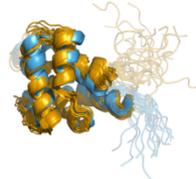 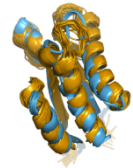

2KHD (1.87Å; 97.29/87.63%)    2KVO (1.87Å; 98.17/90.00%)    2RN7 (1.93Å; 94.35/81.02%)    2FB7 (1.94Å; 95.95/81.18%)

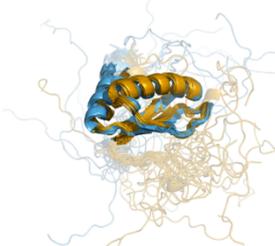 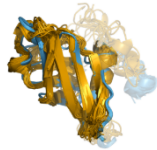 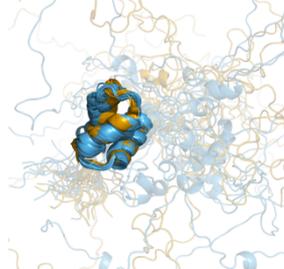 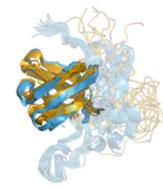

2KD1 (1.99Å; 98.31/90.36%)    6FIP (2.05Å; 97.86/79.15%)    2ERR (2.09Å; 96.36/78.40%)    2KRT (2.09Å; 97.01/82.72%)

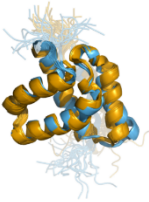 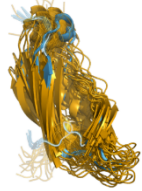 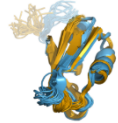 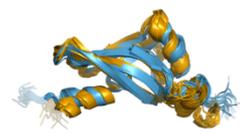

**Supplementary Fig. 2 (continued)**

1SE9 (2.10Å; 88.55/82.13%)  2L1P (2.13Å; 96.29/87.04%)  2M7U (2.14Å; 85.70/73.47%)  2KFP (2.23Å; 97.39/79.51%)

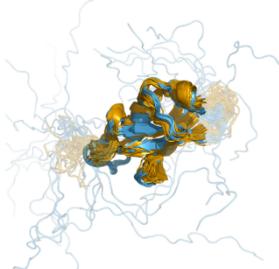 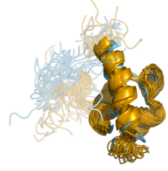 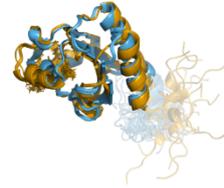 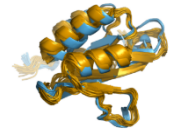

2KOB (2.24Å; 88.18/82.29%)  2LRH (2.30Å; 94.91/77.45%)  2LTL (2.37Å; 96.44/90.46%)  2G0Q (2.38Å; 93.99/84.33%)

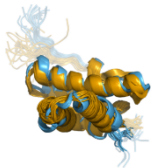 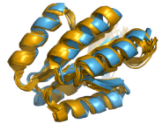 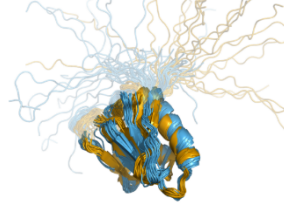 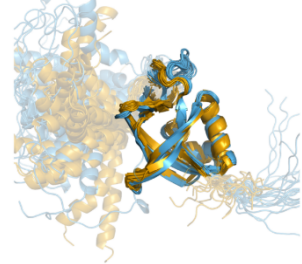

2LTA (2.39Å; 95.17/81.57%)  2LFI (2.42Å; 89.30/76.07%)  2LGH (2.43Å; 98.03/88.92%)  2JXP (2.58Å; 97.08/89.62%)

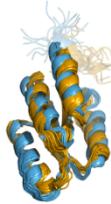 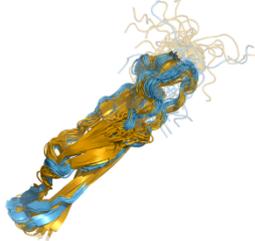 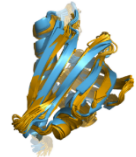 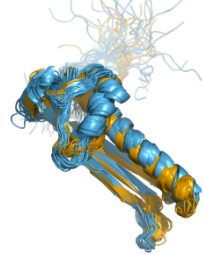

2KL5 (2.58Å; 79.41/68.59%)  2KZV (2.62Å; 95.84/81.67%)  2B3W (2.67Å; 93.39/80.09%)  2LF2 (2.68Å; 97.56/85.86%)

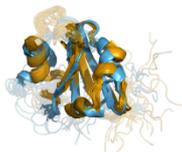 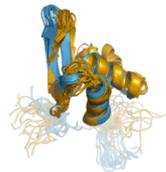 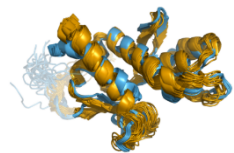 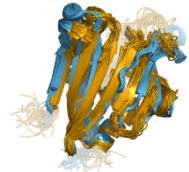

2L8V (2.79Å; 93.82/76.58%)  2KCD (3.13Å; 91.30/79.24%)  2L82 (3.55Å; 97.87/81.05%)  2M47 (4.72Å; 92.86/80.32%)

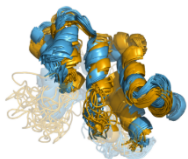 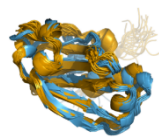 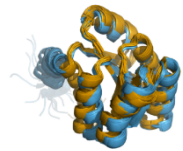 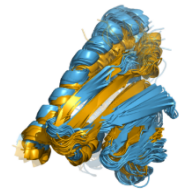

**Supplementary Fig. 3 Robustness of peak picking to background artefacts.** A fragment of four consecutive layers of a $^{13}$C-resolved [$^1$H, $^1$H]-NOESY spectrum of a 20% $^{13}$C-labelled protein of protein 6SOW[39] is shown. Final signal coordinates (after deconvolution) with classifier response > 0.5 are marked in black.

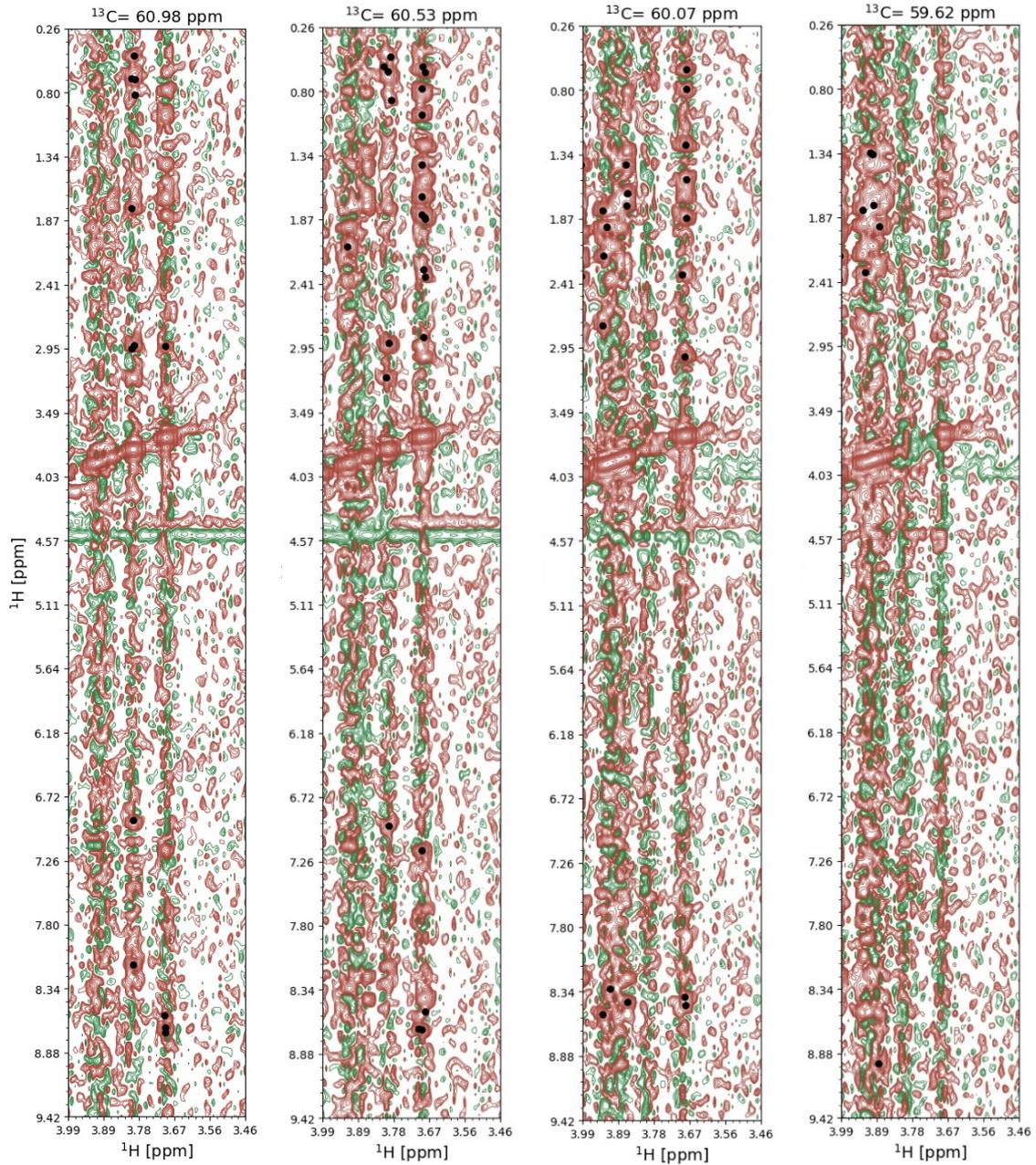

**Supplementary Fig. 4 The three proteins with backbone RMSD > 3 Å between the automatically determined structure (blue) and the PDB deposition (orange). a** Protein 2KCD. The α-helix at residues 105–109 is displaced. **b** Protein 2L82. The α-helix at residues 138–153 is displaced. **c** Protein 2M47. The main reason for the high RMSD is a displacement of the last two α-helices at residues 111–157.

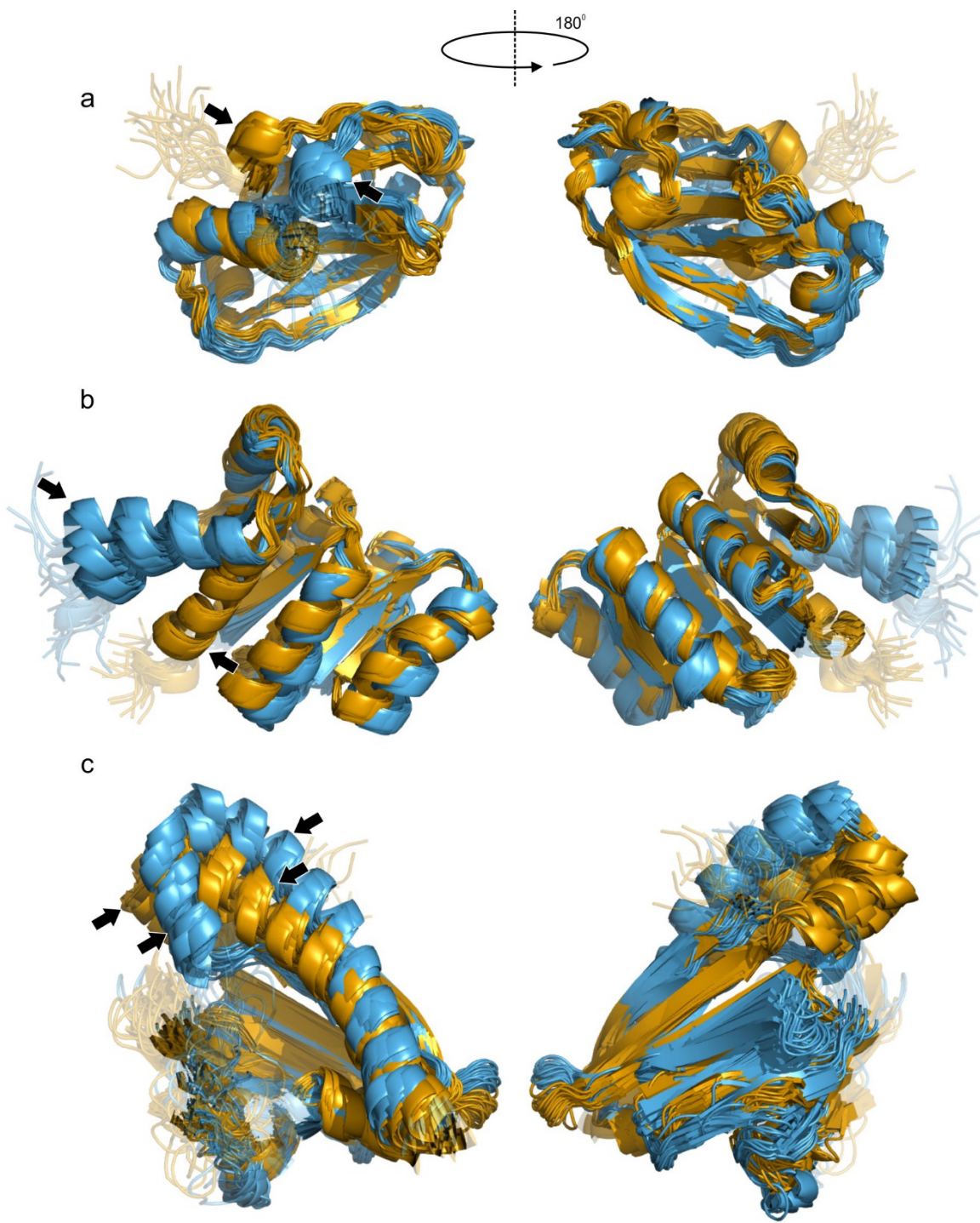

**Supplementary Table 1 Supported NMR spectrum types.**

| Spectrum category | Supported spectrum types |
|---|---|
| 2D | $^{13}$C-HSQC, $^{15}$N-HSQC, NOESY (homonuclear), TOCSY (homonuclear), CACO, HNHA, CBHEaro, CBHDaro |
| 3D for backbone assignment | HNCA, CBCANH, HNcoCA, HNCO, HNcaCO, CBCAcoNH, HBHAcoNH, CBCACOcaHA |
| 3D for side-chain assignment | HCCH-TOCSY, $^{15}$N-TOCSY, HCCH-COSY, CcoNH, HCcoNH, CCH-TOCSY |
| 3D/4D NOESY | 3D $^{13}$C-resolved [$^{1}$H,$^{1}$H]-NOESY, 3D $^{15}$N-resolved [$^{1}$H,$^{1}$H]-NOESY, 4D CC-NOESY |

Further information about spectra can be provided to ARTINA by the use of tags. For instance, the ALI or ARO tags can be used to indicate that a $^{13}$C-NOESY spectrum is expected to contain only aliphatic or aromatic signals. The tags POS and NEG can be used to restrict peak picking to positive and negative signals.

**Supplementary Table 2 Proteins and spectra used for automated structure determination experiments.**

| ID | Protein | Residues | Number of spectra | | | | ID | Protein | Residues | Number of spectra | | | |
|----|---------|----------|----------|------------|-------|-------|----|---------|----------|----------|------------|-------|-------|
| | | | Backbone | Side-chain | NOESY | Total | | | | Backbone | Side-chain | NOESY | Total |
| 1 | 6SVC | 35 | 2 | 1 | 3 | 6 | 51 | 2KIF | 108 | 1 | 1 | 3 | 5 |
| 2 | 2JVD | 54 | 6 | 3 | 2 | 11 | 52 | 2KBN | 109 | 6 | 5 | 4 | 15 |
| 3 | 2K57 | 55 | 6 | 6 | 2 | 14 | 53 | 2MK2 | 109 | 7 | 4 | 3 | 14 |
| 4 | 6SOW | 58 | 5 | 2 | 3 | 10 | 54 | 2K50 | 110 | 8 | 3 | 3 | 14 |
| 5 | 2LX7 | 60 | 7 | 7 | 4 | 18 | 55 | 2KL5 | 110 | 3 | 2 | 3 | 8 |
| 6 | 2MA6 | 61 | 5 | 7 | 4 | 16 | 56 | 2LTA | 110 | 5 | 2 | 3 | 10 |
| 7 | 2JRM | 65 | 5 | 4 | 3 | 12 | 57 | 2KIW | 111 | 7 | 7 | 4 | 18 |
| 8 | 1YEZ | 68 | 8 | 5 | 3 | 16 | 58 | 2LVB | 112 | 4 | 1 | 3 | 8 |
| 9 | 2L9R | 69 | 5 | 1 | 3 | 9 | 59 | 2LND | 112 | 5 | 3 | 3 | 11 |
| 10 | 2K52 | 74 | 7 | 4 | 3 | 14 | 60 | 1WQU | 114 | 8 | 6 | 2 | 16 |
| 11 | 2KRS | 74 | 5 | 7 | 4 | 16 | 61 | 2KL6 | 114 | 8 | 5 | 3 | 16 |
| 12 | 2K53 | 76 | 4 | 4 | 3 | 11 | 62 | 6GT7 | 115 | 5 | 3 | 3 | 11 |
| 13 | 2JT1 | 77 | 6 | 1 | 3 | 10 | 63 | 2JN8 | 115 | 7 | 8 | 3 | 18 |
| 14 | 2JVO | 77 | 4 | 4 | 2 | 10 | 64 | 2K5D | 116 | 7 | 4 | 3 | 14 |
| 15 | 2ERR | 81 | 3 | 2 | 2 | 7 | 65 | 2KD1 | 118 | 6 | 4 | 3 | 13 |
| 16 | 2L1P | 83 | 4 | 2 | 3 | 9 | 66 | 2LTL | 119 | 4 | 3 | 3 | 10 |
| 17 | 2LN3 | 83 | 5 | 3 | 3 | 11 | 67 | 2KVO | 120 | 7 | 7 | 4 | 18 |
| 18 | 2HEQ | 84 | 6 | 3 | 4 | 13 | 68 | 1T0Y | 120 | 6 | 5 | 3 | 14 |
| 19 | 2KK8 | 84 | 5 | 4 | 3 | 12 | 69 | 2KCD | 120 | 8 | 6 | 5 | 19 |
| 20 | 2KD0 | 85 | 6 | 4 | 3 | 13 | 70 | 2KRT | 121 | 7 | 5 | 3 | 15 |
| 21 | 2LML | 86 | 6 | 5 | 4 | 15 | 71 | 2LFI | 122 | 5 | 5 | 3 | 13 |
| 22 | 2K3D | 87 | 6 | 2 | 4 | 12 | 72 | 2JQN | 122 | 5 | 3 | 3 | 11 |
| 23 | 2LK2 | 89 | 7 | 6 | 4 | 17 | 73 | 2L7Q | 124 | 6 | 4 | 3 | 13 |
| 24 | MH04 | 90 | 7 | 3 | 2 | 12 | 74 | 2KFP | 125 | 7 | 6 | 4 | 17 |
| 25 | 1PQX | 91 | 4 | 5 | 3 | 12 | 75 | 1SE9 | 126 | 6 | 3 | 3 | 12 |
| 26 | 2L33 | 91 | 4 | 3 | 3 | 10 | 76 | 2L3G | 126 | 5 | 3 | 3 | 11 |
| 27 | 2KZV | 92 | 7 | 6 | 4 | 17 | 77 | 2L3B | 130 | 7 | 4 | 4 | 15 |
| 28 | 2KCT | 94 | 6 | 6 | 3 | 15 | 78 | 2LRH | 134 | 5 | 3 | 3 | 11 |
| 29 | 2MDR | 94 | 4 | 3 | 3 | 10 | 79 | 1VEE | 134 | 8 | 5 | 3 | 16 |
| 30 | 2FB7 | 95 | 5 | 4 | 2 | 11 | 80 | 2K1G | 136 | 7 | 7 | 3 | 17 |
| 31 | 2MB0 | 95 | 2 | 3 | 3 | 8 | 81 | 2KKZ | 140 | 7 | 5 | 3 | 15 |
| 32 | 2L05 | 95 | 6 | 4 | 3 | 13 | 82 | 1VDY | 140 | 8 | 7 | 2 | 17 |
| 33 | 2KJR | 95 | 5 | 4 | 3 | 12 | 83 | 2KKL | 140 | 7 | 7 | 4 | 18 |
| 34 | 2M5O | 97 | 5 | 1 | 3 | 9 | 84 | 2N4B | 142 | 6 | 5 | 4 | 15 |
| 35 | MDM2 | 97 | 3 | 2 | 2 | 7 | 85 | 2L8V | 143 | 7 | 4 | 4 | 15 |
| 36 | 2LNA | 99 | 6 | 7 | 4 | 17 | 86 | 2LGH | 144 | 7 | 6 | 4 | 17 |
| 37 | 2LA6 | 99 | 5 | 2 | 3 | 10 | 87 | 2K1S | 149 | 6 | 6 | 4 | 16 |
| 38 | 6FIP | 99 | 6 | 5 | 2 | 13 | 88 | 2M4F | 151 | 8 | 6 | 2 | 16 |
| 39 | 2LEA | 100 | 4 | 3 | 3 | 10 | 89 | 2JXP | 155 | 8 | 7 | 2 | 17 |
| 40 | 2LL8 | 101 | 6 | 6 | 3 | 15 | 90 | 2L06 | 155 | 6 | 6 | 3 | 15 |
| 41 | 2KPN | 103 | 7 | 4 | 3 | 14 | 91 | 2LAH | 160 | 4 | 3 | 3 | 10 |
| 42 | 2K0M | 104 | 7 | 4 | 3 | 14 | 92 | 2LAK | 160 | 7 | 6 | 4 | 17 |
| 43 | 2K5V | 104 | 8 | 4 | 3 | 15 | 93 | 2L82 | 162 | 4 | 2 | 3 | 9 |
| 44 | 2MQL | 105 | 2 | 4 | 3 | 9 | 94 | 2M47 | 163 | 7 | 7 | 3 | 17 |
| 45 | 2K75 | 106 | 6 | 6 | 3 | 15 | 95 | 2K3A | 163 | 8 | 6 | 3 | 17 |
| 46 | 2LTM | 107 | 5 | 3 | 3 | 11 | 96 | 2M7U | 165 | 5 | 3 | 3 | 11 |
| 47 | 2KOB | 108 | 7 | 5 | 4 | 16 | 97 | 2B3W | 168 | 8 | 7 | 5 | 20 |
| 48 | 2KHD | 108 | 7 | 7 | 3 | 17 | 98 | KRAS4B | 169 | 6 | 5 | 3 | 14 |
| 49 | 2RN7 | 108 | 5 | 4 | 3 | 12 | 99 | 2G0Q | 173 | 7 | 2 | 3 | 12 |
| 50 | 2LXU | 108 | 5 | 5 | 4 | 14 | 100 | 2LF2 | 175 | 7 | 7 | 4 | 18 |

**Supplementary Table 3 Metadata for PDB reference structures.** Extracted from PDB entries.

| | |
|---|---|
| Protein number | 1 |
| PDB code | 6SVC, doi:10.2210/pdb6SVC/pdb |
| BMRB code | 34432, doi:10.13018/BMR34432 |
| PDB Header | PEPTIDE BINDING PROTEIN |
| Protein name | PROTEIN ALLOSTERY OF WW DOMAIN AT ATOMIC RESOLUTION: APO STRUCTURE |
| Deposition date | 18.09.2019 |
| PDB title | PROTEIN ALLOSTERY OF THE WW DOMAIN AT ATOMIC RESOLUTION: APO STRUCTURE |
| PDB authors | D.STROTZ, J.ORTS, M.FRIEDMANN, P.GUNTERT, B.VOGELI, R.RIEK |
| Last author | RIEK |
| Reference | ANGEW.CHEM.INT.ED.ENGL. 59, 22132 (2020), doi:10.1002/ANIE.202008734 |
| Reference authors | D.STROTZ, J.ORTS, H.KADAVATH, M.FRIEDMANN, D.GHOSH, S.OLSSON, C.N.CHI, A.POKHARNA, P.GUNTERT, B.VOGELI, R.RIEK |
| Reference title | PROTEIN ALLOSTERY AT ATOMIC RESOLUTION |
| Software listed | CCPNMR, CYANA, NMRDRAW, NMRPIPE |
| Spectrometer | BRUKER (700 MHZ) |
| Protein number | 2 |
| PDB code | 2JVD, doi:10.2210/pdb2JVD/pdb |
| BMRB code | 15476, doi:10.13018/BMR15476 |
| PDB Header | STRUCTURAL GENOMICS, UNKNOWN FUNCTION |
| Protein name | FOLDED N-TERMINAL FRAGMENT OF UPF0291 PROTEIN YNZC FROM BACILLUS SUBTILIS |
| Deposition date | 18.09.2007 |
| PDB title | SOLUTION NMR STRUCTURE OF THE FOLDED N-TERMINAL FRAGMENT OF UPF0291 PROTEIN YNZC FROM BACILLUS SUBTILIS. NORTHEAST STRUCTURAL GENOMICS TARGET SR384-1-46 |
| PDB authors | J.M.ARAMINI, S.SHARMA, Y.J.HUANG, L.ZHAO, L.A.OWENS, K.STOKES, M.JIANG, R.XIAO, M.C.BARAN, G.V.T.SWAPNA, T.B.ACTON, G.T.MONTELIONE, NORTHEAST STRUCTURAL GENOMICS CONSORTIUM (NESG) |
| Last author | MONTELIONE |
| Reference | PROTEINS 72, 526 (2008), doi:10.1002/PROT.22064 |
| Reference authors | J.M.ARAMINI, S.SHARMA, Y.J.HUANG, G.V.SWAPNA, C.K.HO, K.SHETTY, K.CUNNINGHAM, L.C.MA, L.ZHAO, L.A.OWENS, M.JIANG, R.XIAO, J.LIU, M.C.BARAN, T.B.ACTON, B.ROST, G.T.MONTELIONE |
| Reference title | SOLUTION NMR STRUCTURE OF THE SOS RESPONSE PROTEIN YNZC FROM BACILLUS SUBTILIS |
| Software listed | AUTOASSIGN, AUTOSTRUCTURE, CYANA, MOLPROBITY, NMRPIPE, PDBSTAT, PROCHECK, PSVS, SPARKY, TOPSPIN, VNMR |
| Spectrometer | BRUKER, VARIAN (800 MHZ, 600 MHZ) |
| Protein number | 3 |
| PDB code | 2K57, doi:10.2210/pdb2K57/pdb |
| BMRB code | 15825, doi:10.13018/BMR15825 |
| PDB Header | STRUCTURAL GENOMICS, UNKNOWN FUNCTION |
| Protein name | PUTATIVE LIPOPROTEIN FROM PSEUDOMONAS SYRINGAE GENE LOCUS PSPTO2350 |
| Deposition date | 25.06.2008 |
| PDB title | SOLUTION NMR STRUCTURE OF PUTATIVE LIPOPROTEIN FROM PSEUDOMONAS SYRINGAE GENE LOCUS PSPTO2350. NORTHEAST STRUCTURAL GENOMICS TARGET PSR76A |
| PDB authors | D.HANG, J.A.ARAMINI, P.ROSSI, D.WANG, M.JIANG, M.MAGLAQUI, R.XIAO, J.LIU, M.C.BARAN, T.B.ACTON, B.ROST, G.T.MONTELIONE, NORTHEAST STRUCTURAL GENOMICS CONSORTIUM (NESG) |
| Last author | MONTELIONE |
| Reference | |
| Reference authors | |
| Reference title | |
| Software listed | AUTOASSIGN, CNS, CYANA, MOLMOL, MOLPROBITY, NMRPIPE, PDBSTAT, PROCHECK, PSVS, RPF(AUTOSTRUCTURE), SPARKY, TALOS, TOPSPIN |
| Spectrometer | BRUKER, VARIAN (800 MHZ, 600 MHZ) |
| Protein number | 4 |
| PDB code | 6SOW, doi:10.2210/pdb6SOW/pdb |
| BMRB code | 34430, doi:10.13018/BMR34430 |
| PDB Header | PROTEIN BINDING |
| Protein name | STAPHYLOCOCCAL PROTEIN A, C DOMAIN |
| Deposition date | 30.08.2019 |
| PDB title | NMR SOLUTION STRUCTURE OF STAPHYLOCOCCAL PROTEIN A, C DOMAIN |
| PDB authors | S.M.BACKLUND, H.IWAI |
| Last author | IWAI |
| Reference | MOLECULES 2, 6 (2021), doi:10.3390/MOLECULES26030747 |
| Reference authors | H.A.HEIKKINEN, S.M.BACKLUND, H.IWAI |
| Reference title | NMR STRUCTURE DETERMINATIONS OF SMALL PROTEINS USING ONLY ONE FRACTIONALLY 20% 13 C- AND UNIFORMLY 100% 15 N-LABELED SAMPLE |
| Software listed | AMBER, CCPNMR, CYANA, PSVS, TALOS |
| Spectrometer | BRUKER (850 MHZ) |
| Protein number | 5 |
| PDB code | 2LX7, doi:10.2210/pdb2LX7/pdb |
| **BMRB code** | 18662, doi:10.13018/BMR18662 |
| **PDB Header** | PROTEIN BINDING |
| **Protein name** | SH3 DOMAIN OF GROWTH ARREST-SPECIFIC PROTEIN 7 (GAS7) (FRAGMENT 1-60) FROM HOMO SAPIENS |
| **Deposition date** | 15.08.2012 |
| **PDB title** | SOLUTION NMR STRUCTURE OF SH3 DOMAIN OF GROWTH ARREST-SPECIFIC PROTEIN 7 (GAS7) (FRAGMENT 1-60) FROM HOMO SAPIENS, NORTHEAST STRUCTURAL GENOMICS CONSORTIUM (NESG) TARGET HR8574A |
| **PDB authors** | Y.YANG, T.A.RAMELOT, L.DAN, E.KOHAN, H.JANJUA, R.XIAO, T.ACTON, J.K.EVERETT, G.T.MONTELIONE, M.A.KENNEDY, NORTHEAST STRUCTURAL GENOMICS CONSORTIUM (NESG) |
| **Last author** | KENNEDY |
| **Reference** | |
| **Reference authors** | |
| **Reference title** | |
| **Software listed** | AUTOASSIGN, AUTOSTRUCTURE, CNS, CYANA, NMRPIPE, PINE, PSVS, SPARKY, TALOS+, TOPSPIN, VNMRJ |
| **Spectrometer** | BRUKER, VARIAN (850 MHZ, 600 MHZ) |
| **Protein number** | 6 |
| **PDB code** | 2MA6, doi:10.2210/pdb2MA6/pdb |
| **BMRB code** | 19329, doi:10.13018/BMR19329 |
| **PDB Header** | LIGASE |
| **Protein name** | RING FINGER DOMAIN FROM KIP1 UBIQUITINATION-PROMOTING E3 COMPLEX PROTEIN 1 (KPC1/RNF123) FROM HOMO SAPIENS |
| **Deposition date** | 28.06.2013 |
| **PDB title** | SOLUTION NMR STRUCTURE OF THE RING FINGER DOMAIN FROM THE KIP1 UBIQUITINATION-PROMOTING E3 COMPLEX PROTEIN 1 (KPC1/RNF123) FROM HOMO SAPIENS, NORTHEAST STRUCTURAL GENOMICS CONSORTIUM (NESG) TARGET HR8700A |
| **PDB authors** | T.A.RAMELOT, Y.YANG, H.JANJUA, E.KOHAN, H.WANG, R.XIAO, T.B.ACTON, J.K.EVERETT, G.T.MONTELIONE, M.A.KENNEDY, NORTHEAST STRUCTURAL GENOMICS CONSORTIUM (NESG) |
| **Last author** | KENNEDY |
| **Reference** | |
| **Reference authors** | |
| **Reference title** | |
| **Software listed** | AUTOSTRUCTURE, CNS, CYANA, FMCGUI, NMRPIPE, PINE, PSVS, SPARKY, TALOS+, TOPSPIN, VNMRJ |
| **Spectrometer** | BRUKER, VARIAN (850 MHZ, 600 MHZ) |
| **Protein number** | 7 |
| **PDB code** | 2JRM, doi:10.2210/pdb2JRM/pdb |
| **BMRB code** | 15339, doi:10.13018/BMR15339 |
| **PDB Header** | STRUCTURAL GENOMICS, UNKNOWN FUNCTION |
| **Protein name** | RIBOSOME MODULATION FACTOR VP1593 FROM VIBRIO PARAHAEMOLYTICUS |
| **Deposition date** | 27.06.2007 |
| **PDB title** | SOLUTION NMR STRUCTURE OF RIBOSOME MODULATION FACTOR VP1593 FROM VIBRIO PARAHAEMOLYTICUS. NORTHEAST STRUCTURAL GENOMICS TARGET VPR55 |
| **PDB authors** | Y.TANG, P.ROSSI, G.SWAPNA, H.WANG, M.JIANG, K.CUNNINGHAM, L.OWENS, L.MA, R.XIAO, J.LIU, M.AC.BARAN, T.B.ACTON, B.ROST, G.T.MONTELIONE, NORTHEAST STRUCTURAL GENOMICS CONSORTIUM (NESG) |
| **Last author** | MONTELIONE |
| **Reference** | |
| **Reference authors** | |
| **Reference title** | |
| **Software listed** | AUTOASSIGN, AUTOSTRUCTURE, CNS, CYANA, NMRPIPE, PDBSTAT, PSVS, SPARKY, TOPSPIN, X-PLOR |
| **Spectrometer** | BRUKER (800 MHZ, 600 MHZ) |
| **Protein number** | 8 |
| **PDB code** | 1YEZ, doi:10.2210/pdb1YEZ/pdb |
| **BMRB code** | 6505, doi:10.13018/BMR6505 |
| **PDB Header** | STRUCTURAL GENOMICS, UNKNOWN FUNCTION |
| **Protein name** | CONSERVED PROTEIN FROM GENE LOCUS MM1357 OF METHANOSARCINA MAZEI |
| **Deposition date** | 29.12.2004 |
| **PDB title** | SOLUTION STRUCTURE OF THE CONSERVED PROTEIN FROM THE GENE LOCUS MM1357 OF METHANOSARCINA MAZEI. NORTHEAST STRUCTURAL GENOMICS TARGET MAR30 |
| **PDB authors** | P.ROSSI, J.M.ARAMINI, G.V.T.SWAPNA, Y.P.HUANG, R.XIAO, C.K.HO, L.C.MA, T.B.ACTON, G.T.MONTELIONE, NORTHEAST STRUCTURAL GENOMICS CONSORTIUM (NESG) |
| **Last author** | MONTELIONE |
| **Reference** | |
| **Reference authors** | |
| **Reference title** | |
| **Software listed** | AUTOASSIGN, AUTOSTRUCTURE, NMRPIPE, SPARKY, VNMR, XWINNMR |
| **Spectrometer** | BRUKER, VARIAN (600 MHZ, 500 MHZ) |
| **Protein number** | 9 |
| **PDB code** | 2L9R, doi:10.2210/pdb2L9R/pdb |
| **BMRB code** | 17484, doi:10.13018/BMR17484 |
| **PDB Header** | TRANSCRIPTION |
| **Protein name** | HOMEOBOX DOMAIN OF HOMEOBOX PROTEIN NKX-3.1 FROM HOMO SAPIENS |

| Deposition date | 22.02.2011 |
|---|---|
| PDB title | SOLUTION NMR STRUCTURE OF HOMEOBOX DOMAIN OF HOMEOBOX PROTEIN NKX-3.1 FROM HOMO SAPIENS, NORTHEAST STRUCTURAL GENOMICS CONSORTIUM TARGET HR6470A |
| PDB authors | G.LIU, R.XIAO, H.-W.LEE, K.HAMILTON, C.CICCOSANTI, H.B.WANG, T.B.ACTON, J.K.EVERETT, Y.J.HUANG, G.T.MONTELIONE, NORTHEAST STRUCTURAL GENOMICS CONSORTIUM (NESG) |
| Last author | MONTELIONE |
| Reference | |
| Reference authors | |
| Reference title | |
| Software listed | AUTOASSIGN, AUTOSTRUCTURE, CNS, CYANA, NMRPIPE, TALOS+, TOPSPIN, VNMRJ, XEASY |
| Spectrometer | BRUKER, VARIAN (800 MHZ, 600 MHZ) |

| Protein number | 10 |
|---|---|
| PDB code | 2K52, doi:10.2210/pdb2K52/pdb |
| BMRB code | 15821, doi:10.13018/BMR15821 |
| PDB Header | STRUCTURAL GENOMICS, UNKNOWN FUNCTION |
| Protein name | UNCHARACTERIZED PROTEIN MJ1198 FROM METHANOCALDOCOCCUS JANNASCHII |
| Deposition date | 24.06.2008 |
| PDB title | STRUCTURE OF UNCHARACTERIZED PROTEIN MJ1198 FROM METHANOCALDOCOCCUS JANNASCHII. NORTHEAST STRUCTURAL GENOMICS TARGET MJR117B |
| PDB authors | P.ROSSI, M.MAGLAQUI, E.L.FOOTE, K.HAMILTON, C.CICCOSANTI, R.XIAO, R.NAIR, G.SWAPNA, J.K.EVERETT, T.B.ACTON, B.ROST, G.T.MONTELIONE, NORTHEAST STRUCTURAL GENOMICS CONSORTIUM (NESG) |
| Last author | MONTELIONE |
| Reference | |
| Reference authors | |
| Reference title | |
| Software listed | AUTOASSIGN, CNS, CYANA, MOLMOL, MOLPROBITY, NMRPIPE, PROCHECK, PSVS, RPF(AUTOSTRUCTURE), SPARKY, TALOS, TOPSPIN |
| Spectrometer | BRUKER (800 MHZ, 600 MHZ) |

| Protein number | 11 |
|---|---|
| PDB code | 2KRS, doi:10.2210/pdb2KRS/pdb |
| BMRB code | 16647, doi:10.13018/BMR16647 |
| PDB Header | STRUCTURAL GENOMICS, UNKNOWN FUNCTION |
| Protein name | SH3 DOMAIN FROM CPF_0587 (FRAGMENT 415-479) FROM CLOSTRIDIUM PERFRINGENS |
| Deposition date | 22.12.2009 |
| PDB title | SOLUTION NMR STRUCTURE OF SH3 DOMAIN FROM CPF_0587 (FRAGMENT 415-479) FROM CLOSTRIDIUM PERFRINGENS. NORTHEAST STRUCTURAL GENOMICS CONSORTIUM (NESG) TARGET CPR74A |
| PDB authors | T.A.RAMELOT, J.R.CORT, M.MAGLAQUI, C.CICCOSANTI, H.JANJUA, R.NAIR, B.ROST, T.B.ACTON, R.XIAO, J.K.EVERETT, G.T.MONTELIONE, M.A.KENNEDY, NORTHEAST STRUCTURAL GENOMICS CONSORTIUM (NESG) |
| Last author | KENNEDY |
| Reference | |
| Reference authors | |
| Reference title | |
| Software listed | AUTOASSIGN, AUTOSTRUCTURE, CNS, NMRPIPE, PDBSTAT, PSVS, SPARKY, TOPSPIN, VNMR, X-PLOR |
| Spectrometer | BRUKER, VARIAN (850 MHZ, 600 MHZ) |

| Protein number | 12 |
|---|---|
| PDB code | 2K53, doi:10.2210/pdb2K53/pdb |
| BMRB code | 15822, doi:10.13018/BMR15822 |
| PDB Header | STRUCTURAL GENOMICS, UNKNOWN FUNCTION |
| Protein name | A3DK08 PROTEIN FROM CLOSTRIDIUM THERMOCELLUM |
| Deposition date | 24.06.2008 |
| PDB title | NMR SOLUTION STRUCTURE OF A3DK08 PROTEIN FROM CLOSTRIDIUM THERMOCELLUM: NORTHEAST STRUCTURAL GENOMICS CONSORTIUM TARGET CMR9 |
| PDB authors | G.V.T.SWAPNA, W.HUANG, M.JIANG, E.L.FOOTE, R.XIAO, R.NAIR, J.EVERETT, T.B.ACTON, B.ROST, G.T.MONTELIONE, NORTHEAST STRUCTURAL GENOMICS CONSORTIUM (NESG) |
| Last author | MONTELIONE |
| Reference | |
| Reference authors | |
| Reference title | |
| Software listed | AUTOASSIGN, AUTOSTRUCTURE, CNS, CYANA |
| Spectrometer | BRUKER, VARIAN (800 MHZ, 600 MHZ) |

| Protein number | 13 |
|---|---|
| PDB code | 2JT1, doi:10.2210/pdb2JT1/pdb |
| BMRB code | 15386, doi:10.13018/BMR15386 |
| PDB Header | TRANSCRIPTION |
| Protein name | PEFI (PLASMID-ENCODED FIMBRIAE REGULATORY) PROTEIN FROM SALMONELLA TYPHIMURIUM |
| Deposition date | 17.07.2007 |
| PDB title | SOLUTION NMR STRUCTURE OF PEFI (PLASMID-ENCODED FIMBRIAE REGULATORY) PROTEIN FROM SALMONELLA TYPHIMURIUM. NORTHEAST STRUCTURAL GENOMICS TARGET STR82 |

| PDB authors | J.M.ARAMINI, P.ROSSI, H.WANG, C.NWOSU, K.CUNNINGHAM, L.-C.MA, R.XIAO, J.LIU, M.C.BARAN, G.V.T.SWAPNA, T.B.ACTON, B.ROST, G.T.MONTELIONE, NORTHEAST STRUCTURAL GENOMICS CONSORTIUM (NESG) |
|---|---|
| Last author | MONTELIONE |
| Reference | PROTEINS 79, 335 (2011), doi:10.1002/PROT.22869 |
| Reference authors | J.M.ARAMINI, P.ROSSI, J.R.CORT, L.C.MA, R.XIAO, T.B.ACTON, G.T.MONTELIONE |
| Reference title | SOLUTION NMR STRUCTURE OF THE PLASMID-ENCODED FIMBRIAE REGULATORY PROTEIN PEFI FROM SALMONELLA ENTERICA SEROVAR TYPHIMURIUM |
| Software listed | AUTOASSIGN, AUTOSTRUCTURE, CNS, CYANA, MOLMOL, MOLPROBITY, NMRPIPE, PDBSTAT, PROCHECK, PSVS, SPARKY, TOPSPIN |
| Spectrometer | BRUKER (800 MHZ, 600 MHZ) |

| Protein number | 14 |
|---|---|
| PDB code | 2JVO, doi:10.2210/pdb2JVO/pdb |
| BMRB code | 15485, doi:10.13018/BMR15485 |
| PDB Header | RNA BINDING PROTEIN |
| Protein name | SEGMENTAL ISOTOPE LABELING OF NPL3 |
| Deposition date | 24.09.2007 |
| PDB title | SEGMENTAL ISOTOPE LABELING OF NPL3 |
| PDB authors | L.SKRISOVSKA, F.H.-T.ALLAIN |
| Last author | ALLAIN |
| Reference | J.MOL.BIOL. 375, 151 (2008), doi:10.1016/J.JMB.2007.09.030 |
| Reference authors | L.SKRISOVSKA, F.H.ALLAIN |
| Reference title | IMPROVED SEGMENTAL ISOTOPE LABELING METHODS FOR THE NMR STUDY OF MULTIDOMAIN OR LARGE PROTEINS: APPLICATION TO THE RRMS OF NPL3P AND HNRNP L |
| Software listed | ATNOS/CANDID, SPARKY, XWINNMR |
| Spectrometer | BRUKER (900 MHZ, 600 MHZ, 500 MHZ) |

| Protein number | 15 |
|---|---|
| PDB code | 2ERR, doi:10.2210/pdb2ERR/pdb |
| BMRB code | 6895, doi:10.13018/BMR6895 |
| PDB Header | RNA BINDING PROTEIN |
| Protein name | RNA BINDING DOMAIN OF HUMAN FOX-1 IN COMPLEX WITH UGCAUGU |
| Deposition date | 25.10.2005 |
| PDB title | NMR STRUCTURE OF THE RNA BINDING DOMAIN OF HUMAN FOX-1 IN COMPLEX WITH UGCAUGU |
| PDB authors | F.H.ALLAIN, S.D.AUWETER |
| Last author | AUWETER |
| Reference | EMBO J. 25, 163 (2006), doi:10.1038/SJ.EMBOJ.7600918 |
| Reference authors | S.D.AUWETER, R.FASAN, L.REYMOND, J.G.UNDERWOOD, D.L.BLACK, S.PITSCH, F.H.ALLAIN |
| Reference title | MOLECULAR BASIS OF RNA RECOGNITION BY THE HUMAN ALTERNATIVE SPLICING FACTOR FOX-1 |
| Software listed | AMBER, CYANA |
| Spectrometer | BRUKER (900 MHZ, 600 MHZ, 500 MHZ) |

| Protein number | 16 |
|---|---|
| PDB code | 2L1P, doi:10.2210/pdb2L1P/pdb |
| BMRB code | 17092, doi:10.13018/BMR17092 |
| PDB Header | DNA BINDING PROTEIN |
| Protein name | N-TERMINAL DOMAIN OF DNA-BINDING PROTEIN SATB1 FROM HOMO SAPIENS |
| Deposition date | 02.08.2010 |
| PDB title | NMR SOLUTION STRUCTURE OF THE N-TERMINAL DOMAIN OF DNA-BINDING PROTEIN SATB1 FROM HOMO SAPIENS: NORTHEAST STRUCTURAL GENOMICS TARGET HR4435B(179-250) |
| PDB authors | G.V.T.SWAPNA, A.F.MONTELIONE, R.SHASTRY, C.CICCOSANTI, H.JANJUA, R.XIAO, T.B.ACTON, J.K.EVERETT, G.T.MONTELIONE, NORTHEAST STRUCTURAL GENOMICS CONSORTIUM (NESG) |
| Last author | MONTELIONE |
| Reference | |
| Reference authors | |
| Reference title | |
| Software listed | AUTOASSIGN, AUTOSTRUCTURE, CNS, CYANA |
| Spectrometer | BRUKER (800 MHZ, 600 MHZ) |

| Protein number | 17 |
|---|---|
| PDB code | 2LN3, doi:10.2210/pdb2LN3/pdb |
| BMRB code | 18145, doi:10.13018/BMR18145 |
| PDB Header | DE NOVO PROTEIN |
| Protein name | DE NOVO DESIGNED PROTEIN, IF3-LIKE FOLD |
| Deposition date | 15.12.2011 |
| PDB title | SOLUTION NMR STRUCTURE OF DE NOVO DESIGNED PROTEIN, IF3-LIKE FOLD, NORTHEAST STRUCTURAL GENOMICS CONSORTIUM TARGET OR135 (CASD TARGET) |
| PDB authors | G.LIU, R.KOGA, N.KOGA, R.XIAO, H.LEE, H.JANJUA, E.KOHAN, T.B.ACTON, J.K.EVERETT, D.BAKER, G.T.MONTELIONE, NORTHEAST STRUCTURAL GENOMICS CONSORTIUM (NESG) |
| Last author | MONTELIONE |
| Reference | NATURE 491, 222 (2012), doi:10.1038/NATURE11600 |
| Reference authors | N.KOGA, R.TATSUMI-KOGA, G.LIU, R.XIAO, T.B.ACTON, G.T.MONTELIONE, D.BAKER |
| Reference title | PRINCIPLES FOR DESIGNING IDEAL PROTEIN STRUCTURES |

| Software listed | AUTOASSIGN, AUTOSTRUCTURE, CNS, CYANA, NMRPIPE, REDCAT, SPARKY, TALOS+, TOPSPIN, VNMRJ, XEASY |
|---|---|
| Spectrometer | BRUKER, VARIAN (800 MHZ, 600 MHZ) |

| Protein number | 18 |
|---|---|
| PDB code | 2HEQ, doi:10.2210/pdb2HEQ/pdb |
| BMRB code | 7175, doi:10.13018/BMR7175 |
| PDB Header | STRUCTURAL GENOMICS, UNKNOWN FUNCTION |
| Protein name | BACILLUS SUBTILIS PROTEIN YORP |
| Deposition date | 21.06.2006 |
| PDB title | NMR STRUCTURE OF BACILLUS SUBTILIS PROTEIN YORP, NORTHEAST STRUCTURAL GENOMICS TARGET SR399 |
| PDB authors | T.A.RAMELOT, J.R.CORT, D.WANG, H.JANJUA, K.CUNNINGHAM, L.-C.MA, R.XIAO, J.LIU, M.BARAN, G.V.T.SWAPNA, T.B.ACTON, B.ROST, G.M.MONTELIONE, M.A.KENNEDY, NORTHEAST STRUCTURAL GENOMICS CONSORTIUM (NESG) |
| Last author | KENNEDY |
| Reference | |
| Reference authors | |
| Reference title | |
| Software listed | AUTOSTRUCTURE, CNS, NMRPIPE, SPARKY, VNMR, X-PLOR_NIH |
| Spectrometer | VARIAN (750 MHZ, 600 MHZ) |

| Protein number | 19 |
|---|---|
| PDB code | 2KK8, doi:10.2210/pdb2KK8/pdb |
| BMRB code | 16355, doi:10.13018/BMR16355 |
| PDB Header | STRUCTURAL GENOMICS, UNKNOWN FUNCTION |
| Protein name | A PUTATIVE UNCHARACTERIZED PROTEIN OBTAINED FROM ARABIDOPSIS THALIANA |
| Deposition date | 16.06.2009 |
| PDB title | NMR SOLUTION STRUCTURE OF A PUTATIVE UNCHARACTERIZED PROTEIN OBTAINED FROM ARABIDOPSIS THALIANA: NORTHEAST STRUCTURAL GENOMICS CONSORTIUM TARGET AR3449A |
| PDB authors | R.MANI, S.V.T.GURLA, R.SHASTRY, C.CICCOSANTI, E.FOOTE, M.JIANG, R.XIAO, R.NAIR, J.EVERETT, Y.HUANG, T.ACTON, B.ROST, G.T.MONTELIONE, NORTHEAST STRUCTURAL GENOMICS CONSORTIUM (NESG) |
| Last author | MONTELIONE |
| Reference | |
| Reference authors | |
| Reference title | |
| Software listed | AUTOASSIGN, AUTOSTRUCTURE, CNS, CYANA |
| Spectrometer | BRUKER, VARIAN (800 MHZ, 600 MHZ, 500 MHZ) |

| Protein number | 20 |
|---|---|
| PDB code | 2KD0, doi:10.2210/pdb2KD0/pdb |
| BMRB code | 16101, doi:10.13018/BMR16101 |
| PDB Header | SIGNALING PROTEIN |
| Protein name | O64736 PROTEIN FROM ARABIDOPSIS THALIANA |
| Deposition date | 31.12.2008 |
| PDB title | NMR SOLUTION STRUCTURE OF O64736 PROTEIN FROM ARABIDOPSIS THALIANA. NORTHEAST STRUCTURAL GENOMICS CONSORTIUM MEGA TARGET AR3445A |
| PDB authors | G.V.T.SWAPNA, R.SHASTRY, E.FOOTE, C.CICCOSANTI, M.JIANG, R.XIAO, R.NAIR, J.EVERETT, Y.HUANG, T.B.ACTON, B.ROST, G.T.MONTELIONE, NORTHEAST STRUCTURAL GENOMICS CONSORTIUM (NESG) |
| Last author | MONTELIONE |
| Reference | |
| Reference authors | |
| Reference title | |
| Software listed | AUTOASSIGN, AUTOSTRUCTURE, CNS, CYANA, NMRPIPE, SPARKY |
| Spectrometer | BRUKER (800 MHZ, 600 MHZ) |

| Protein number | 21 |
|---|---|
| PDB code | 2LML, doi:10.2210/pdb2LML/pdb |
| BMRB code | 16860, doi:10.13018/BMR16860 |
| PDB Header | TRANSPORT PROTEIN |
| Protein name | HOLO ACYL CARRIER PROTEIN FROM GEOBACTER METALLIREDUCENS REFINED WITH NH RDCS |
| Deposition date | 05.12.2011 |
| PDB title | SOLUTION NMR STRUCTURE OF HOLO ACYL CARRIER PROTEIN FROM GEOBACTER METALLIREDUCENS REFINED WITH NH RDCS, NORTHEAST STRUCTURAL GENOMICS CONSORTIUM TARGET GMR141 |
| PDB authors | T.A.RAMELOT, M.J.SMOLA, H.LEE, L.ZHAO, C.CICCOSANTI, E.L.FOOTE, K.HAMILTON, R.NAIR, B.ROST, G.SWAPNA, T.B.ACTON, J.K.EVERETT, J.H.PRESTEGARD, G.T.MONTELIONE, M.A.KENNEDY, NORTHEAST STRUCTURAL GENOMICS CONSORTIUM (NESG) |
| Last author | KENNEDY |
| Reference | BIOCHEMISTRY 50, 1442 (2011), doi:10.1021/BI101932S |
| Reference authors | T.A.RAMELOT, M.J.SMOLA, H.W.LEE, C.CICCOSANTI, K.HAMILTON, T.B.ACTON, R.XIAO, J.K.EVERETT, J.H.PRESTEGARD, G.T.MONTELIONE, M.A.KENNEDY |
| Reference title | SOLUTION STRUCTURE OF 4'-PHOSPHOPANTETHEINE - GMACP3 FROM GEOBACTER METALLIREDUCENS: A SPECIALIZED ACYL CARRIER PROTEIN WITH ATYPICAL STRUCTURAL FEATURES AND A PUTATIVE ROLE IN LIPOPOLYSACCHARIDE BIOSYNTHESIS |
| Software listed | AUTOASSIGN, AUTOSTRUCTURE, CNS, CYANA, FMCGUI, NMRPIPE, PDBSTAT, PINE_SERVER, PSVS, SPARKY, TOPSPIN, VNMR, X-PLOR_NIH |
| Spectrometer | BRUKER, VARIAN (850 MHZ, 600 MHZ) |

| Protein number | 22 |
|---|---|
| PDB code | 2K3D, doi:10.2210/pdb2K3D/pdb |
| BMRB code | 15750, doi:10.13018/BMR15750 |
| PDB Header | STRUCTURAL GENOMICS, UNKNOWN FUNCTION |
| Protein name | FOLDED 79 RESIDUE FRAGMENT OF LIN0334 FROM LISTERIA INNOCUA |
| Deposition date | 02.05.2008 |
| PDB title | SOLUTION NMR STRUCTURE OF THE FOLDED 79 RESIDUE FRAGMENT OF LIN0334 FROM LISTERIA INNOCUA. NORTHEAST STRUCTURAL GENOMICS CONSORTIUM TARGET LKR15 |
| PDB authors | T.A.RAMELOT, L.ZHAO, M.JIANG, E.L.FOOTE, R.XIAO, J.LIU, M.C.BARAN, G.V.T.SWAPNA, T.B.ACTON, B.ROST, G.T.MONTELIONE, M.A.KENNEDY, NORTHEAST STRUCTURAL GENOMICS CONSORTIUM (NESG) |
| Last author | KENNEDY |
| Reference | |
| Reference authors | |
| Reference title | |
| Software listed | AUTOASSIGN, AUTOSTRUCTURE, NMRPIPE, PSVS, SPARKY, TOPSPIN, VNMR, X-PLOR |
| Spectrometer | BRUKER, VARIAN (850 MHZ, 600 MHZ) |

| Protein number | 23 |
|---|---|
| PDB code | 2LK2, doi:10.2210/pdb2LK2/pdb |
| BMRB code | 17971, doi:10.13018/BMR17971 |
| PDB Header | TRANSCRIPTION |
| Protein name | HOMEOBOX DOMAIN (171-248) OF HUMAN HOMEOBOX PROTEIN TGIF1 |
| Deposition date | 30.09.2011 |
| PDB title | SOLUTION NMR STRUCTURE OF HOMEOBOX DOMAIN (171-248) OF HUMAN HOMEOBOX PROTEIN TGIF1, NORTHEAST STRUCTURAL GENOMICS CONSORTIUM TARGET HR4411B |
| PDB authors | Y.YANG, T.A.RAMELOT, J.R.CORT, R.SHASTRY, C.CICCOSANTI, K.HAMILTON, T.B.ACTON, R.XIAO, J.K.EVERETT, G.T.MONTELIONE, M.A.KENNEDY, NORTHEAST STRUCTURAL GENOMICS CONSORTIUM (NESG) |
| Last author | KENNEDY |
| Reference | |
| Reference authors | |
| Reference title | |
| Software listed | ASDP, AUTOASSIGN, CNS, CYANA, NMRPIPE, PDBSTAT, PINE_SERVER, PSVS, SPARKY, TOPSPIN, VNMR, X-PLOR_NIH |
| Spectrometer | BRUKER, VARIAN (850 MHZ, 600 MHZ) |

| Protein number | 24 |
|---|---|
| PDB code | (MH04) |
| BMRB code | |
| PDB Header | |
| Protein name | |
| Deposition date | |
| PDB title | |
| PDB authors | |
| Last author | |
| Reference | |
| Reference authors | |
| Reference title | |
| Software listed | |
| Spectrometer | |

| Protein number | 25 |
|---|---|
| PDB code | 1PQX, doi:10.2210/pdb1PQX/pdb |
| BMRB code | 5844, doi:10.13018/BMR5844 |
| PDB Header | STRUCTURAL GENOMICS, UNKNOWN FUNCTION |
| Protein name | STAPHYLOCOCCUS AUREUS PROTEIN SAV1430 |
| Deposition date | 19.06.2003 |
| PDB title | SOLUTION NMR STRUCTURE OF STAPHYLOCOCCUS AUREUS PROTEIN SAV1430. NORTHEAST STRUCTURAL GENOMICS CONSORTIUM TARGET ZR18 |
| PDB authors | M.C.BARAN, J.M.ARAMINI, R.XIAO, Y.J.HUANG, T.B.ACTON, L.SHIH, G.T.MONTELIONE, NORTHEAST STRUCTURAL GENOMICS CONSORTIUM (NESG) |
| Last author | MONTELIONE |
| Reference | |
| Reference authors | |
| Reference title | |
| Software listed | AUTOASSIGN, AUTOPROC, AUTOSTRUCTURE, NMRPIPE, SPARKY, VNMR, X-PLOR |
| Spectrometer | VARIAN (600 MHZ, 500 MHZ) |

| Protein number | 26 |
|---|---|
| PDB code | 2L33, doi:10.2210/pdb2L33/pdb |
| BMRB code | 17169, doi:10.13018/BMR17169 |
| PDB Header | TRANSCRIPTION REGULATOR |
| Protein name | DRBM 2 DOMAIN OF INTERLEUKIN ENHANCER- BINDING FACTOR 3 FROM HOMO SAPIENS |
| Deposition date | 03.09.2010 |
| PDB title | SOLUTION NMR STRUCTURE OF DRBM 2 DOMAIN OF INTERLEUKIN ENHANCER- BINDING FACTOR 3 FROM HOMO SAPIENS, NORTHEAST STRUCTURAL GENOMICS CONSORTIUM TARGET HR4527E |

| PDB authors | G.LIU, H.JANJUA, R.XIAO, T.B.ACTON, A.CICCOSANTI, R.B.SHASTRY, J.EVERETT, G.T.MONTELIONE, NORTHEAST STRUCTURAL GENOMICS CONSORTIUM, NORTHEAST STRUCTURAL GENOMICS CONSORTIUM (NESG) |
|---|---|
| Last author | MONTELIONE |
| Reference | |
| Reference authors | |
| Reference title | |
| Software listed | AUTOASSIGN, AUTOSTRUCTURE, CNS, CYANA, NMRPIPE, TALOS+, TOPSPIN, VNMRJ, XEASY |
| Spectrometer | BRUKER, VARIAN (800 MHZ, 600 MHZ) |

| Protein number | 27 |
|---|---|
| PDB code | 2KZV, doi:10.2210/pdb2KZV/pdb |
| BMRB code | 17020, doi:10.13018/BMR17020 |
| PDB Header | STRUCTURAL GENOMICS, UNKNOWN FUNCTION |
| Protein name | CV_0373(175-257) PROTEIN FROM CHROMOBACTERIUM VIOLACEUM |
| Deposition date | 25.06.2010 |
| PDB title | SOLUTION NMR STRUCTURE OF CV_0373(175-257) PROTEIN FROM CHROMOBACTERIUM VIOLACEUM, NORTHEAST STRUCTURAL GENOMICS CONSORTIUM TARGET CVR118A |
| PDB authors | Y.YANG, T.A.RAMELOT, D.WANG, C.CICCOSANTI, L.MAO, H.JANJUA, T.B.ACTON, R.XIAO, J.K.EVERETT, G.T.MONTELIONE, M.A.KENNEDY, NORTHEAST STRUCTURAL GENOMICS CONSORTIUM (NESG) |
| Last author | KENNEDY |
| Reference | |
| Reference authors | |
| Reference title | |
| Software listed | AUTOASSIGN, AUTOSTRUCTURE, CNS, CYANA, NMRPIPE, PDBSTAT, PINE, PSVS, SPARKY, TOPSPIN, VNMR, X-PLOR |
| Spectrometer | BRUKER, VARIAN (850 MHZ, 600 MHZ) |

| Protein number | 28 |
|---|---|
| PDB code | 2KCT, doi:10.2210/pdb2KCT/pdb |
| BMRB code | 16096, doi:10.13018/BMR16096 |
| PDB Header | CHAPERONE |
| Protein name | OB-FOLD DOMAIN OF HEME CHAPERONE CCME FROM DESULFOVIBRIO VULGARIS |
| Deposition date | 29.12.2008 |
| PDB title | SOLUTION NMR STRUCTURE OF THE OB-FOLD DOMAIN OF HEME CHAPERONE CCME FROM DESULFOVIBRIO VULGARIS. NORTHEAST STRUCTURAL GENOMICS TARGET DVR115G |
| PDB authors | J.M.ARAMINI, P.ROSSI, H.LEE, A.LEMAK, H.WANG, E.L.FOOTE, M.JIANG, R.XIAO, R.NAIR, G.V.T.SWAPNA, T.B.ACTON, B.ROST, J.K.EVERETT, G.T.MONTELIONE, NORTHEAST STRUCTURAL GENOMICS CONSORTIUM (NESG) |
| Last author | MONTELIONE |
| Reference | |
| Reference authors | |
| Reference title | |
| Software listed | AUTOASSIGN, AUTOSTRUCTURE, CNS, CYANA, NMRPIPE, PALES, PDBSTAT, PINE, PSVS, SPARKY, TOPSPIN, VNMRJ |
| Spectrometer | BRUKER, VARIAN (800 MHZ, 600 MHZ) |

| Protein number | 29 |
|---|---|
| PDB code | 2MDR, doi:10.2210/pdb2MDR/pdb |
| BMRB code | 19502, doi:10.13018/BMR19502 |
| PDB Header | HYDROLASE |
| Protein name | THIRD DOUBLE-STRANDED RNA-BINDING DOMAIN (DSRBD3) OF HUMAN ADENOSINE-DEAMINASE ADAR1 |
| Deposition date | 17.09.2013 |
| PDB title | SOLUTION STRUCTURE OF THE THIRD DOUBLE-STRANDED RNA-BINDING DOMAIN (DSRBD3) OF HUMAN ADENOSINE-DEAMINASE ADAR1 |
| PDB authors | P.BARRAUD, S.BANERJEE, W.I.MOHAMED, M.F.JANTSCH, F.H.ALLAIN |
| Last author | ALLAIN |
| Reference | PROC.NATL.ACAD.SCI.USA 111, E1852 (2014), doi:10.1073/PNAS.1323698111 |
| Reference authors | P.BARRAUD, S.BANERJEE, W.I.MOHAMED, M.F.JANTSCH, F.H.ALLAIN |
| Reference title | A BIMODULAR NUCLEAR LOCALIZATION SIGNAL ASSEMBLED VIA AN EXTENDED DOUBLE-STRANDED RNA-BINDING DOMAIN ACTS AS AN RNA-SENSING SIGNAL FOR TRANSPORTIN 1 |
| Software listed | ATNOS, CING, CNS, CYANA, PROCHECKNMR, SPARKY, TOPSPIN |
| Spectrometer | BRUKER (900 MHZ, 750 MHZ, 700 MHZ, 600 MHZ, 500 MHZ) |

| Protein number | 30 |
|---|---|
| PDB code | 2FB7, doi:10.2210/pdb2FB7/pdb |
| BMRB code | 7084, doi:10.13018/BMR7084 |
| PDB Header | STRUCTURAL GENOMICS, UNKNOWN FUNCTION |
| Protein name | PROTEIN FROM ZEBRA FISH DR.13312 |
| Deposition date | 08.12.2005 |
| PDB title | NMR SOLUTION STRUCTURE OF PROTEIN FROM ZEBRA FISH DR.13312 |
| PDB authors | R.C.TYLER, J.SONG, J.L.MARKLEY, CENTER FOR EUKARYOTIC STRUCTURAL GENOMICS (CESG) |
| Last author | MARKLEY |
| Reference | |
| Reference authors | |
| Reference title | |
| Software listed | ARIA, CNS, NMRPIPE, NMRVIEW |
| Spectrometer | VARIAN (600 MHZ) |
| Protein number | 31 |
| PDB code | 2MB0, doi:10.2210/pdb2MB0/pdb |
| BMRB code | 19382, doi:10.13018/BMR19382 |
| PDB Header | SPLICING/RNA |
| Protein name | HNRNP G RRM IN COMPLEX WITH RNA 5'-AUCAAA-3' |
| Deposition date | 22.07.2013 |
| PDB title | SOLUTION STRUCTURE OF HNRNP G RRM IN COMPLEX WITH THE RNA 5'-AUCAAA-3' |
| PDB authors | A.MOURSY, F.H.-T.ALLAIN, A.CLERY |
| Last author | CLERY |
| Reference | NUCLEIC ACIDS RES. 42, 6659 (2014), doi:10.1093/NAR/GKU244 |
| Reference authors | A.MOURSY, F.H.ALLAIN, A.CLERY |
| Reference title | CHARACTERIZATION OF THE RNA RECOGNITION MODE OF HNRNP G EXTENDS ITS ROLE IN SMN2 SPLICING REGULATION |
| Software listed | AMBER |
| Spectrometer | BRUKER (900 MHZ, 700 MHZ, 600 MHZ, 500 MHZ) |
| Protein number | 32 |
| PDB code | 2L05, doi:10.2210/pdb2L05/pdb |
| BMRB code | 17030, doi:10.13018/BMR17030 |
| PDB Header | TRANSFERASE |
| Protein name | RAS-BINDING DOMAIN OF SERINE/THREONINE- PROTEIN KINASE B-RAF FROM HOMO SAPIENS |
| Deposition date | 30.06.2010 |
| PDB title | SOLUTION NMR STRUCTURE OF THE RAS-BINDING DOMAIN OF SERINE/THREONINE- PROTEIN KINASE B-RAF FROM HOMO SAPIENS, NORTHEAST STRUCTURAL GENOMICS CONSORTIUM TARGET HR4694F |
| PDB authors | J.M.ARAMINI, H.JANJUA, C.CICCOSANTI, R.SHASTRY, Y.J.HUANG, T.B.ACTON, R.XIAO, J.K.EVERETT, G.T.MONTELIONE, NORTHEAST STRUCTURAL GENOMICS CONSORTIUM (NESG) |
| Last author | MONTELIONE |
| Reference | |
| Reference authors | |
| Reference title | |
| Software listed | AUTOSTRUCTURE, CNS, CYANA, MOLPROBITY, NMRPIPE, PDBSTAT, PINE, PSVS, SPARKY, TALOS+, TOPSPIN |
| Spectrometer | BRUKER (800 MHZ, 600 MHZ) |
| Protein number | 33 |
| PDB code | 2KJR, doi:10.2210/pdb2KJR/pdb |
| BMRB code | 16338, doi:10.13018/BMR16338 |
| PDB Header | CHAPERONE |
| Protein name | N-TERMINAL UBIQUITIN-LIKE DOMAIN FROM TUBULIN-BINDING COFACTOR B, CG11242, FROM DROSOPHILA MELANOGASTER |
| Deposition date | 08.06.2009 |
| PDB title | SOLUTION NMR STRUCTURE OF THE N-TERMINAL UBIQUITIN-LIKE DOMAIN FROM TUBULIN-BINDING COFACTOR B, CG11242, FROM DROSOPHILA MELANOGASTER. NORTHEAST STRUCTURAL GENOMICS CONSORTIUM TARGET FR629A (RESIDUES 8- 92) |
| PDB authors | T.A.RAMELOT, J.R.CORT, R.SHASTRY, C.CICCOSANTI, M.JIANG, R.NAIR, B.ROST, G.SWAPNA, T.B.ACTON, R.XIAO, J.K.EVERETT, G.T.MONTELIONE, M.A.KENNEDY, NORTHEAST STRUCTURAL GENOMICS CONSORTIUM (NESG) |
| Last author | KENNEDY |
| Reference | |
| Reference authors | |
| Reference title | |
| Software listed | AUTOASSIGN, AUTOSTRUCTURE, NMRPIPE, PDBSTAT, PSVS, SPARKY, TOPSPIN, VNMR, X-PLOR |
| Spectrometer | BRUKER, VARIAN (850 MHZ, 500 MHZ) |
| Protein number | 34 |
| PDB code | 2M5O, doi:10.2210/pdb2M5O/pdb |
| BMRB code | 19068, doi:10.13018/BMR19068 |
| PDB Header | BIOSYNTHETIC PROTEIN |
| Protein name | SOLUTION NMR STRUCTURE CTD DOMAIN OF NFU1 IRON-SULFUR CLUSTER SCAFFOLD HOMOLOG FROM HOMO SAPIENS |
| Deposition date | 01.03.2013 |
| PDB title | SOLUTION NMR STRUCTURE CTD DOMAIN OF NFU1 IRON-SULFUR CLUSTER SCAFFOLD HOMOLOG FROM HOMO SAPIENS, NORTHEAST STRUCTURAL GENOMICS CONSORTIUM (NESG) TARGET HR2876C |
| PDB authors | G.LIU, R.XIAO, H.JANJUA, K.HAMILTON, R.SHASTRY, E.KOHAN, T.B.ACTON, J.K.EVERETT, K.PEDERSON, Y.J.HUANG, G.T.MONTELIONE, NORTHEAST STRUCTURAL GENOMICS CONSORTIUM (NESG), MITOCHONDRIAL PROTEIN PARTNERSHIP (MPP) |
| Last author | MONTELIONE |
| Reference | |
| Reference authors | |
| Reference title | |
| Software listed | AUTOASSIGN, AUTOSTRUCTURE, CNS, CYANA, NMRPIPE, PSVS, REDCAT, SPARKY, TALOS+, TOPSPIN, VNMRJ, XEASY |
| Spectrometer | BRUKER, VARIAN (800 MHZ, 600 MHZ) |

| Protein number | 35 |
|---|---|
| PDB code | (MDM2) |
| BMRB code | |
| PDB Header | |
| Protein name | |
| Deposition date | |
| PDB title | |
| PDB authors | |
| Last author | |
| Reference | |
| Reference authors | |
| Reference title | |
| Software listed | |
| Spectrometer | |

| Protein number | 36 |
|---|---|
| PDB code | 2LNA, doi:10.2210/pdb2LNA/pdb |
| BMRB code | 18156, doi:10.13018/BMR18156 |
| PDB Header | HYDROLASE |
| Protein name | MITOCHONDRIAL INNER MEMBRANE DOMAIN (RESIDUES 164–251), FTSH_EXT, FROM PARAPLEGIN–LIKE PROTEIN AFG3L2 FROM HOMO SAPIENS |
| Deposition date | 20.12.2011 |
| PDB title | SOLUTION NMR STRUCTURE OF THE MITOCHONDRIAL INNER MEMBRANE DOMAIN (RESIDUES 164–251), FTSH_EXT, FROM THE PARAPLEGIN–LIKE PROTEIN AFG3L2 FROM HOMO SAPIENS, NORTHEAST STRUCTURAL GENOMICS CONSORTIUM TARGET HR6741A |
| PDB authors | T.A.RAMELOT, Y.YANG, H.LEE, H.JANUA, E.KOHAN, R.SHASTRY, T.B.ACTON, R.XIAO, J.K.EVERETT, J.H.PRESTEGARD, G.T.MONTELIONE, M.A.KENNEDY, NORTHEAST STRUCTURAL GENOMICS CONSORTIUM (NESG), MITOCHONDRIAL PROTEIN PARTNERSHIP (MPP) |
| Last author | KENNEDY |
| Reference | FEBS LETT. 587, 3522 (2013), doi:10.1016/J.FEBSLET.2013.09.009 |
| Reference authors | T.A.RAMELOT, Y.YANG, I.D.SAHU, H.W.LEE, R.XIAO, G.A.LORIGAN, G.T.MONTELIONE, M.A.KENNEDY |
| Reference title | NMR STRUCTURE AND MD SIMULATIONS OF THE AAA PROTEASE INTERMEMBRANE SPACE DOMAIN INDICATES PERIPHERAL MEMBRANE LOCALIZATION WITHIN THE HEXAOLIGOMER |
| Software listed | AUTOSTRUCTURE, CNS, CYANA, NMRPIPE, PALES, PINE, PSVS, SPARKY, TALOS+, TOPSPIN, VNMRJ, X-PLOR NIH |
| Spectrometer | BRUKER, VARIAN (850 MHZ, 600 MHZ) |

| Protein number | 37 |
|---|---|
| PDB code | 2LA6, doi:10.2210/pdb2LA6/pdb |
| BMRB code | 17508, doi:10.13018/BMR17508 |
| PDB Header | RNA BINDING PROTEIN |
| Protein name | RRM DOMAIN OF RNA–BINDING PROTEIN FUS FROM HOMO SAPIENS |
| Deposition date | 04.03.2011 |
| PDB title | SOLUTION NMR STRUCTURE OF RRM DOMAIN OF RNA-BINDING PROTEIN FUS FROM HOMO SAPIENS, NORTHEAST STRUCTURAL GENOMICS CONSORTIUM (NESG) |
| PDB authors | G.LIU, R.XIAO, H.JANJUA, C.CICCOSANTI, H.WANG, H.LEE, T.B.ACTON, J.K.EVERETT, Y.J.HUANG, G.T.MONTELIONE, NORTHEAST STRUCTURAL GENOMICS CONSORTIUM (NESG) |
| Last author | MONTELIONE |
| Reference | |
| Reference authors | |
| Reference title | |
| Software listed | AUTOASSIGN, AUTOSTRUCTURE, CNS, CYANA, NMRPIPE, SPARKY, TALOS+, TOPSPIN, VNMRJ, XEASY |
| Spectrometer | BRUKER, VARIAN (800 MHZ, 600 MHZ) |

| Protein number | 38 |
|---|---|
| PDB code | 6FIP, doi:10.2210/pdb6FIP/pdb |
| BMRB code | 34235, doi:10.13018/BMR34235 |
| PDB Header | TRANSPORT PROTEIN |
| Protein name | PSEUDOMONAS AERUGINOSA TONB CTD |
| Deposition date | 19.01.2018 |
| PDB title | SOLUTION NMR STRUCTURE OF PSEUDOMONAS AERUGINOSA TONB CTD |
| PDB authors | J.S.OEEMIG, O.H.SAMULI OLLILA, H.IWAI |
| Last author | IWAI |
| Reference | PEERJ 6, E5412 (2018), doi:10.7717/PEERJ.5412 |
| Reference authors | J.S.OEEMIG, O.H.S.OLLILA, H.IWAI |
| Reference title | NMR STRUCTURE OF THE C-TERMINAL DOMAIN OF TONB PROTEIN FROMPSEUDOMONAS AERUGINOSA |
| Software listed | AMBER, CCPNMR, CYANA, NMRPIPE |
| Spectrometer | VARIAN (800 MHZ) |

| Protein number | 39 |
|---|---|
| PDB code | 2LEA, doi:10.2210/pdb2LEA/pdb |
| BMRB code | 17705, doi:10.13018/BMR17705 |
| PDB Header | RNA BINDING PROTEIN |
| Protein name | HUMAN SRSF2 (SC35) RRM |
| Deposition date | 15.06.2011 |
| PDB title | SOLUTION STRUCTURE OF HUMAN SRSF2 (SC35) RRM |

| PDB authors | G.M.DAUBNER, A.CLERY, S.JAYNE, J.STEVENIN, F.H.–T.ALLAIN |
|---|---|
| Last author | ALLAIN |
| Reference | EMBO J. 31, 162 (2012), doi:10.1038/EMBOJ.2011.367 |
| Reference authors | G.M.DAUBNER, A.CLERY, S.JAYNE, J.STEVENIN, F.H.ALLAIN |
| Reference title | A SYN-ANTI CONFORMATIONAL DIFFERENCE ALLOWS SRSF2 TO RECOGNIZE GUANINES AND CYTOSINES EQUALLY WELL |
| Software listed | AMBER, CYANA, SPARKY |
| Spectrometer | BRUKER (900 MHZ, 700 MHZ, 600 MHZ, 500 MHZ) |

| Protein number | 40 |
|---|---|
| PDB code | 2LL8, doi:10.2210/pdb2LL8/pdb |
| BMRB code | 18032, doi:10.13018/BMR18032 |
| PDB Header | TRANSFERASE |
| Protein name | SPECIALIZED HOLO-ACYL CARRIER PROTEIN RPA2022 FROM RHODOPSEUDOMONAS PALUSTRIS REFINED WITH NH RDCS |
| Deposition date | 31.10.2011 |
| PDB title | SOLUTION NMR STRUCTURE OF THE SPECIALIZED HOLO-ACYL CARRIER PROTEIN RPA2022 FROM RHODOPSEUDOMONAS PALUSTRIS REFINED WITH NH RDCS, NORTHEAST STRUCTURAL GENOMICS CONSORTIUM TARGET RPR324 |
| PDB authors | T.A.RAMELOT, S.NI, P.ROSSI, Y.YANG, H.WANG, C.CICCOSANTI, M.MAGLAQUI, H.JANJUA, R.NAIR, B.ROSET, T.B.ACTON, R.XIAO, J.K.EVERETT, J.H.PRESTEGARD, G.T.MONTELIONE, M.A.KENNEDY, NORTHEAST STRUCTURAL GENOMICS CONSORTIUM (NESG) |
| Last author | KENNEDY |
| Reference | BIOCHEMISTRY 51, 7239 (2012), doi:10.1021/BI300546B |
| Reference authors | T.A.RAMELOT, P.ROSSI, F.FOROUHAR, H.W.LEE, Y.YANG, S.NI, S.UNSER, S.LEW, J.SEETHARAMAN, R.XIAO, T.B.ACTON, J.K.EVERETT, J.H.PRESTEGARD, J.F.HUNT, G.T.MONTELIONE, M.A.KENNEDY |
| Reference title | STRUCTURE OF A SPECIALIZED ACYL CARRIER PROTEIN ESSENTIAL FOR LIPID A BIOSYNTHESIS WITH VERY LONG-CHAIN FATTY ACIDS IN OPEN AND CLOSED CONFORMATIONS |
| Software listed | AUTOASSIGN, AUTOSTRUCTURE, CNS, CYANA, FMCGUI, NMRPIPE, PDBSTAT, PINE_SERVER, PSVS, SPARKY, TOPSPIN, VNMR, X-PLOR_NIH |
| Spectrometer | BRUKER, VARIAN (850 MHZ, 600 MHZ) |

| Protein number | 41 |
|---|---|
| PDB code | 2KPN, doi:10.2210/pdb2KPN/pdb |
| BMRB code | 16561, doi:10.13018/BMR16561 |
| PDB Header | HYDROLASE |
| Protein name | A BACTERIAL IG-LIKE (BIG_3) DOMAIN FROM BACILLUS CEREUS |
| Deposition date | 16.10.2009 |
| PDB title | SOLUTION NMR STRUCTURE OF A BACTERIAL IG-LIKE (BIG_3) DOMAIN FROM BACILLUS CEREUS. NORTHEAST STRUCTURAL GENOMICS CONSORTIUM TARGET BCR147A |
| PDB authors | J.M.ARAMINI, D.WANG, C.T.CICCOSANTI, H.JANJUA, B.ROST, T.B.ACTON, R.XIAO, G.V.T.SWAPNA, J.K.EVERETT, G.T.MONTELIONE, NORTHEAST STRUCTURAL GENOMICS CONSORTIUM (NESG) |
| Last author | MONTELIONE |
| Reference |  |
| Reference authors |  |
| Reference title |  |
| Software listed | AUTOSTRUCTURE, CNS, CYANA, MOLPROBITY, NMRPIPE, PDBSTAT, PINE, PSVS, SPARKY, TALOS, TOPSPIN |
| Spectrometer | BRUKER (800 MHZ, 600 MHZ) |

| Protein number | 42 |
|---|---|
| PDB code | 2K0M, doi:10.2210/pdb2K0M/pdb |
| BMRB code | 15652, doi:10.13018/BMR15652 |
| PDB Header | STRUCTURAL GENOMICS, UNKNOWN FUNCTION |
| Protein name | UNCHARACTERIZED PROTEIN FROM RHODOSPIRILLUM RUBRUM GENE LOCUS RRU_A0810 |
| Deposition date | 04.02.2008 |
| PDB title | SOLUTION NMR STRUCTURE OF THE UNCHARACTERIZED PROTEIN FROM RHODOSPIRILLUM RUBRUM GENE LOCUS RRU_A0810. NORTHEAST STRUCTURAL GENOMICS CONSORTIUM TARGET RRR43 |
| PDB authors | P.ROSSI, H.WANG, M.JIANG, E.L.FOOTE, R.XIAO, J.LIU, G.SWAPNA, T.B.ACTON, M.C.BARAN, B.ROST, G.T.MONTELIONE, NORTHEAST STRUCTURAL GENOMICS CONSORTIUM (NESG) |
| Last author | MONTELIONE |
| Reference |  |
| Reference authors |  |
| Reference title |  |
| Software listed | AUTOASSIGN, CNS, CYANA, MOLMOL, MOLPROBITY, NMRPIPE, PROCHECKNMR, PSVS, SPARKY, TALOS, TOPSPIN |
| Spectrometer | BRUKER (800 MHZ) |

| Protein number | 43 |
|---|---|
| PDB code | 2K5V, doi:10.2210/pdb2K5V/pdb |
| BMRB code | 15849, doi:10.13018/BMR15849 |
| PDB Header | DNA BINDING PROTEIN |
| Protein name | SECOND OB-FOLD DOMAIN OF REPLICATION PROTEIN A FROM METHANOCOCCUS MARIPALUDIS |
| Deposition date | 30.06.2008 |
| PDB title | SOLUTION NMR STRUCTURE OF THE SECOND OB-FOLD DOMAIN OF REPLICATION PROTEIN A FROM METHANOCOCCUS MARIPALUDIS. NORTHEAST STRUCTURAL GENOMICS TARGET MRR110B |

| PDB authors | J.M.ARAMINI, M.MAGLAQUI, M.JIANG, C.CICCOSANTI, R.XIAO, R.NAIR, J.K.EVERETT, G.VT.SWAPNA, T.B.ACTON, B.ROST, G.T.MONTELIONE, NORTHEAST STRUCTURAL GENOMICS CONSORTIUM (NESG) |
|---|---|
| Last author | MONTELIONE |
| Reference | |
| Reference authors | |
| Reference title | |
| Software listed | AUTOASSIGN, AUTOSTRUCTURE, CNS, CYANA, NMRPIPE, PDBSTAT, PSVS, SPARKY, TOPSPIN |
| Spectrometer | BRUKER (800 MHZ, 600 MHZ) |

| Protein number | 44 |
|---|---|
| PDB code | 2MQL, doi:10.2210/pdb2MQL/pdb |
| BMRB code | 25038, doi:10.13018/BMR25038 |
| PDB Header | RNA BINDING PROTEIN |
| Protein name | STRUCTURAL INVESTIGATION OF HNRNP L |
| Deposition date | 24.06.2014 |
| PDB title | STRUCTURAL INVESTIGATION OF HNRNP L |
| PDB authors | M.BLATTER, F.ALLAIN |
| Last author | ALLAIN |
| Reference | J.MOL.BIOL. 427, 3001 (2015), doi:10.1016/J.JMB.2015.05.020 |
| Reference authors | M.BLATTER, S.DUNIN-HORKAWICZ, I.GRISHINA, C.MARIS, S.THORE, T.MAIER, A.BINDEREIF, J.M.BUJNICKI, F.H.ALLAIN |
| Reference title | THE SIGNATURE OF THE FIVE-STRANDED VRRM FOLD DEFINED BY FUNCTIONAL, STRUCTURAL AND COMPUTATIONAL ANALYSIS OF THE HNRNP L PROTEIN |
| Software listed | AMBER, CYANA, SPARKY, TOPSPIN |
| Spectrometer | BRUKER (900 MHZ, 700 MHZ) |

| Protein number | 45 |
|---|---|
| PDB code | 2K75, doi:10.2210/pdb2K75/pdb |
| BMRB code | 15902, doi:10.13018/BMR15902 |
| PDB Header | DNA BINDING PROTEIN |
| Protein name | OB DOMAIN OF TA0387 FROM THERMOPLASMA ACIDOPHILUM |
| Deposition date | 01.08.2008 |
| PDB title | SOLUTION NMR STRUCTURE OF THE OB DOMAIN OF TA0387 FROM THERMOPLASMA ACIDOPHILUM. NORTHEAST STRUCTURAL GENOMICS CONSORTIUM TARGET TAR80B |
| PDB authors | T.A.RAMELOT, K.DING, D.LEE, M.JIANG, C.CICCOSANTI, R.XIAO, R.NAIR, J.K.EVERETT, G.SWAPNA, T.B.ACTON, B.ROST, G.T.MONTELIONE, M.A.KENNEDY, NORTHEAST STRUCTURAL GENOMICS CONSORTIUM (NESG) |
| Last author | KENNEDY |
| Reference | |
| Reference authors | |
| Reference title | |
| Software listed | AUTOASSIGN, AUTOSTRUCTURE, NMRPIPE, PSVS, SPARKY, TOPSPIN, VNMR, X-PLOR |
| Spectrometer | BRUKER, VARIAN (850 MHZ, 600 MHZ) |

| Protein number | 46 |
|---|---|
| PDB code | 2LTM, doi:10.2210/pdb2LTM/pdb |
| BMRB code | 18489, doi:10.13018/BMR18489 |
| PDB Header | ELECTRON TRANSPORT |
| Protein name | NFU1 IRON-SULFUR CLUSTER SCAFFOLD HOMOLOG FROM HOMO SAPIENS |
| Deposition date | 29.05.2012 |
| PDB title | SOLUTION NMR STRUCTURE OF NFU1 IRON-SULFUR CLUSTER SCAFFOLD HOMOLOG FROM HOMO SAPIENS, NORTHEAST STRUCTURAL GENOMICS CONSORTIUM (NESG) TARGET HR2876B |
| PDB authors | G.LIU, R.XIAO, H.JANJUA, K.HAMILTON, R.SHASTRY, E.KOHAN, T.B.ACTON, J.K.EVERETT, H.LEE, Y.J.HUANG, G.T.MONTELIONE, NORTHEAST STRUCTURAL GENOMICS CONSORTIUM (NESG), MITOCHONDRIAL PROTEIN PARTNERSHIP (MPP) |
| Last author | MONTELIONE |
| Reference | |
| Reference authors | |
| Reference title | |
| Software listed | AUTOASSIGN, AUTOSTRUCTURE, CNS, CYANA, NMRPIPE, PALES, PSVS, REDCAT, SPARKY, TALOS+, TOPSPIN, VNMRJ, XEASY |
| Spectrometer | BRUKER, VARIAN (800 MHZ, 600 MHZ) |

| Protein number | 47 |
|---|---|
| PDB code | 2KOB, doi:10.2210/pdb2KOB/pdb |
| BMRB code | 16498, doi:10.13018/BMR16498 |
| PDB Header | STRUCTURAL GENOMICS, UNKNOWN FUNCTION |
| Protein name | CLOLEP_01837 (FRAGMENT 61-160) FROM CLOSTRIDIUM LEPTUM |
| Deposition date | 15.09.2009 |
| PDB title | SOLUTION NMR STRUCTURE OF CLOLEP_01837 (FRAGMENT 61-160) FROM CLOSTRIDIUM LEPTUM. NORTHEAST STRUCTURAL GENOMICS CONSORTIUM TARGET QLR8A |
| PDB authors | T.A.RAMELOT, D.LEE, C.CICCOSANTI, M.JIANG, R.NAIR, B.ROST, T.B.ACTON, R.XIAO, J.K.EVERETT, G.T.MONTELIONE, M.A.KENNEDY, NORTHEAST STRUCTURAL GENOMICS CONSORTIUM (NESG) |
| Last author | KENNEDY |
| Reference | |
| Reference authors | |
| Reference title | |

| Software listed | AUTOASSIGN, AUTOSTRUCTURE, CNS, NMRPIPE, PDBSTAT, PSVS, SPARKY, TOPSPIN, VNMR, X-PLOR |
|---|---|
| Spectrometer | BRUKER, VARIAN (850 MHZ, 600 MHZ) |

| Protein number | 48 |
|---|---|
| PDB code | 2KHD, doi:10.2210/pdb2KHD/pdb |
| BMRB code | 16238, doi:10.13018/BMR16238 |
| PDB Header | STRUCTURAL GENOMICS, UNKNOWN FUNCTION |
| Protein name | VC_A0919 FROM VIBRIO CHOLERAE |
| Deposition date | 02.04.2009 |
| PDB title | SOLUTION NMR STRUCTURE OF VC_A0919 FROM VIBRIO CHOLERAE. NORTHEAST STRUCTURAL GENOMICS CONSORTIUM TARGET VCR52 |
| PDB authors | T.A.RAMELOT, J.R.CORT, H.WANG, C.CICCOSANTI, M.JIANG, J.LIU, B.ROST, G.V.T.SWAPNA, T.B.ACTON, R.XIAO, J.K.EVERETT, G.T.MONTELIONE, M.A.KENNEDY, NORTHEAST STRUCTURAL GENOMICS CONSORTIUM (NESG) |
| Last author | KENNEDY |
| Reference | |
| Reference authors | |
| Reference title | |
| Software listed | AUTOASSIGN, AUTOSTRUCTURE, NMRPIPE, PDBSTAT, PSVS, SPARKY, TOPSPIN, VNMR, X-PLOR |
| Spectrometer | BRUKER, VARIAN (850 MHZ, 750 MHZ, 600 MHZ) |

| Protein number | 49 |
|---|---|
| PDB code | 2RN7, doi:10.2210/pdb2RN7/pdb |
| BMRB code | 11017, doi:10.13018/BMR11017 |
| PDB Header | UNKNOWN FUNCTION |
| Protein name | TNPE PROTEIN FROM SHIGELLA FLEXNERI |
| Deposition date | 08.12.2007 |
| PDB title | NMR SOLUTION STRUCTURE OF TNPE PROTEIN FROM SHIGELLA FLEXNERI. NORTHEAST STRUCTURAL GENOMICS TARGET SFR125 |
| PDB authors | T.A.RAMELOT, J.R.CORT, A.SEMESI, M.GARCIA, A.A.YEE, C.H.ARROWSMITH, M.A.KENNEDY, NORTHEAST STRUCTURAL GENOMICS CONSORTIUM (NESG) |
| Last author | KENNEDY |
| Reference | |
| Reference authors | |
| Reference title | |
| Software listed | AUTOSTRUCTURE, CNS, NMRPIPE, SPARKY, VNMR, X-PLOR |
| Spectrometer | VARIAN (750 MHZ, 600 MHZ) |

| Protein number | 50 |
|---|---|
| PDB code | 2LXU, doi:10.2210/pdb2LXU/pdb |
| BMRB code | 18698, doi:10.13018/BMR18698 |
| PDB Header | RNA BINDING PROTEIN |
| Protein name | EUKARYOTIC RNA RECOGNITION MOTIF, RRM1, FROM HETEROGENEOUS NUCLEAR RIBONUCLEOPROTEIN H FROM HOMO SAPIENS |
| Deposition date | 31.08.2012 |
| PDB title | SOLUTION NMR STRUCTURE OF THE EUKARYOTIC RNA RECOGNITION MOTIF, RRM1, FROM THE HETEROGENEOUS NUCLEAR RIBONUCLEOPROTEIN H FROM HOMO SAPIENS, NORTHEAST STRUCTURAL GENOMICS CONSORTIUM (NESG) TARGET HR8614A |
| PDB authors | T.A.RAMELOT, Y.YANG, K.PEDERSON, R.SHASTRY, E.KOHAN, H.JANJUA, R.XIAO, T.B.ACTON, J.K.EVERETT, J.H.PRESTEGARD, G.T.MONTELIONE, M.A.KENNEDY, NORTHEAST STRUCTURAL GENOMICS CONSORTIUM (NESG) |
| Last author | KENNEDY |
| Reference | |
| Reference authors | |
| Reference title | |
| Software listed | AUTOSTRUCTURE, CNS, CYANA, FMCGUI, NMRPIPE, PALES, PINE, PSVS, SPARKY, TALOS+, TOPSPIN, VNMRJ |
| Spectrometer | BRUKER, VARIAN (800 MHZ, 600 MHZ) |

| Protein number | 51 |
|---|---|
| PDB code | 2KIF, doi:10.2210/pdb2KIF/pdb |
| BMRB code | 16272, doi:10.13018/BMR16272 |
| PDB Header | TRANSFERASE |
| Protein name | AN O6-METHYLGUANINE DNA METHYLTRANSFERASE FAMILY PROTEIN FROM VIBRIO PARAHAEMOLYTICUS |
| Deposition date | 03.05.2009 |
| PDB title | SOLUTION NMR STRUCTURE OF AN O6-METHYLGUANINE DNA METHYLTRANSFERASE FAMILY PROTEIN FROM VIBRIO PARAHAEMOLYTICUS. NORTHEAST STRUCTURAL GENOMICS CONSORTIUM TARGET VPR247 |
| PDB authors | J.M.ARAMINI, R.L.BELOTE, C.T.CICCOSANTI, M.JIANG, B.ROST, R.NAIR, G.V.T.SWAPNA, T.B.ACTON, R.XIAO, J.K.EVERETT, G.T.MONTELIONE, NORTHEAST STRUCTURAL GENOMICS CONSORTIUM (NESG) |
| Last author | MONTELIONE |
| Reference | J.BIOL.CHEM. 285, 13736 (2010), doi:10.1074/JBC.M109.093591 |
| Reference authors | J.M.ARAMINI, J.L.TUBBS, S.KANUGULA, P.ROSSI, A.ERTEKIN, M.MAGLAQUI, K.HAMILTON, C.T.CICCOSANTI, M.JIANG, R.XIAO, T.T.SOONG, B.ROST, T.B.ACTON, J.K.EVERETT, A.E.PEGG, J.A.TAINER, G.T.MONTELIONE |

| Reference title | STRUCTURAL BASIS OF O6-ALKYLGUANINE RECOGNITION BY A BACTERIAL ALKYLTRANSFERASE-LIKE DNA REPAIR PROTEIN |
|---|---|
| Software listed | AUTOASSIGN, AUTOSTRUCTURE, CNS, CYANA, MOLPROBITY, NMRPIPE, PDBSTAT, PINE, PSVS, SPARKY, TOPSPIN |
| Spectrometer | BRUKER (800 MHZ, 600 MHZ) |

| Protein number | 52 |
|---|---|
| PDB code | 2KBN, doi:10.2210/pdb2KBN/pdb |
| BMRB code | 16051, doi:10.13018/BMR16051 |
| PDB Header | STRUCTURAL GENOMICS, UNKNOWN FUNCTION |
| Protein name | OB DOMAIN (RESIDUES 67-166) OF MM0293 FROM METHANOSARCINA MAZEI |
| Deposition date | 03.12.2008 |
| PDB title | SOLUTION NMR STRUCTURE OF THE OB DOMAIN (RESIDUES 67-166) OF MM0293 FROM METHANOSARCINA MAZEI. NORTHEAST STRUCTURAL GENOMICS CONSORTIUM TARGET MAR214A |
| PDB authors | T.A.RAMELOT, K.DING, M.MAGLIQUI, M.JIANG, C.CICCOSANTI, R.XIAO, J.LUI, J.K.EVERETT, G.SWAPNA, T.B.ACTON, B.ROST, G.T.MONTELIONE, M.A.KENNEDY, NORTHEAST STRUCTURAL GENOMICS CONSORTIUM (NESG) |
| Last author | KENNEDY |
| Reference |  |
| Reference authors |  |
| Reference title |  |
| Software listed | AUTOASSIGN, AUTOSTRUCTURE, CNS, CYANA, PSVS, SPARKY, TOPSPIN, X-PLOR |
| Spectrometer | BRUKER, VARIAN (850 MHZ, 600 MHZ) |

| Protein number | 53 |
|---|---|
| PDB code | 2MK2, doi:10.2210/pdb2MK2/pdb |
| BMRB code | 19749, doi:10.13018/BMR19749 |
| PDB Header | HYDROLASE |
| Protein name | N-TERMINAL DOMAIN (SH2 DOMAIN) OF HUMAN INOSITOL POLYPHOSPHATE PHOSPHATASE-LIKE PROTEIN 1 (INPPL1) (FRAGMENT 20-117) |
| Deposition date | 23.01.2014 |
| PDB title | SOLUTION NMR STRUCTURE OF N-TERMINAL DOMAIN (SH2 DOMAIN) OF HUMAN INOSITOL POLYPHOSPHATE PHOSPHATASE-LIKE PROTEIN 1 (INPPL1) (FRAGMENT 20-117), NORTHEAST STRUCTURAL GENOMICS CONSORTIUM TARGET HR9134A |
| PDB authors | Y.YANG, T.A.RAMELOT, H.JANJUA, R.XIAO, J.K.EVERETT, G.T.MONTELIONE, M.A.KENNEDY, NORTHEAST STRUCTURAL GENOMICS CONSORTIUM (NESG) |
| Last author | KENNEDY |
| Reference |  |
| Reference authors |  |
| Reference title |  |
| Software listed | AUTOASSIGN, AUTOSTRUCTURE, CNS, CYANA, NMRPIPE, PALES, PINE, PSVS, REDCAT, SPARKY, TALOS+, TOPSPIN, VNMRJ, XEASY |
| Spectrometer | BRUKER, VARIAN (850 MHZ, 600 MHZ) |

| Protein number | 54 |
|---|---|
| PDB code | 2K50, doi:10.2210/pdb2K50/pdb |
| BMRB code | 15819, doi:10.13018/BMR15819 |
| PDB Header | STRUCTURAL GENOMICS, UNKNOWN FUNCTION |
| Protein name | REPLICATION FACTOR A RELATED PROTEIN FROM METHANOBACTERIUM THERMOAUTOTROPHICUM |
| Deposition date | 23.06.2008 |
| PDB title | SOLUTION NMR STRUCTURE OF THE REPLICATION FACTOR A RELATED PROTEIN FROM METHANOBACTERIUM THERMOAUTOTROPHICUM. NORTHEAST STRUCTURAL GENOMICS TARGET TR91A |
| PDB authors | P.ROSSI, R.XIAO, M.MAGLAQUI, E.L.FOOTE, C.CICCOSANTI, G.SWAPNA, T.B.ACTON, B.ROST, J.K.EVERETT, M.JIANG, R.NAIR, G.T.MONTELIONE, NORTHEAST STRUCTURAL GENOMICS CONSORTIUM (NESG) |
| Last author | MONTELIONE |
| Reference |  |
| Reference authors |  |
| Reference title |  |
| Software listed | AUTOASSIGN, CNS, CYANA, MOLMOL, MOLPROBITY, NMRPIPE, PROCHECK, PSVS, RPF(AUTOSTRUCTURE), SPARKY, TALOS, TOPSPIN |
| Spectrometer | BRUKER (800 MHZ, 600 MHZ) |

| Protein number | 55 |
|---|---|
| PDB code | 2KL5, doi:10.2210/pdb2KL5/pdb |
| BMRB code | 16384, doi:10.13018/BMR16384 |
| PDB Header | STRUCTURAL GENOMICS, UNKNOWN FUNCTION |
| Protein name | PROTEIN YUTD FROM B.SUBTILIS |
| Deposition date | 30.06.2009 |
| PDB title | SOLUTION NMR STRUCTURE OF PROTEIN YUTD FROM B.SUBTILIS, NORTHEAST STRUCTURAL GENOMICS CONSORTIUM TARGET SR232 |
| PDB authors | G.LIU, K.HAMILTON, R.XIAO, C.CICCOSANTI, C.J.HO, J.EVERETT, R.NAIR, T.ACTON, B.ROST, G.T.MONTELIONE, NORTHEAST STRUCTURAL GENOMICS CONSORTIUM (NESG) |
| Last author | MONTELIONE |
| Reference |  |
| Reference authors |  |
| Reference title |  |
| Software listed | AUTOASSIGN, AUTOSTRUCTURE, CNS, CYANA, NMRPIPE, TOPSPIN, VNMRJ, XEASY |

| Spectrometer | BRUKER, VARIAN (800 MHZ, 600 MHZ) |
|---|---|

| Protein number | 56 |
|---|---|
| PDB code | 2LTA, doi:10.2210/pdb2LTA/pdb |
| BMRB code | 18465, doi:10.13018/BMR18465 |
| PDB Header | DE NOVO PROTEIN |
| Protein name | DE NOVO DESIGNED PROTEIN, ROSSMANN 3X1 FOLD |
| Deposition date | 15.05.2012 |
| PDB title | SOLUTION NMR STRUCTURE OF DE NOVO DESIGNED PROTEIN, ROSSMANN 3X1 FOLD, NORTHEAST STRUCTURAL GENOMICS CONSORTIUM TARGET OR157 |
| PDB authors | G.LIU, R.KOGA, N.KOGA, R.XIAO, K.PEDERSON, K.HAMILTON, E.KOHAN, T.B.ACTON, G.KORNHABER, J.K.EVERETT, D.BAKER, G.T.MONTELIONE, NORTHEAST STRUCTURAL GENOMICS CONSORTIUM (NESG) |
| Last author | MONTELIONE |
| Reference | NATURE 491, 222 (2012), doi:10.1038/NATURE11600 |
| Reference authors | N.KOGA, R.TATSUMI-KOGA, G.LIU, R.XIAO, T.B.ACTON, G.T.MONTELIONE, D.BAKER |
| Reference title | PRINCIPLES FOR DESIGNING IDEAL PROTEIN STRUCTURES |
| Software listed | AUTOASSIGN, AUTOSTRUCTURE, CNS, CYANA, NMRPIPE, PSVS, REDCAT, SPARKY, TALOS+, TOPSPIN, VNMRJ, XEASY |
| Spectrometer | BRUKER, VARIAN (800 MHZ, 600 MHZ) |

| Protein number | 57 |
|---|---|
| PDB code | 2KIW, doi:10.2210/pdb2KIW/pdb |
| BMRB code | 16298, doi:10.13018/BMR16298 |
| PDB Header | DNA BINDING PROTEIN |
| Protein name | DOMAIN N-TERMINAL TO INTEGRASE DOMAIN OF SH1003 FROM STAPHYLOCOCCUS HAEMOLYTICUS |
| Deposition date | 12.05.2009 |
| PDB title | SOLUTION NMR STRUCTURE OF THE DOMAIN N-TERMINAL TO THE INTEGRASE DOMAIN OF SH1003 FROM STAPHYLOCOCCUS HAEMOLYTICUS. NORTHEAST STRUCTURAL GENOMICS CONSORTIUM TARGET SHR105F (64-166) |
| PDB authors | Y.YANG, T.A.RAMELOT, R.L.BELOTE, E.L.FOOTE, H.JANJUA, R.NAIR, B.ROST, G.SWAPNA, T.B.ACTON, R.XIAO, J.K.EVERETT, G.T.MONTELIONE, M.A.KENNEDY, NORTHEAST STRUCTURAL GENOMICS CONSORTIUM (NESG) |
| Last author | KENNEDY |
| Reference | |
| Reference authors | |
| Reference title | |
| Software listed | AUTOASSIGN, AUTOSTRUCTURE, CNS, NMRPIPE, PDBSTAT, PSVS, SPARKY, TOPSPIN, VNMR, X-PLOR |
| Spectrometer | BRUKER, VARIAN (850 MHZ, 600 MHZ) |

| Protein number | 58 |
|---|---|
| PDB code | 2LVB, doi:10.2210/pdb2LVB/pdb |
| BMRB code | 18561, doi:10.13018/BMR18561 |
| PDB Header | DE NOVO PROTEIN |
| Protein name | SOLUTION NMR STRUCTURE DE NOVO DESIGNED PFK FOLD PROTEIN |
| Deposition date | 30.06.2012 |
| PDB title | SOLUTION NMR STRUCTURE OF DE NOVO DESIGNED PFK FOLD PROTEIN, NORTHEAST STRUCTURAL GENOMICS CONSORTIUM (NESG) TARGET OR250 |
| PDB authors | G.LIU, N.KOGA, R.KOGA, R.XIAO, K.HAMILTON, E.KOHAN, T.B.ACTON, G.KORNHABER, J.K.EVERETT, D.BAKER, G.T.MONTELIONE, NORTHEAST STRUCTURAL GENOMICS CONSORTIUM (NESG) |
| Last author | MONTELIONE |
| Reference | NATURE 491, 222 (2012), doi:10.1038/NATURE11600 |
| Reference authors | N.KOGA, R.TATSUMI-KOGA, G.LIU, R.XIAO, T.B.ACTON, G.T.MONTELIONE, D.BAKER |
| Reference title | PRINCIPLES FOR DESIGNING IDEAL PROTEIN STRUCTURES |
| Software listed | AUTOASSIGN, AUTOSTRUCTURE, CNS, CYANA, NMRPIPE, PSVS, SPARKY, TALOS+, TOPSPIN, VNMRJ, XEASY |
| Spectrometer | BRUKER, VARIAN (800 MHZ, 600 MHZ) |

| Protein number | 59 |
|---|---|
| PDB code | 2LND, doi:10.2210/pdb2LND/pdb |
| BMRB code | 18161, doi:10.13018/BMR18161 |
| PDB Header | DE NOVO PROTEIN |
| Protein name | DE NOVO DESIGNED PROTEIN, PFK FOLD |
| Deposition date | 23.12.2011 |
| PDB title | SOLUTION NMR STRUCTURE OF DE NOVO DESIGNED PROTEIN, PFK FOLD, NORTHEAST STRUCTURAL GENOMICS CONSORTIUM TARGET OR134 |
| PDB authors | G.LIU, N.KOGA, R.KOGA, R.XIAO, H.LEE, H.JANJUA, E.KOHAN, T.B.ACTON, J.K.EVERETT, D.BAKER, G.T.MONTELIONE, NORTHEAST STRUCTURAL GENOMICS CONSORTIUM (NESG) |
| Last author | MONTELIONE |
| Reference | |
| Reference authors | |
| Reference title | |
| Software listed | AUTOASSIGN, AUTOSTRUCTURE, CNS, CYANA, NMRPIPE, PSVS, SPARKY, TALOS+, TOPSPIN, VNMRJ, XEASY |
| Spectrometer | BRUKER, VARIAN (800 MHZ, 600 MHZ) |

| Protein number | 60 |
|---|---|
| PDB code | 1WQU, doi:10.2210/pdb1WQU/pdb |
| BMRB code | 6331, doi:10.13018/BMR6331 |
| PDB Header | TRANSFERASE |
| Protein name | HUMAN FES SH2 DOMAIN |
| Deposition date | 02.10.2004 |
| PDB title | SOLUTION STRUCTURE OF THE HUMAN FES SH2 DOMAIN |
| PDB authors | A.SCOTT, D.PANTOJA-UCEDA, S.KOSHIBA, M.INOUE, T.KIGAWA, T.TERADA, M.SHIROUZU, A.TANAKA, S.SUGANO, S.YOKOYAMA, P.GUNTERT, RIKEN STRUCTURAL GENOMICS/PROTEOMICS INITIATIVE (RSGI) |
| Last author | GUNTERT |
| Reference | J.BIOMOL.NMR 31, 357 (2005), doi:10.1007/S10858-005-0946-6 |
| Reference authors | A.SCOTT, D.PANTOJA-UCEDA, S.KOSHIBA, M.INOUE, T.KIGAWA, T.TERADA, M.SHIROUZU, A.TANAKA, S.SUGANO, S.YOKOYAMA, P.GUNTERT |
| Reference title | SOLUTION STRUCTURE OF THE SRC HOMOLOGY 2 DOMAIN FROM THE HUMAN FELINE SARCOMA ONCOGENE FES |
| Software listed | CYANA, NMRPIPE, NMRVIEW, OPALP |
| Spectrometer | BRUKER (800 MHZ, 600 MHZ) |

| Protein number | 61 |
|---|---|
| PDB code | 2KL6, doi:10.2210/pdb2KL6/pdb |
| BMRB code | 16385, doi:10.13018/BMR16385 |
| PDB Header | STRUCTURAL GENOMICS, UNKNOWN FUNCTION |
| Protein name | CARDB DOMAIN OF PF1109 FROM PYROCOCCUS FURIOSUS |
| Deposition date | 30.06.2009 |
| PDB title | SOLUTION NMR STRUCTURE OF THE CARDB DOMAIN OF PF1109 FROM PYROCOCCUS FURIOSUS. NORTHEAST STRUCTURAL GENOMICS CONSORTIUM TARGET PFR193A |
| PDB authors | J.M.ARAMINI, D.LEE, C.CICCOSANTI, K.HAMILTON, R.NAIR, B.ROST, T.B.ACTON, R.XIAO, G.V.T.SWAPNA, J.K.EVERETT, G.T.MONTELIONE, NORTHEAST STRUCTURAL GENOMICS CONSORTIUM (NESG) |
| Last author | MONTELIONE |
| Reference | |
| Reference authors | |
| Reference title | |
| Software listed | AUTOSTRUCTURE, CNS, CYANA, MOLPROBITY, NMRPIPE, PDBSTAT, PINE, PSVS, SPARKY, TALOS, TOPSPIN |
| Spectrometer | BRUKER (800 MHZ, 600 MHZ) |

| Protein number | 62 |
|---|---|
| PDB code | 6GT7, doi:10.2210/pdb6GT7/pdb |
| BMRB code | 34287, doi:10.13018/BMR34287 |
| PDB Header | TRANSFERASE |
| Protein name | FREE HELIX BUNDLE DOMAIN FROM FUNCTIONAL PRN1 PRIMASE |
| Deposition date | 15.06.2018 |
| PDB title | NMR STRUCTURE OF THE FREE HELIX BUNDLE DOMAIN FROM THE FUNCTIONAL PRN1 PRIMASE |
| PDB authors | J.BOUDET, G.LIPPS, F.ALLAIN |
| Last author | ALLAIN |
| Reference | CELL 176, 154 (2019), doi:10.1016/J.CELL.2018.11.031 |
| Reference authors | J.BOUDET, J.C.DEVILLIER, T.WIEGAND, L.SALMON, B.H.MEIER, G.LIPPS, F.H.ALLAIN |
| Reference title | A SMALL HELICAL BUNDLE PREPARES PRIMER SYNTHESIS BY BINDING TWO NUCLEOTIDES THAT ENHANCE SEQUENCE-SPECIFIC RECOGNITION OF THE DNA TEMPLATE |
| Software listed | AMBER, CANDID, CYANA, SPARKY |
| Spectrometer | BRUKER (900 MHZ, 700 MHZ, 600 MHZ) |

| Protein number | 63 |
|---|---|
| PDB code | 2JN8, doi:10.2210/pdb2JN8/pdb |
| BMRB code | 15089, doi:10.13018/BMR15089 |
| PDB Header | STRUCTURAL GENOMICS, UNKNOWN FUNCTION |
| Protein name | Q8ZRJ2 FROM SALMONELLA TYPHIMURIUM |
| Deposition date | 29.12.2006 |
| PDB title | SOLUTION NMR STRUCTURE OF Q8ZRJ2 FROM SALMONELLA TYPHIMURIUM. NORTHEAST STRUCTURAL GENOMICS TARGET STR65 |
| PDB authors | J.M.ARAMINI, J.R.CORT, C.K.HO, K.CUNNINGHAM, L.-C.MA, R.XIAO, J.LIU, M.C.BARAN, G.V.T.SWAPNA, T.B.ACTON, B.ROST, G.T.MONTELIONE, NORTHEAST STRUCTURAL GENOMICS CONSORTIUM (NESG) |
| Last author | MONTELIONE |
| Reference | |
| Reference authors | |
| Reference title | |
| Software listed | AUTOASSIGN, AUTOSTRUCTURE, BRUKER, NMRPIPE, PDBSTAT, PSVS, SPARKY, VNMR, X-PLOR |
| Spectrometer | BRUKER, VARIAN (750 MHZ, 600 MHZ, 500 MHZ) |

| Protein number | 64 |
|---|---|
| PDB code | 2K5D, doi:10.2210/pdb2K5D/pdb |
| BMRB code | 15829, doi:10.13018/BMR15829 |
| PDB Header | STRUCTURAL GENOMICS, UNKNOWN FUNCTION |
| Protein name | SAG0934 FROM STREPTOCOCCUS AGALACTIAE |
| Deposition date | 26.06.2008 |

| PDB title | SOLUTION NMR STRUCTURE OF SAG0934 FROM STREPTOCOCCUS AGALACTIAE. NORTHEAST STRUCTURAL GENOMICS TARGET SAR32[1-108] |
|---|---|
| PDB authors | J.M.ARAMINI, P.ROSSI, L.ZHAO, E.L.FOOTE, M.JIANG, R.XIAO, S.SHARMA, G.VT.SWAPNA, R.NAIR, J.K.EVERETT, T.B.ACTON, B.ROST, G.T.MONTELIONE, NORTHEAST STRUCTURAL GENOMICS CONSORTIUM (NESG) |
| Last author | MONTELIONE |
| Reference | |
| Reference authors | |
| Reference title | |
| Software listed | AUTOASSIGN, AUTOSTRUCTURE, CNS, CYANA, NMRPIPE, PDBSTAT, PSVS, SPARKY, TOPSPIN |
| Spectrometer | BRUKER (800 MHZ, 600 MHZ) |

| Protein number | 65 |
|---|---|
| PDB code | 2KD1, doi:10.2210/pdb2KD1/pdb |
| BMRB code | 16102, doi:10.13018/BMR16102 |
| PDB Header | STRUCTURAL GENOMICS, UNKNOWN FUNCTION |
| Protein name | INTEGRASE-LIKE DOMAIN FROM BACILLUS CEREUS ORDERED LOCUS BC_1272 |
| Deposition date | 31.12.2008 |
| PDB title | SOLUTION NMR STRUCTURE OF THE INTEGRASE-LIKE DOMAIN FROM BACILLUS CEREUS ORDERED LOCUS BC_1272. NORTHEAST STRUCTURAL GENOMICS CONSORTIUM TARGET BCR268F |
| PDB authors | P.ROSSI, H.LEE, M.MAGLAQUI, E.L.FOOTE, W.A.BUCHWALD, M.JIANG, G.V.T.SWAPNA, R.NAIR, R.XIAO, T.B.ACTON, B.ROST, J.H.PRESTEGARD, G.T.MONTELIONE, NORTHEAST STRUCTURAL GENOMICS CONSORTIUM (NESG) |
| Last author | MONTELIONE |
| Reference | |
| Reference authors | |
| Reference title | |
| Software listed | CNS, CYANA, MOLMOL, MOLPROBITY, NMRPIPE, PDBSTAT, PINE, PROCHECK, PSVS, SPARKY, TALOS, TOPSPIN |
| Spectrometer | BRUKER, VARIAN (800 MHZ, 600 MHZ) |

| Protein number | 66 |
|---|---|
| PDB code | 2LTL, doi:10.2210/pdb2LTL/pdb |
| BMRB code | 18487, doi:10.13018/BMR18487 |
| PDB Header | ELECTRON TRANSPORT |
| Protein name | NIFU-LIKE PROTEIN FROM SACCHAROMYCES CEREVISIAE |
| Deposition date | 29.05.2012 |
| PDB title | SOLUTION NMR STRUCTURE OF NIFU-LIKE PROTEIN FROM SACCHAROMYCES CEREVISIAE, NORTHEAST STRUCTURAL GENOMICS CONSORTIUM (NESG) TARGET YR313A |
| PDB authors | G.LIU, R.XIAO, K.HAMILTON, H.JANJUA, R.SHASTRY, E.KOHAN, T.B.ACTON, J.K.EVERETT, H.LEE, Y.J.HUANG, G.T.MONTELIONE, NORTHEAST STRUCTURAL GENOMICS CONSORTIUM (NESG), MITOCHONDRIAL PROTEIN PARTNERSHIP (MPP) |
| Last author | MONTELIONE |
| Reference | |
| Reference authors | |
| Reference title | |
| Software listed | AUTOASSIGN, AUTOSTRUCTURE, CNS, CYANA, NMRPIPE, PALES, PINE, PSVS, REDCAT, SPARKY, TALOS+, TOPSPIN, VNMRJ, XEASY |
| Spectrometer | BRUKER, VARIAN (800 MHZ, 600 MHZ) |

| Protein number | 67 |
|---|---|
| PDB code | 2KVO, doi:10.2210/pdb2KVO/pdb |
| BMRB code | 16782, doi:10.13018/BMR16782 |
| PDB Header | PHOTOSYNTHESIS |
| Protein name | PHOTOSYSTEM II REACTION CENTER PSB28 PROTEIN FROM SYNECHOCYSTIS SP.(STRAIN PCC 6803) |
| Deposition date | 22.03.2010 |
| PDB title | SOLUTION NMR STRUCTURE OF PHOTOSYSTEM II REACTION CENTER PSB28 PROTEIN FROM SYNECHOCYSTIS SP.(STRAIN PCC 6803), NORTHEAST STRUCTURAL GENOMICS CONSORTIUM TARGET SGR171 |
| PDB authors | Y.YANG, T.A.RAMELOT, J.R.CORT, D.WANG, C.CICCOSANTI, K.HAMILTON, R.NAIR, B.ROST, T.B.ACTON, R.XIAO, J.K.EVERETT, G.T.MONTELIONE, M.A.KENNEDY, NORTHEAST STRUCTURAL GENOMICS CONSORTIUM (NESG) |
| Last author | KENNEDY |
| Reference | PROTEINS 79, 340 (2011), doi:10.1002/PROT.22876 |
| Reference authors | Y.YANG, T.A.RAMELOT, J.R.CORT, D.WANG, C.CICCOSANTI, K.HAMILTON, R.NAIR, B.ROST, T.B.ACTON, R.XIAO, J.K.EVERETT, G.T.MONTELIONE, M.A.KENNEDY |
| Reference title | SOLUTION NMR STRUCTURE OF PHOTOSYSTEM II REACTION CENTER PROTEIN PSB28 FROM SYNECHOCYSTIS SP. STRAIN PCC 6803 |
| Software listed | AUTOASSIGN, AUTOSTRUCTURE, CNS, NMRPIPE, PDBSTAT, PINE_SERVER, PSVS, SPARKY, TOPSPIN, VNMR, X-PLOR_NIH |
| Spectrometer | BRUKER, VARIAN (850 MHZ, 600 MHZ) |

| Protein number | 68 |
|---|---|
| PDB code | 1T0Y, doi:10.2210/pdb1T0Y/pdb |
| BMRB code | 6176, doi:10.13018/BMR6176 |
| PDB Header | CHAPERONE |
| Protein name | A UBIQUITIN-LIKE DOMAIN FROM TUBULIN-BINDING COFACTOR B |

| Deposition date | 13.04.2004 |
|---|---|
| PDB title | SOLUTION STRUCTURE OF A UBIQUITIN-LIKE DOMAIN FROM TUBULIN-BINDING COFACTOR B |
| PDB authors | B.L.LYTLE, F.C.PETERSON, S.H.QUI, M.LUO, B.F.VOLKMAN, J.L.MARKLEY, CENTER FOR EUKARYOTIC STRUCTURAL GENOMICS (CESG) |
| Last author | MARKLEY |
| Reference | J.BIOL.CHEM. 279, 46787 (2004), doi:10.1074/JBC.M409422200 |
| Reference authors | B.L.LYTLE, F.C.PETERSON, S.H.QIU, M.LUO, Q.ZHAO, J.L.MARKLEY, B.F.VOLKMAN |
| Reference title | SOLUTION STRUCTURE OF A UBIQUITIN-LIKE DOMAIN FROM TUBULIN-BINDING COFACTOR B |
| Software listed | CYANA, NMRPIPE, SPSCAN, X-PLOR_NIH, XEASY, XWINNMR |
| Spectrometer | BRUKER (600 MHZ) |

| Protein number | 69 |
|---|---|
| PDB code | 2KCD, doi:10.2210/pdb2KCD/pdb |
| BMRB code | 16072, doi:10.13018/BMR16072 |
| PDB Header | STRUCTURAL GENOMICS, UNKNOWN FUNCTION |
| Protein name | SSP0047 FROM STAPHYLOCOCCUS SAPROPHYTICUS |
| Deposition date | 19.12.2008 |
| PDB title | SOLUTION NMR STRUCTURE OF SSP0047 FROM STAPHYLOCOCCUS SAPROPHYTICUS. NORTHEAST STRUCTURAL GENOMICS CONSORTIUM TARGET SYR6 |
| PDB authors | T.A.RAMELOT, K.DING, C.X.CHEN, M.JIANG, C.CICCOSANTI, R.XIAO, J.LIU, M.C.BARAN, G.SWAPNA, T.B.ACTON, B.ROST, G.T.MONTELIONE, M.A.KENNEDY, NORTHEAST STRUCTURAL GENOMICS CONSORTIUM (NESG) |
| Last author | KENNEDY |
| Reference | |
| Reference authors | |
| Reference title | |
| Software listed | AUTOASSIGN, AUTOSTRUCTURE, NMRPIPE, PDBSTAT, PSVS, SPARKY, TOPSPIN, VNMR, X-PLOR |
| Spectrometer | BRUKER, VARIAN (850 MHZ, 750 MHZ, 600 MHZ) |

| Protein number | 70 |
|---|---|
| PDB code | 2KRT, doi:10.2210/pdb2KRT/pdb |
| BMRB code | 16648, doi:10.13018/BMR16648 |
| PDB Header | LIPID BINDING PROTEIN |
| Protein name | A CONSERVED HYPOTHETICAL MEMBRANE LIPOPROTEIN OBTAINED FROM UREAPLASMA PARVUM |
| Deposition date | 22.12.2009 |
| PDB title | SOLUTION NMR STRUCTURE OF A CONSERVED HYPOTHETICAL MEMBRANE LIPOPROTEIN OBTAINED FROM UREAPLASMA PARVUM: NORTHEAST STRUCTURAL GENOMICS CONSORTIUM TARGET UUR17A (139-239) |
| PDB authors | R.MANI, G.SWAPNA, H.JANJUA, C.CICCOSANTI, Y.HUANG, D.PATEL, R.XIAO, T.ACTON, J.EVERETT, G.T.MONTELIONE, NORTHEAST STRUCTURAL GENOMICS CONSORTIUM (NESG) |
| Last author | MONTELIONE |
| Reference | |
| Reference authors | |
| Reference title | |
| Software listed | AUTOASSIGN, AUTOSTRUCTURE, CNS, CYANA, PINE |
| Spectrometer | BRUKER, VARIAN (800 MHZ, 600 MHZ, 500 MHZ) |

| Protein number | 71 |
|---|---|
| PDB code | 2LFI, doi:10.2210/pdb2LFI/pdb |
| BMRB code | 17754, doi:10.13018/BMR17754 |
| PDB Header | METAL BINDING PROTEIN |
| Protein name | A MUCBP DOMAIN (FRAGMENT 187-294) OF PROTEIN LBA1460 FROM LACTOBACILLUS ACIDOPHILUS |
| Deposition date | 30.06.2011 |
| PDB title | SOLUTION NMR STRUCTURE OF A MUCBP DOMAIN (FRAGMENT 187-294) OF THE PROTEIN LBA1460 FROM LACTOBACILLUS ACIDOPHILUS, NORTHEAST STRUCTURAL GENOMICS CONSORTIUM TARGET LAR80A |
| PDB authors | E.A.FELDMANN, T.A.RAMELOT, Y.YANG, H.LEE, C.CICCOSANTI, H.JANJUA, R.NAIR, T.B.ACTON, R.XIAO, J.K.EVERETT, J.H.PRESTEGARD, G.T.MONTELIONE, M.A.KENNEDY, NORTHEAST STRUCTURAL GENOMICS CONSORTIUM (NESG) |
| Last author | KENNEDY |
| Reference | |
| Reference authors | |
| Reference title | |
| Software listed | AUTOSTRUCTURE, CNS, CYANA, NMRPIPE, PDBSTAT, PINE_SERVER, PSVS, SPARKY, TOPSPIN, VNMR, X-PLOR_NIH |
| Spectrometer | BRUKER, VARIAN (850 MHZ, 600 MHZ) |

| Protein number | 72 |
|---|---|
| PDB code | 2JQN, doi:10.2210/pdb2JQN/pdb |
| BMRB code | 15281, doi:10.13018/BMR15281 |
| PDB Header | STRUCTURAL GENOMICS |
| Protein name | CC0527 FROM CAULOBACTER CRESCENTUS |
| Deposition date | 05.06.2007 |
| PDB title | SOLUTION NMR STRUCTURE OF CC0527 FROM CAULOBACTER CRESCENTUS. NORTHEAST STRUCTURAL GENOMICS TARGET CCR55 |
| PDB authors | J.M.ARAMINI, P.ROSSI, H.N.B.MOSELEY, D.WANG, C.NWOSU, K.CUNNINGHAM, L.MA, R.XIAO, J.LIU, M.C.BARAN, G.V.T.SWAPNA, T.B.ACTON, B.ROST, G.T.MONTELIONE, NORTHEAST STRUCTURAL GENOMICS CONSORTIUM (NESG) |
| Last author | MONTELIONE |
| Reference | |
| Reference authors | |
| Reference title | |
| Software listed | AUTOASSIGN, AUTOSTRUCTURE, CNS, NMRPIPE, PDBSTAT, PSVS, SPARKY, TOPSPIN, VNMR, X-PLOR |
| Spectrometer | BRUKER, VARIAN (800 MHZ, 600 MHZ) |
| Protein number | 73 |
| PDB code | 2L7Q, doi:10.2210/pdb2L7Q/pdb |
| BMRB code | 17370, doi:10.13018/BMR17370 |
| PDB Header | STRUCTURAL GENOMICS, UNKNOWN FUNCTION |
| Protein name | CONJUGATE TRANSPOSON PROTEIN BVU_1572(27- 141) FROM BACTEROIDES VULGATUS |
| Deposition date | 20.12.2010 |
| PDB title | SOLUTION NMR STRUCTURE OF CONJUGATE TRANSPOSON PROTEIN BVU_1572(27- 141) FROM BACTEROIDES VULGATUS, NORTHEAST STRUCTURAL GENOMICS CONSORTIUM TARGET BVR155 |
| PDB authors | Y.YANG, T.A.RAMELOT, J.R.CORT, D.WANG, C.CICCOSANTI, H.JANJUA, T.B.ACTON, R.XIAO, J.K.EVERETT, G.T.MONTELIONE, M.A.KENNEDY, NORTHEAST STRUCTURAL GENOMICS CONSORTIUM (NESG) |
| Last author | KENNEDY |
| Reference | PROTEINS 80, 667 (2012), doi:10.1002/PROT.23235 |
| Reference authors | T.A.RAMELOT, Y.YANG, R.XIAO, T.B.ACTON, J.K.EVERETT, G.T.MONTELIONE, M.A.KENNEDY |
| Reference title | SOLUTION NMR STRUCTURE OF BT_0084, A CONJUGATIVE TRANSPOSON LIPOPROTEIN FROM BACTEROIDES THETAIOTAMICRON |
| Software listed | AUTOASSIGN, AUTOSTRUCTURE, CNS, CYANA, NMRPIPE, PDBSTAT, PINE_SERVER, PSVS, SPARKY, TOPSPIN, VNMR, X-PLOR_NIH |
| Spectrometer | BRUKER, VARIAN (850 MHZ, 600 MHZ) |
| Protein number | 74 |
| PDB code | 2KFP, doi:10.2210/pdb2KFP/pdb |
| BMRB code | 16186, doi:10.13018/BMR16186 |
| PDB Header | STRUCTURAL GENOMICS, UNKNOWN FUNCTION |
| Protein name | PSPTO 3016 FROM PSEUDOMONAS SYRINGAE |
| Deposition date | 24.02.2009 |
| PDB title | SOLUTION NMR STRUCTURE OF PSPTO_3016 FROM PSEUDOMONAS SYRINGAE. NORTHEAST STRUCTURAL GENOMICS CONSORTIUM TARGET PSR293 |
| PDB authors | E.A.FELDMANN, T.A.RAMELOT, L.ZHAO, K.HAMILTON, C.CICCOSANTI, R.XIAO, R.NAIR, J.K.EVERETT, G.SWAPNA, T.B.ACTON, B.ROST, G.T.MONTELIONE, M.A.KENNEDY, NORTHEAST STRUCTURAL GENOMICS CONSORTIUM (NESG) |
| Last author | KENNEDY |
| Reference | J.STRUCT.FUNCT.GENOM. 13, 155 (2012), doi:10.1007/S10969-012-9140-8 |
| Reference authors | E.A.FELDMANN, J.SEETHARAMAN, T.A.RAMELOT, S.LEW, L.ZHAO, K.HAMILTON, C.CICCOSANTI, R.XIAO, T.B.ACTON, J.K.EVERETT, L.TONG, G.T.MONTELIONE, M.A.KENNEDY |
| Reference title | SOLUTION NMR AND X-RAY CRYSTAL STRUCTURES OF PSEUDOMONAS SYRINGAE PSPTO_3016 FROM PROTEIN DOMAIN FAMILY PF04237 (DUF419) ADOPT A "DOUBLE WING" DNA BINDING MOTIF |
| Software listed | AUTOASSIGN, AUTOSTRUCTURE, NMRPIPE, PSVS, SPARKY, TOPSPIN, VNMR, X-PLOR |
| Spectrometer | BRUKER, VARIAN (850 MHZ, 600 MHZ) |
| Protein number | 75 |
| PDB code | 1SE9, doi:10.2210/pdb1SE9/pdb |
| BMRB code | 6128, doi:10.13018/BMR6128 |
| PDB Header | PLANT PROTEIN |
| Protein name | AT3G01050, A UBIQUITIN-FOLD PROTEIN FROM ARABIDOPSIS THALIANA |
| Deposition date | 16.02.2004 |
| PDB title | STRUCTURE OF AT3G01050, A UBIQUITIN-FOLD PROTEIN FROM ARABIDOPSIS THALIANA |
| PDB authors | B.F.VOLKMAN, B.L.LYTLE, F.C.PETERSON, CENTER FOR EUKARYOTIC STRUCTURAL GENOMICS (CESG) |
| Last author | PETERSON |
| Reference | NAT.METHODS 1, 149 (2004), doi:10.1038/NMETH716 |
| Reference authors | D.A.VINAROV, B.L.LYTLE, F.C.PETERSON, E.M.TYLER, B.F.VOLKMAN, J.L.MARKLEY |
| Reference title | CELL-FREE PROTEIN PRODUCTION AND LABELING PROTOCOL FOR NMR-BASED STRUCTURAL PROTEOMICS |
| Software listed | CYANA, NMRPIPE, SPSCAN, X-PLOR_NIH, XEASY, XWINNMR |
| Spectrometer | BRUKER (600 MHZ) |
| Protein number | 76 |
| PDB code | 2L3G, doi:10.2210/pdb2L3G/pdb |
| BMRB code | 17192, doi:10.13018/BMR17192 |
| PDB Header | SIGNALING PROTEIN |
| Protein name | CH DOMAIN OF RHO GUANINE NUCLEOTIDE EXCHANGE FACTOR 7 FROM HOMO SAPIENS |
| Deposition date | 13.09.2010 |
| PDB title | SOLUTION NMR STRUCTURE OF CH DOMAIN OF RHO GUANINE NUCLEOTIDE EXCHANGE FACTOR 7 FROM HOMO SAPIENS, NORTHEAST STRUCTURAL GENOMICS CONSORTIUM TARGET HR4495E |
| PDB authors | G.LIU, R.XIAO, H.JANJUA, T.B.ACTON, A.CICCOSANTI, R.SHASTRY, J.EVERETT, G.T.MONTELIONE, NORTHEAST STRUCTURAL GENOMICS CONSORTIUM (NESG) |
| Last author | MONTELIONE |
| Reference | |

| Reference authors | |
|---|---|
| Reference title | |
| Software listed | AUTOASSIGN, AUTOSTRUCTURE, CNS, CYANA, NMRPIPE, TALOS+, TOPSPIN, VNMRJ, XEASY |
| Spectrometer | BRUKER, VARIAN (800 MHZ, 600 MHZ) |

| Protein number | 77 |
|---|---|
| PDB code | 2L3B, doi:10.2210/pdb2L3B/pdb |
| BMRB code | 17176, doi:10.13018/BMR17176 |
| PDB Header | STRUCTURAL GENOMICS, UNKNOWN FUNCTION |
| Protein name | BT_0084 LIPOPROTEIN FROM BACTEROIDES THETAIOTAOMICRON |
| Deposition date | 10.09.2010 |
| PDB title | SOLUTION NMR STRUCTURE OF THE BT_0084 LIPOPROTEIN FROM BACTEROIDES THETAIOTAOMICRON, NORTHEAST STRUCTURAL GENOMICS CONSORTIUM TARGET BTR376 |
| PDB authors | T.A.RAMELOT, Y.YANG, D.WANG, C.CICCOSANTI, H.JANJUA, T.B.ACTON, R.XIAO, J.K.EVERETT, G.T.MONTELIONE, M.A.KENNEDY, NORTHEAST STRUCTURAL GENOMICS CONSORTIUM (NESG) |
| Last author | KENNEDY |
| Reference | PROTEINS    867 2012, doi:10.1002/PROT.23235 |
| Reference authors | T.A.RAMELOT, Y.YANG, R.XIAO, T.B.ACTON, J.K.EVERETT, G.T.MONTELIONE, M.A.KENNEDY |
| Reference title | SOLUTION NMR STRUCTURE OF BT_0084, A CONJUGATIVE TRANSPOSON LIPOPROTEIN FROM BACTEROIDES THETAIOTAMICRON |
| Software listed | AUTOASSIGN, AUTOSTRUCTURE, CYANA, NMRPIPE, PDBSTAT, PSVS, SPARKY, TOPSPIN, VNMR, X-PLOR |
| Spectrometer | BRUKER, VARIAN (850 MHZ, 600 MHZ) |

| Protein number | 78 |
|---|---|
| PDB code | 2LRH, doi:10.2210/pdb2LRH/pdb |
| BMRB code | 18372, doi:10.13018/BMR18372 |
| PDB Header | DE NOVO PROTEIN |
| Protein name | DE NOVO DESIGNED PROTEIN, P-LOOP NTPASE FOLD |
| Deposition date | 30.03.2012 |
| PDB title | SOLUTION NMR STRUCTURE OF DE NOVO DESIGNED PROTEIN, P-LOOP NTPASE FOLD, NORTHEAST STRUCTURAL GENOMICS CONSORTIUM TARGET OR137 |
| PDB authors | G.LIU, N.KOGA, R.KOGA, R.XIAO, H.LEE, H.JANJUA, E.KOHAN, T.B.ACTON, J.K.EVERETT, D.BAKER, G.T.MONTELIONE, NORTHEAST STRUCTURAL GENOMICS CONSORTIUM (NESG) |
| Last author | MONTELIONE |
| Reference | |
| Reference authors | |
| Reference title | |
| Software listed | AUTOASSIGN, AUTOSTRUCTURE, CNS, CYANA, NMRPIPE, PALES, PSVS, REDCAT, SPARKY, TALOS+, TOPSPIN, VNMRJ, XEASY |
| Spectrometer | BRUKER, VARIAN (800 MHZ, 600 MHZ) |

| Protein number | 79 |
|---|---|
| PDB code | 1VEE, doi:10.2210/pdb1VEE/pdb |
| BMRB code | 5929, doi:10.13018/BMR5929 |
| PDB Header | STRUCTURAL GENOMICS, UNKNOWN FUNCTION |
| Protein name | HYPOTHETICAL RHODANESE DOMAIN AT4G01050 FROM ARABIDOPSIS THALIANA |
| Deposition date | 30.03.2004 |
| PDB title | NMR STRUCTURE OF THE HYPOTHETICAL RHODANESE DOMAIN AT4G01050 FROM ARABIDOPSIS THALIANA |
| PDB authors | D.PANTOJA-UCEDA, B.LOPEZ-MENDEZ, S.KOSHIBA, M.INOUE, T.KIGAWA, T.TERADA, M.SHIROUZU, A.TANAKA, M.SEKI, K.SHINOZAKI, S.YOKOYAMA, P.GUNTERT, RIKEN STRUCTURAL GENOMICS/PROTEOMICS INITIATIVE (RSGI) |
| Last author | GUNTERT |
| Reference | PROTEIN SCI. 14, 224 (2005), doi:10.1110/PS.041138705 |
| Reference authors | D.PANTOJA-UCEDA, B.LOPEZ-MENDEZ, S.KOSHIBA, M.INOUE, T.KIGAWA, T.TERADA, M.SHIROUZU, A.TANAKA, M.SEKI, K.SHINOZAKI, S.YOKOYAMA, P.GUNTERT |
| Reference title | SOLUTION STRUCTURE OF THE RHODANESE HOMOLOGY DOMAIN AT4G01050(175-295) FROM ARABIDOPSIS THALIANA |
| Software listed | CYANA, OPALP |
| Spectrometer | BRUKER (800 MHZ, 600 MHZ) |

| Protein number | 80 |
|---|---|
| PDB code | 2K1G, doi:10.2210/pdb2K1G/pdb |
| BMRB code | 15603, doi:10.13018/BMR15603 |
| PDB Header | LIPOPROTEIN |
| Protein name | LIPOPROTEIN SPR FROM ESCHERICHIA COLI K12 |
| Deposition date | 03.03.2008 |
| PDB title | SOLUTION NMR STRUCTURE OF LIPOPROTEIN SPR FROM ESCHERICHIA COLI K12. NORTHEAST STRUCTURAL GENOMICS TARGET ER541-37-162 |
| PDB authors | J.M.ARAMINI, P.ROSSI, L.ZHAO, M.JIANG, M.MAGLAQUI, R.XIAO, J.LIU, M.C.BARAN, G.V.T.SWAPNA, Y.J.HUANG, T.B.ACTON, B.ROST, G.T.MONTELIONE, NORTHEAST STRUCTURAL GENOMICS CONSORTIUM (NESG) |
| Last author | MONTELIONE |
| Reference | BIOCHEMISTRY 47, 9715 (2008), doi:10.1021/BI8010779 |
| Reference authors | J.M.ARAMINI, P.ROSSI, Y.J.HUANG, L.ZHAO, M.JIANG, M.MAGLAQUI, R.XIAO, J.LOCKE, R.NAIR, B.ROST, T.B.ACTON, M.INOUYE, G.T.MONTELIONE |

| Reference title | SOLUTION NMR STRUCTURE OF THE NLPC/P60 DOMAIN OF LIPOPROTEIN SPR FROM ESCHERICHIA COLI: STRUCTURAL EVIDENCE FOR A NOVEL CYSTEINE PEPTIDASE CATALYTIC TRIAD |
| --- | --- |
| Software listed | AUTOASSIGN, AUTOSTRUCTURE, CNS, CYANA, NMRPIPE, PDBSTAT, PSVS, SPARKY, TOPSPIN, VNMR |
| Spectrometer | BRUKER, VARIAN (800 MHZ, 600 MHZ) |

| Protein number | 81 |
| --- | --- |
| PDB code | 2KKZ, doi:10.2210/pdb2KKZ/pdb |
| BMRB code | 16376, doi:10.13018/BMR16376 |
| PDB Header | ANTIVIRAL PROTEIN |
| Protein name | MONOMERIC W187R MUTANT OF A/UDORN NS1 EFFECTOR DOMAIN |
| Deposition date | 29.06.2009 |
| PDB title | SOLUTION NMR STRUCTURE OF THE MONOMERIC W187R MUTANT OF A/UDORN NS1 EFFECTOR DOMAIN. NORTHEAST STRUCTURAL GENOMICS TARGET OR8C[W187R] |
| PDB authors | J.M.ARAMINI, L.MA, H.LEE, L.ZHAO, K.CUNNINGHAM, C.CICCOSANTI, H.JANJUA, Y.FANG, R.XIAO, R.M.KRUG, G.T.MONTELIONE, NORTHEAST STRUCTURAL GENOMICS CONSORTIUM (NESG) |
| Last author | MONTELIONE |
| Reference | |
| Reference authors | |
| Reference title | |
| Software listed | AUTOASSIGN, AUTOSTRUCTURE, CNS, CYANA, MOLPROBITY, NMRPIPE, PDBSTAT, PSVS, SPARKY, TALOS, TOPSPIN |
| Spectrometer | BRUKER (800 MHZ, 600 MHZ) |

| Protein number | 82 |
| --- | --- |
| PDB code | 1VDY, doi:10.2210/pdb1VDY/pdb |
| BMRB code | 5928, doi:10.13018/BMR5928 |
| PDB Header | STRUCTURAL GENOMICS, UNKNOWN FUNCTION |
| Protein name | HYPOTHETICAL ENTH-VHS DOMAIN AT3G16270 FROM ARABIDOPSIS THALIANA |
| Deposition date | 25.03.2004 |
| PDB title | NMR STRUCTURE OF THE HYPOTHETICAL ENTH-VHS DOMAIN AT3G16270 FROM ARABIDOPSIS THALIANA |
| PDB authors | B.LOPEZ-MENDEZ, D.PANTOJA-UCEDA, T.TOMIZAWA, S.KOSHIBA, T.KIGAWA, M.SHIROUZU, T.TERADA, M.INOUE, T.YABUKI, M.AOKI, E.SEKI, T.MATSUDA, H.HIROTA, M.YOSHIDA, A.TANAKA, T.OSANAI, M.SEKI, K.SHINOZAKI, S.YOKOYAMA, P.GUNTERT, RIKEN STRUCTURAL GENOMICS/PROTEOMICS INITIATIVE (RSGI) |
| Last author | GUNTERT |
| Reference | |
| Reference authors | |
| Reference title | |
| Software listed | CYANA, OPALP |
| Spectrometer | BRUKER (800 MHZ, 600 MHZ) |

| Protein number | 83 |
| --- | --- |
| PDB code | 2KKL, doi:10.2210/pdb2KKL/pdb |
| BMRB code | 16364, doi:10.13018/BMR16364 |
| PDB Header | STRUCTURAL GENOMICS, UNKNOWN FUNCTION |
| Protein name | FHA DOMAIN OF MB1858 FROM MYCOBACTERIUM BOVIS |
| Deposition date | 25.06.2009 |
| PDB title | SOLUTION NMR STRUCTURE OF FHA DOMAIN OF MB1858 FROM MYCOBACTERIUM BOVIS. NORTHEAST STRUCTURAL GENOMICS CONSORTIUM TARGET MBR243C (24- 155) |
| PDB authors | Y.YANG, T.A.RAMELOT, D.WANG, E.L.FOOTE, M.JIANG, R.NAIR, B.ROST, G.SWAPNA, T.B.ACTON, R.XIAO, J.K.EVERETT, G.T.MONTELIONE, M.A.KENNEDY, NORTHEAST STRUCTURAL GENOMICS CONSORTIUM (NESG) |
| Last author | KENNEDY |
| Reference | |
| Reference authors | |
| Reference title | |
| Software listed | AUTOASSIGN, AUTOSTRUCTURE, CNS, NMRPIPE, PDBSTAT, PSVS, SPARKY, TOPSPIN, VNMR, X-PLOR |
| Spectrometer | BRUKER, VARIAN (850 MHZ, 600 MHZ) |

| Protein number | 84 |
| --- | --- |
| PDB code | 2N4B, doi:10.2210/pdb2N4B/pdb |
| BMRB code | 17611, doi:10.13018/BMR17611 |
| PDB Header | STRUCTURAL GENOMICS, UNKNOWN FUNCTION |
| Protein name | RALSTONIA METALLIDURANS RMET_5065 DETERMINED BY COMBINING EVOLUTIONARY COUPLINGS (EC) AND SPARSE NMR DATA |
| Deposition date | 17.06.2015 |
| PDB title | EC-NMR STRUCTURE OF RALSTONIA METALLIDURANS RMET_5065 DETERMINED BY COMBINING EVOLUTIONARY COUPLINGS (EC) AND SPARSE NMR DATA. NORTHEAST STRUCTURAL GENOMICS CONSORTIUM TARGET CRR115 |
| PDB authors | Y.TANG, Y.J.HUANG, T.A.HOPF, C.SANDER, D.MARKS, G.T.MONTELIONE, NORTHEAST STRUCTURAL GENOMICS CONSORTIUM (NESG) |
| Last author | MONTELIONE |
| Reference | NAT.METHODS 12, 751 (2015), doi:10.1038/NMETH.3455 |
| Reference authors | Y.TANG, Y.J.HUANG, T.A.HOPF, C.SANDER, D.S.MARKS, G.T.MONTELIONE |

| Reference title | PROTEIN STRUCTURE DETERMINATION BY COMBINING SPARSE NMR DATA WITH EVOLUTIONARY COUPLINGS |
|---|---|
| Software listed | ASDP, CYANA, EC-NMR, EVFOLD-PLM, REDUCE, ROSETTA, TALOS+ |
| Spectrometer | |

| Protein number | 85 |
|---|---|
| PDB code | 2L8V, doi:10.2210/pdb2L8V/pdb |
| BMRB code | 17429, doi:10.13018/BMR17429 |
| PDB Header | PHOTOSYNTHESIS |
| Protein name | PHYCOBILISOME LINKER POLYPEPTIDE DOMAIN OF CPCC (20-153) FROM THERMOSYNECHOCOCCUS ELONGATUS |
| Deposition date | 26.01.2011 |
| PDB title | SOLUTION NMR STRUCTURE OF THE PHYCOBILISOME LINKER POLYPEPTIDE DOMAIN OF CPCC (20-153) FROM THERMOSYNECHOCOCCUS ELONGATUS, NORTHEAST STRUCTURAL GENOMICS CONSORTIUM TARGET TER219A |
| PDB authors | T.A.RAMELOT, Y.YANG, J.R.CORT, D.LEE, C.CICCOSANTI, K.HAMILTON, T.B.ACTON, R.XIAO, J.K.EVERETT, G.T.MONTELIONE, M.A.KENNEDY, NORTHEAST STRUCTURAL GENOMICS CONSORTIUM (NESG) |
| Last author | KENNEDY |
| Reference | |
| Reference authors | |
| Reference title | |
| Software listed | AUTOASSIGN, AUTOSTRUCTURE, CNS, CYANA, NMRPIPE, PDBSTAT, PINE_SERVER, PSVS, SPARKY, TOPSPIN, VNMR, X-PLOR_NIH |
| Spectrometer | BRUKER, VARIAN (850 MHZ, 600 MHZ) |

| Protein number | 86 |
|---|---|
| PDB code | 2LGH, doi:10.2210/pdb2LGH/pdb |
| BMRB code | 17809, doi:10.13018/BMR17809 |
| PDB Header | STRUCTURAL GENOMICS, UNKNOWN FUNCTION |
| Protein name | AHSA1-LIKE PROTEIN AHA_2358 FROM AEROMONAS HYDROPHILA REFINED WITH NH RDCS |
| Deposition date | 26.07.2011 |
| PDB title | SOLUTION NMR STRUCTURE OF THE AHSA1-LIKE PROTEIN AHA_2358 FROM AEROMONAS HYDROPHILA REFINED WITH NH RDCS, NORTHEAST STRUCTURAL GENOMICS CONSORTIUM TARGET AHR99 |
| PDB authors | T.A.RAMELOT, Y.YANG, H.LEE, D.WANG, C.CICCOSANTI, H.JANJUA, R.NAIR, T.B.ACTON, R.XIAO, J.K.EVERETT, J.H.PRESTEGARD, G.T.MONTELIONE, M.A.KENNEDY, NORTHEAST STRUCTURAL GENOMICS CONSORTIUM (NESG) |
| Last author | KENNEDY |
| Reference | |
| Reference authors | |
| Reference title | |
| Software listed | AUTOASSIGN, AUTOSTRUCTURE, CNS, CYANA, NMRPIPE, PDBSTAT, PINE_SERVER, PSVS, SPARKY, TOPSPIN, VNMR, X-PLOR_NIH |
| Spectrometer | BRUKER, VARIAN (850 MHZ, 600 MHZ) |

| Protein number | 87 |
|---|---|
| PDB code | 2K1S, doi:10.2210/pdb2K1S/pdb |
| BMRB code | 15683, doi:10.13018/BMR15683 |
| PDB Header | LIPOPROTEIN |
| Protein name | FOLDED C-TERMINAL FRAGMENT OF YIAD FROM ESCHERICHIA COLI |
| Deposition date | 14.03.2008 |
| PDB title | SOLUTION NMR STRUCTURE OF THE FOLDED C-TERMINAL FRAGMENT OF YIAD FROM ESCHERICHIA COLI. NORTHEAST STRUCTURAL GENOMICS CONSORTIUM TARGET ER553 |
| PDB authors | T.A.RAMELOT, L.ZHAO, K.HAMILTON, M.MAGLAQUI, R.XIAO, J.LIU, M.C.BARAN, G.SWAPNA, T.B.ACTON, B.ROST, G.T.MONTELIONE, M.A.KENNEDY, NORTHEAST STRUCTURAL GENOMICS CONSORTIUM (NESG) |
| Last author | KENNEDY |
| Reference | |
| Reference authors | |
| Reference title | |
| Software listed | AUTOASSIGN, AUTOSTRUCTURE, CNS, NMRPIPE, PSVS, SPARKY, TOPSPIN, VNMR, X-PLOR |
| Spectrometer | BRUKER, VARIAN (850 MHZ, 600 MHZ) |

| Protein number | 88 |
|---|---|
| PDB code | 2M4F, doi:10.2210/pdb2M4F/pdb |
| BMRB code | 19001, doi:10.13018/BMR19001 |
| PDB Header | IMMUNE SYSTEM |
| Protein name | OUTER SURFACE PROTEIN E |
| Deposition date | 05.02.2013 |
| PDB title | SOLUTION STRUCTURE OF OUTER SURFACE PROTEIN E |
| PDB authors | A.BHATTACHARJEE, J.S.OEEMIG, R.KOLODZIEJCZYK, T.MERI, T.KAJANDER, H.IWAI, T.JOKIRANTA, A.GOLDMAN |
| Last author | GOLDMAN |
| Reference | J.BIOL.CHEM. 288, 18685 (2013), doi:10.1074/JBC.M113.459040 |
| Reference authors | A.BHATTACHARJEE, J.S.OEEMIG, R.KOLODZIEJCZYK, T.MERI, T.KAJANDER, M.J.LEHTINEN, H.IWAI, T.S.JOKIRANTA, A.GOLDMAN |
| Reference title | STRUCTURAL BASIS FOR COMPLEMENT EVASION BY LYME DISEASE PATHOGEN BORRELIA BURGDORFERI |
| Software listed | AMBER, CCPNMR_ANALYSIS, CING, CYANA, NMRPIPE, VNMRJ |

| Spectrometer | VARIAN (800 MHZ, 600 MHZ) |
|---|---|

| Protein number | 89 |
|---|---|
| PDB code | 2JXP, doi:10.2210/pdb2JXP/pdb |
| BMRB code | 15568, doi:10.13018/BMR15568 |
| PDB Header | LIPOPROTEIN |
| Protein name | UNCHARACTERIZED LIPOPROTEIN B FROM NITROSOMONAS EUROPAEA |
| Deposition date | 27.11.2007 |
| PDB title | SOLUTION NMR STRUCTURE OF UNCHARACTERIZED LIPOPROTEIN B FROM NITROSOMONAS EUROPAEA. NORTHEAST STRUCTURAL GENOMICS TARGET NER45A |
| PDB authors | P.ROSSI, D.WANG, H.JANJUA, L.OWENS, R.XIAO, M.C.BARAN, G.SWAPNA, T.B.ACTON, B.ROST, G.T.MONTELIONE, NORTHEAST STRUCTURAL GENOMICS CONSORTIUM (NESG) |
| Last author | MONTELIONE |
| Reference | |
| Reference authors | |
| Reference title | |
| Software listed | AUTOASSIGN, CNS, CYANA, MOLMOL, MOLPROBITY, NMRPIPE, PROCHECK, PSVS, RPF, SPARKY, TOPSPIN, XEASY |
| Spectrometer | BRUKER (800 MHZ) |

| Protein number | 90 |
|---|---|
| PDB code | 2L06, doi:10.2210/pdb2L06/pdb |
| BMRB code | 17031, doi:10.13018/BMR17031 |
| PDB Header | PROTEIN BINDING |
| Protein name | PBS LINKER POLYPEPTIDE DOMAIN (FRAGMENT 254-400) OF PHYCOBILISOME LINKER PROTEIN APCE FROM SYNECHOCYSTIS SP. PCC 6803 |
| Deposition date | 30.06.2010 |
| PDB title | SOLUTION NMR STRUCTURE OF THE PBS LINKER POLYPEPTIDE DOMAIN (FRAGMENT 254-400) OF PHYCOBILISOME LINKER PROTEIN APCE FROM SYNECHOCYSTIS SP. PCC 6803. NORTHEAST STRUCTURAL GENOMICS CONSORTIUM TARGET SGR209C |
| PDB authors | T.A.RAMELOT, Y.YANG, J.R.CORT, K.HAMILTON, C.CICCOSANTI, D.LEE, T.B.ACTON, R.XIAO, J.K.EVERETT, G.T.MONTELIONE, M.A.KENNEDY, NORTHEAST STRUCTURAL GENOMICS CONSORTIUM (NESG) |
| Last author | KENNEDY |
| Reference | |
| Reference authors | |
| Reference title | |
| Software listed | AUTOASSIGN, AUTOSTRUCTURE, CYANA, NMRPIPE, PDBSTAT, PSVS, SPARKY, TOPSPIN, VNMR, X-PLOR |
| Spectrometer | BRUKER, VARIAN (850 MHZ, 750 MHZ, 600 MHZ, 500 MHZ) |

| Protein number | 91 |
|---|---|
| PDB code | 2LAH, doi:10.2210/pdb2LAH/pdb |
| BMRB code | 17524, doi:10.13018/BMR17524 |
| PDB Header | CELL CYCLE, APOPTOSIS |
| Protein name | MITOTIC CHECKPOINT SERINE/THREONINE-PROTEIN KINASE BUB1 N-TERMINAL DOMAIN FROM HOMO SAPIENS |
| Deposition date | 14.03.2011 |
| PDB title | SOLUTION NMR STRUCTURE OF MITOTIC CHECKPOINT SERINE/THREONINE-PROTEIN KINASE BUB1 N-TERMINAL DOMAIN FROM HOMO SAPIENS, NORTHEAST STRUCTURAL GENOMICS CONSORTIUM TARGET HR5460A (METHODS DEVELOPMENT) |
| PDB authors | G.LIU, R.XIAO, H.LEE, K.HAMILTON, T.B.ACTON, C.CICCOSANTI, J.K.EVERETT, R.T.SHASTRY, Y.J.HUANG, G.T.MONTELIONE, N.NORTHEAST STRUCTURAL GENOMICS CONSORTIUM, NORTHEAST STRUCTURAL GENOMICS CONSORTIUM (NESG) |
| Last author | MONTELIONE |
| Reference | |
| Reference authors | |
| Reference title | |
| Software listed | AUTOASSIGN, AUTOSTRUCTURE, CNS, CYANA, NMRPIPE, SPARKY, TALOS+, TOPSPIN, VNMRJ, XEASY |
| Spectrometer | BRUKER, VARIAN (800 MHZ, 600 MHZ) |

| Protein number | 92 |
|---|---|
| PDB code | 2LAK, doi:10.2210/pdb2LAK/pdb |
| BMRB code | 17530, doi:10.13018/BMR17530 |
| PDB Header | STRUCTURE GENOMICS, UNKNOWN FUNCTION |
| Protein name | AHSA1-LIKE PROTEIN RHE_CH02687 (1-152) FROM RHIZOBIUM ETLI |
| Deposition date | 16.03.2011 |
| PDB title | SOLUTION NMR STRUCTURE OF THE AHSA1-LIKE PROTEIN RHE_CH02687 (1-152) FROM RHIZOBIUM ETLI, NORTHEAST STRUCTURAL GENOMICS CONSORTIUM TARGET RER242 |
| PDB authors | Y.YANG, T.A.RAMELOT, J.R.CORT, D.WANG, C.CICCOSANTI, H.JANJUA, R.NAIR, B.ROST, T.B.ACTON, R.XIAO, J.K.EVERETT, G.T.MONTELIONE, M.A.KENNEDY, NORTHEAST STRUCTURAL GENOMICS CONSORTIUM (NESG) |
| Last author | KENNEDY |
| Reference | |
| Reference authors | |
| Reference title | |
| Software listed | AUTOASSIGN, AUTOSTRUCTURE, CNS, CYANA, NMRPIPE, PDBSTAT, PINE_SERVER, PSVS, SPARKY, TOPSPIN, VNMR, X-PLOR_NIH |
| Spectrometer | BRUKER, VARIAN (850 MHZ, 600 MHZ) |

| Protein number | 93 |
|---|---|
| PDB code | 2L82, doi:10.2210/pdb2L82/pdb |
| BMRB code | 17390, doi:10.13018/BMR17390 |
| PDB Header | DE NOVO PROTEIN |
| Protein name | DE NOVO DESIGNED PROTEIN, P-LOOP NTPASE FOLD |
| Deposition date | 31.12.2010 |
| PDB title | SOLUTION NMR STRUCTURE OF DE NOVO DESIGNED PROTEIN, P-LOOP NTPASE FOLD, NORTHEAST STRUCTURAL GENOMICS CONSORTIUM TARGET OR32 |
| PDB authors | G.LIU, N.KOGA, R.KOGA, R.XIAO, K.HAMILTON, H.JANJUA, S.TONG, T.B.ACTON, J.EVERETT, D.BAKER, G.T.MONTELIONE, NORTHEAST STRUCTURAL GENOMICS CONSORTIUM (NESG) |
| Last author | MONTELIONE |
| Reference | |
| Reference authors | |
| Reference title | |
| Software listed | AUTOASSIGN, AUTOSTRUCTURE, CNS, CYANA, NMRPIPE, PALES, PINE, REDCAT, SPARKY, TALOS+, TOPSPIN, VNMRJ, XEASY |
| Spectrometer | BRUKER, VARIAN (800 MHZ, 600 MHZ) |

| Protein number | 94 |
|---|---|
| PDB code | 2M47, doi:10.2210/pdb2M47/pdb |
| BMRB code | 18989, doi:10.13018/BMR18989 |
| PDB Header | STRUCTURAL GENOMICS, UNKNOWN FUNCTION |
| Protein name | POLYKETIDE_CYC-LIKE PROTEIN CGL2372 FROM CORYNEBACTERIUM GLUTAMICUM |
| Deposition date | 30.01.2013 |
| PDB title | SOLUTION NMR STRUCTURE OF THE POLYKETIDE_CYC-LIKE PROTEIN CGL2372 FROM CORYNEBACTERIUM GLUTAMICUM, NORTHEAST STRUCTURAL GENOMICS CONSORTIUM TARGET CGR160 |
| PDB authors | Y.YANG, T.A.RAMELOT, D.LEE, C.CICCOSANTI, A.SAPIN, H.JANJUA, R.NAIR, B.ROST, T.B.ACTON, R.XIAO, J.K.EVERETT, G.T.MONTELIONE, M.A.KENNEDY, NORTHEAST STRUCTURAL GENOMICS CONSORTIUM (NESG) |
| Last author | KENNEDY |
| Reference | |
| Reference authors | |
| Reference title | |
| Software listed | AUTOASSIGN, AUTOSTRUCTURE, CNS, CYANA, NMRPIPE, PINE, PSVS, SPARKY, TALOS+, TOPSPIN, VNMRJ, XEASY |
| Spectrometer | BRUKER, VARIAN (850 MHZ, 600 MHZ) |

| Protein number | 95 |
|---|---|
| PDB code | 2K3A, doi:10.2210/pdb2K3A/pdb |
| BMRB code | 15335, doi:10.13018/BMR15335 |
| PDB Header | HYDROLASE |
| Protein name | STAPHYLOCOCCUS SAPROPHYTICUS CHAP (CYSTEINE, HISTIDINE-DEPENDENT AMIDOHYDROLASES/PEPTIDASES) DOMAIN PROTEIN |
| Deposition date | 29.04.2008 |
| PDB title | NMR SOLUTION STRUCTURE OF STAPHYLOCOCCUS SAPROPHYTICUS CHAP (CYSTEINE, HISTIDINE-DEPENDENT AMIDOHYDROLASES/PEPTIDASES) DOMAIN PROTEIN. NORTHEAST STRUCTURAL GENOMICS CONSORTIUM TARGET SYR11 |
| PDB authors | P.ROSSI, J.M.ARAMINI, C.X.CHEN, C.NWOSU, K.C.CUNNINGHAM, L.A.OWENS, R.XIAO, J.LIU, M.C.BARAN, G.SWAPNA, T.B.ACTON, B.ROST, G.T.MONTELIONE, NORTHEAST STRUCTURAL GENOMICS CONSORTIUM (NESG) |
| Last author | MONTELIONE |
| Reference | PROTEINS 74, 515 (2008), doi:10.1002/PROT.22267 |
| Reference authors | P.ROSSI, J.M.ARAMINI, R.XIAO, C.X.CHEN, C.NWOSU, L.A.OWENS, M.MAGLAQUI, R.NAIR, M.FISCHER, T.B.ACTON, B.HONIG, B.ROST, G.T.MONTELIONE |
| Reference title | STRUCTURAL ELUCIDATION OF THE CYS-HIS-GLU-ASN PROTEOLYTIC RELAY IN THE SECRETED CHAP DOMAIN ENZYME FROM THE HUMAN PATHOGEN STAPHYLOCOCCUS SAPROPHYTICUS |
| Software listed | AUTOASSIGN, AUTOSTRUCTURE, CNS, CYANA, MOLMOL, MOLPROBITY, NMRPIPE, PDBSTAT, PROCHECK, PROSA, PSVS, SPARKY, TOPSPIN, VERIFY3D, X-PLOR |
| Spectrometer | BRUKER (800 MHZ, 600 MHZ) |

| Protein number | 96 |
|---|---|
| PDB code | 2M7U, doi:10.2210/pdb2M7U/pdb |
| BMRB code | 19213, doi:10.13018/BMR19213 |
| PDB Header | SIGNALING PROTEIN |
| Protein name | BLUE LIGHT-ABSORBING STATE OF TEPIXJ, AN ACTIVE CYANOBACTERIOCHROME DOMAIN |
| Deposition date | 01.05.2013 |
| PDB title | BLUE LIGHT-ABSORBING STATE OF TEPIXJ, AN ACTIVE CYANOBACTERIOCHROME DOMAIN |
| PDB authors | G.CORNILESCU, C.C.CORNILESCU, S.E.BURGIE, J.M.WALKER, J.L.MARKLEY, A.T.ULIJASZ, R.D.VIERSTRA |
| Last author | VIERSTRA |
| Reference | |
| Reference authors | |
| Reference title | |
| Software listed | NMRPIPE, PIPP, X-PLOR_NIH |
| Spectrometer | BRUKER, VARIAN (900 MHZ, 800 MHZ, 700 MHZ, 600 MHZ) |
| Protein number | 97 |
| PDB code | 2B3W, doi:10.2210/pdb2B3W/pdb |
| BMRB code | 6782, doi:10.13018/BMR6782 |
| PDB Header | STRUCTURAL GENOMICS, UNKNOWN FUNCTION |
| Protein name | E.COLI PROTEIN YBIA |
| Deposition date | 21.09.2005 |
| PDB title | NMR STRUCTURE OF THE E.COLI PROTEIN YBIA, NORTHEAST STRUCTURAL GENOMICS TARGET ET24 |
| PDB authors | T.A.RAMELOT, J.R.CORT, R.XIAO, L.Y.SHIH, T.B.ACTON, G.T.MONTELIONE, M.A.KENNEDY, NORTHEAST STRUCTURAL GENOMICS CONSORTIUM (NESG) |
| Last author | KENNEDY |
| Reference | |
| Reference authors | |
| Reference title | |
| Software listed | AUTOSTRUCTURE, CNS, NMRPIPE, SPARKY, VNMR, X-PLOR |
| Spectrometer | VARIAN (800 MHZ, 750 MHZ, 600 MHZ) |
| Protein number | 98 |
| PDB code | (KRAS4B) |
| BMRB code | |
| PDB Header | |
| Protein name | |
| Deposition date | |
| PDB title | |
| PDB authors | |
| Last author | |
| Reference | |
| Reference authors | |
| Reference title | |
| Software listed | |
| Spectrometer | |
| Protein number | 99 |
| PDB code | 2G0Q, doi:10.2210/pdb2G0Q/pdb |
| BMRB code | 7007, doi:10.13018/BMR7007 |
| PDB Header | STRUCTURAL GENOMICS, UNKNOWN FUNCTION |
| Protein name | AT5G39720.1 FROM ARABIDOPSIS THALIANA |
| Deposition date | 13.02.2006 |
| PDB title | SOLUTION STRUCTURE OF AT5G39720.1 FROM ARABIDOPSIS THALIANA |
| PDB authors | B.F.VOLKMAN, F.C.PETERSON, B.L.LYTLE, CENTER FOR EUKARYOTIC STRUCTURAL GENOMICS (CESG) |
| Last author | LYTLE |
| Reference | ACTA CRYSTALLOGR.,SECT.F 62, 490 (2006), doi:10.1107/S1744309106015946 |
| Reference authors | B.L.LYTLE, F.C.PETERSON, E.M.TYLER, C.L.NEWMAN, D.A.VINAROV, J.L.MARKLEY, B.F.VOLKMAN |
| Reference title | SOLUTION STRUCTURE OF ARABIDOPSIS THALIANA PROTEIN AT5G39720.1, A MEMBER OF THE AIG2-LIKE PROTEIN FAMILY |
| Software listed | GARANT, NMRPIPE, SPSCAN, X-PLOR_NIH, XEASY, XWINNMR |
| Spectrometer | BRUKER (600 MHZ) |
| Protein number | 100 |
| PDB code | 2LF2, doi:10.2210/pdb2LF2/pdb |
| BMRB code | 17736, doi:10.13018/BMR17736 |
| PDB Header | STRUCTURAL GENOMICS, UNKNOWN FUNCTION |
| Protein name | AHSA1-LIKE PROTEIN CHU_1110 FROM CYTOPHAGA HUTCHINSONII |
| Deposition date | 28.06.2011 |
| PDB title | SOLUTION NMR STRUCTURE OF THE AHSA1-LIKE PROTEIN CHU_1110 FROM CYTOPHAGA HUTCHINSONII, NORTHEAST STRUCTURAL GENOMICS CONSORTIUM TARGET CHR152 |
| PDB authors | Y.YANG, T.A.RAMELOT, D.LEE, C.CICCOSANTI, T.B.ACTON, R.XIAO, J.K.EVERETT, G.T.MONTELIONE, M.A.KENNEDY, NORTHEAST STRUCTURAL GENOMICS CONSORTIUM (NESG) |
| Last author | KENNEDY |
| Reference | |
| Reference authors | |
| Reference title | |
| Software listed | AUTOASSIGN, AUTOSTRUCTURE, CNS, CYANA, NMRPIPE, PDBSTAT, PINE_SERVER, PSVS, SPARKY, TOPSPIN, VNMR, X-PLOR_NIH |
| Spectrometer | BRUKER, VARIAN (850 MHZ, 600 MHZ) |

**Supplementary Table 4 Results of automated structure determination of 100 proteins.**

| ID | Protein | Chemical shift assignment | | Ramachandran plot statistics | | | | Structure calculation and comparison to PDB reference | | | | |
|---|---|---|---|---|---|---|---|---|---|---|---|---|
| | | Backbone accuracy [%] | Side-chain accuracy [%] | Most favoured [%] | Additionally allowed [%] | Generously allowed [%] | Disallowed [%] | Number of distance restraints | CYANA target function [Å²] | Residue range for RMSD calculation | Backbone RMSD to reference [Å] | Side-chain RMSD to reference [Å] |
| 1 | 6SVC | 97.52 | 83.97 | 96.40 | 3.60 | 0.00 | 0.00 | 1066 | 1.80 | 7-29 | 0.83 | 1.53 |
| 2 | 2JVD | 99.57 | 92.94 | 89.80 | 10.20 | 0.00 | 0.00 | 1182 | 0.30 | 4-37 | 0.71 | 1.29 |
| 3 | 2K57 | 99.62 | 95.38 | 88.20 | 11.80 | 0.00 | 0.00 | 1176 | 0.35 | 5-52 | 0.71 | 1.22 |
| 4 | 6SOW | 88.97 | 78.57 | 87.70 | 12.30 | 0.00 | 0.00 | 1111 | 2.32 | 8-55 | 1.16 | 2.04 |
| 5 | 2LX7 | 98.54 | 87.84 | 91.40 | 8.60 | 0.00 | 0.00 | 484 | 0.32 | 5-59 | 1.41 | 2.02 |
| 6 | 2MA6 | 98.94 | 89.36 | 67.40 | 30.00 | 2.60 | 0.00 | 953 | 3.25 | 10-57 | 1.56 | 2.09 |
| 7 | 2JRM | 97.21 | 92.14 | 87.30 | 12.60 | 0.00 | 0.10 | 1312 | 0.50 | 6-47 | 1.43 | 2.55 |
| 8 | 1YEZ | 99.06 | 90.95 | 89.10 | 10.90 | 0.00 | 0.00 | 1063 | 0.58 | 15-25,29-66 | 0.73 | 1.20 |
| 9 | 2L9R | 97.93 | 90.44 | 89.20 | 10.80 | 0.00 | 0.00 | 1322 | 0.37 | 13-56 | 0.59 | 1.18 |
| 10 | 2K52 | 97.17 | 93.54 | 91.00 | 9.00 | 0.00 | 0.00 | 2018 | 2.68 | 7-70 | 1.10 | 1.81 |
| 11 | 2KRS | 98.47 | 96.18 | 88.00 | 12.00 | 0.00 | 0.00 | 1378 | 0.35 | 2-61 | 1.26 | 1.50 |
| 12 | 2K53 | 98.53 | 95.30 | 93.40 | 6.60 | 0.00 | 0.00 | 1183 | 0.60 | 8-28,39-66 | 0.88 | 1.29 |
| 13 | 2JT1 | 88.66 | 90.75 | 91.60 | 8.40 | 0.00 | 0.00 | 1240 | 0.11 | 5-57,66-69 | 0.94 | 1.39 |
| 14 | 2JVO | 96.64 | 76.42 | 77.90 | 22.00 | 0.10 | 0.00 | 2263 | 10.44 | 6-71 | 1.77 | 2.49 |
| 15 | 2ERR | 96.36 | 78.40 | 87.30 | 12.70 | 0.00 | 0.00 | 1578 | 13.61 | 2-75 | 2.09 | 3.16 |
| 16 | 2L1P | 96.29 | 87.04 | 83.90 | 13.40 | 2.60 | 0.00 | 2075 | 2.10 | 19-78 | 2.13 | 2.94 |
| 17 | 2LN3 | 98.37 | 92.52 | 87.20 | 12.70 | 0.10 | 0.00 | 1947 | 1.15 | 6-72 | 0.89 | 1.34 |
| 18 | 2HEQ | 94.48 | 84.63 | 80.40 | 17.80 | 1.80 | 0.00 | 1822 | 3.20 | 17-20,34-68 | 0.60 | 1.48 |
| 19 | 2KK8 | 99.18 | 90.26 | 84.40 | 15.60 | 0.00 | 0.00 | 1818 | 2.24 | 10-82 | 1.25 | 2.04 |
| 20 | 2KD0 | 97.24 | 91.96 | 85.80 | 14.20 | 0.00 | 0.00 | 2215 | 4.43 | 13-81 | 1.37 | 2.02 |
| 21 | 2LML | 97.44 | 91.71 | 91.30 | 8.70 | 0.00 | 0.00 | 1197 | 0.67 | 3-78 | 1.53 | 1.90 |
| 22 | 2K3D | 98.96 | 89.76 | 85.90 | 14.10 | 0.00 | 0.00 | 1922 | 2.00 | 2-81 | 1.44 | 2.13 |
| 23 | 2LK2 | 96.08 | 86.08 | 82.90 | 17.10 | 0.00 | 0.00 | 1152 | 0.94 | 14-65 | 1.42 | 1.98 |
| 24 | MH04 | 98.58 | 88.57 | 89.00 | 11.00 | 0.00 | 0.00 | 1522 | 0.86 | 6-87 | 1.57 | 2.35 |
| 25 | 1PQX | 99.03 | 86.51 | 81.70 | 16.00 | 2.30 | 0.00 | 1977 | 10.18 | 13-19,28-34,38-65,70-81 | 1.40 | 2.02 |
| 26 | 2L33 | 98.68 | 91.76 | 79.70 | 20.10 | 0.20 | 0.00 | 2092 | 1.37 | 19-36,46-79 | 0.79 | 1.32 |
| 27 | 2KZV | 95.84 | 81.67 | 85.50 | 14.50 | 0.00 | 0.00 | 1440 | 0.41 | 9-80 | 2.62 | 3.13 |
| 28 | 2KCT | 96.59 | 93.42 | 83.20 | 16.70 | 0.10 | 0.10 | 2431 | 4.45 | 11-37,44-83 | 0.77 | 1.32 |
| 29 | 2MDR | 93.17 | 82.89 | 85.00 | 13.90 | 0.80 | 0.30 | 2307 | 3.06 | 9-89 | 1.72 | 3.04 |
| 30 | 2FB7 | 95.95 | 81.18 | 79.90 | 20.00 | 0.10 | 0.00 | 1218 | 1.51 | 20-52,74-87 | 1.94 | 2.52 |
| 31 | 2MB0 | 97.69 | 88.46 | 83.50 | 16.50 | 0.00 | 0.00 | 2701 | 9.63 | 8-41,49-85 | 1.11 | 1.81 |
| 32 | 2L05 | 99.50 | 91.56 | 85.10 | 14.10 | 0.10 | 0.80 | 2336 | 2.07 | 19-89 | 0.74 | 1.23 |
| 33 | 2KJR | 99.76 | 91.62 | 79.90 | 20.10 | 0.00 | 0.00 | 2109 | 1.04 | 15-23,28-94 | 1.02 | 1.35 |
| 34 | 2M5O | 92.48 | 90.20 | 77.40 | 22.60 | 0.10 | 0.00 | 2476 | 7.18 | 17-91 | 1.08 | 1.59 |
| 35 | MDM2 | 98.08 | 84.88 | 87.40 | 12.30 | 0.30 | 0.00 | 1452 | 1.29 | 9-92 | 1.24 | 1.87 |
| 36 | 2LNA | 99.30 | 90.78 | 78.80 | 19.90 | 1.20 | 0.00 | 2569 | 4.06 | 16-49,59-94 | 0.86 | 1.45 |
| 37 | 2LA6 | 98.78 | 92.60 | 83.60 | 16.40 | 0.00 | 0.00 | 2370 | 6.62 | 15-97 | 0.81 | 1.34 |
| 38 | 6FIP | 97.86 | 79.15 | 85.20 | 14.80 | 0.10 | 0.00 | 2381 | 17.06 | 11-93 | 2.05 | 2.88 |
| 39 | 2LEA | 96.17 | 83.33 | 75.40 | 24.50 | 0.10 | 0.00 | 2542 | 7.98 | 15-45,55-88 | 1.45 | 2.00 |
| 40 | 2LL8 | 98.04 | 91.42 | 89.90 | 10.10 | 0.00 | 0.00 | 2237 | 2.64 | 4-90 | 1.42 | 1.79 |
| 41 | 2KPN | 96.95 | 91.16 | 79.80 | 20.00 | 0.20 | 0.00 | 2287 | 2.33 | 12-84 | 0.97 | 1.58 |
| 42 | 2K0M | 96.87 | 92.08 | 88.00 | 12.00 | 0.00 | 0.10 | 1508 | 0.69 | 7-70,76-93 | 1.60 | 2.09 |
| 43 | 2K5V | 99.37 | 94.25 | 89.20 | 10.80 | 0.00 | 0.00 | 1974 | 1.10 | 2-29,38-78,83-94 | 0.94 | 1.50 |
| 44 | 2MQL | 83.59 | 70.55 | 72.10 | 25.10 | 2.80 | 0.10 | 2836 | 22.17 | 15-84 | 0.98 | 1.56 |
| 45 | 2K75 | 98.77 | 88.36 | 90.10 | 9.90 | 0.10 | 0.00 | 1881 | 0.77 | 3-92 | 1.65 | 1.98 |
| 46 | 2LTM | 99.31 | 92.79 | 85.60 | 14.40 | 0.00 | 0.00 | 2696 | 2.68 | 14-99 | 0.67 | 1.46 |
| 47 | 2KOB | 88.18 | 82.29 | 84.90 | 15.10 | 0.00 | 0.00 | 1840 | 2.66 | 3-93 | 2.24 | 2.93 |
| 48 | 2KHD | 97.29 | 87.63 | 87.90 | 12.00 | 0.10 | 0.00 | 1059 | 0.04 | 31-97 | 1.87 | 2.22 |
| 49 | 2RN7 | 94.35 | 81.02 | 84.10 | 15.80 | 0.10 | 0.00 | 459 | 0.54 | 10-55 | 1.93 | 2.26 |
| 50 | 2LXU | 99.40 | 91.33 | 82.90 | 17.10 | 0.00 | 0.00 | 2115 | 1.40 | 9-95 | 1.19 | 1.81 |

**Supplementary Table 4 continued**

| ID | Protein | Chemical shift assignment | | Ramachandran plot statistics | | | | Structure calculation and comparison to PDB reference | | | | |
|---|---|---|---|---|---|---|---|---|---|---|---|---|
| | | Backbone accuracy [%] | Side-chain accuracy [%] | Most favored [%] | Additionally allowed [%] | Generously allowed [%] | Disallowed [%] | Number of distance restraints | CYANA target function [Å²] | Residue range for RMSD calculation | Backbone RMSD to reference [Å] | Side-chain RMSD to reference [Å] |
| 51 | 2KIF | 94.43 | 87.78 | 81.80 | 18.20 | 0.10 | 0.00 | 3056 | 3.64 | 3-97 | 0.89 | 1.51 |
| 52 | 2KBN | 97.98 | 88.49 | 88.80 | 11.20 | 0.00 | 0.00 | 2347 | 0.45 | 5-29,34-54,58-76,81-94 | 0.92 | 1.46 |
| 53 | 2MK2 | 99.58 | 91.97 | 85.50 | 14.40 | 0.10 | 0.00 | 2049 | 2.62 | 14-108 | 1.56 | 2.13 |
| 54 | 2K50 | 97.39 | 89.74 | 84.40 | 15.40 | 0.00 | 0.20 | 2157 | 0.86 | 10-34,42-85,92-105 | 1.00 | 1.57 |
| 55 | 2KL5 | 79.41 | 68.59 | 72.40 | 25.50 | 2.20 | 0.00 | 1878 | 5.90 | 12-53,58-66,76-86,93-99 | 2.58 | 3.38 |
| 56 | 2LTA | 95.17 | 81.57 | 91.10 | 8.90 | 0.00 | 0.00 | 1989 | 0.73 | 4-98 | 2.39 | 2.90 |
| 57 | 2KIW | 93.99 | 80.94 | 87.50 | 12.50 | 0.00 | 0.00 | 1964 | 1.10 | 2-81 | 1.59 | 2.08 |
| 58 | 2LVB | 89.27 | 84.11 | 92.70 | 7.30 | 0.00 | 0.00 | 1711 | 0.86 | 3-102 | 1.56 | 2.17 |
| 59 | 2LND | 95.52 | 88.24 | 92.90 | 7.10 | 0.00 | 0.00 | 2605 | 1.29 | 3-48,52-101 | 0.85 | 1.44 |
| 60 | 1WQU | 98.05 | 86.80 | 79.60 | 20.20 | 0.20 | 0.10 | 3785 | 9.51 | 8-106 | 0.99 | 1.53 |
| 61 | 2KL6 | 98.64 | 93.52 | 86.00 | 14.00 | 0.00 | 0.00 | 2752 | 1.99 | 6-106 | 0.87 | 1.38 |
| 62 | 6GT7 | 96.93 | 79.31 | 89.10 | 10.90 | 0.00 | 0.00 | 3069 | 4.21 | 7-30,40-87,94-113 | 1.39 | 2.00 |
| 63 | 2JN8 | 95.45 | 89.68 | 83.00 | 16.80 | 0.30 | 0.00 | 1863 | 1.12 | 12-26,31-109 | 1.83 | 2.36 |
| 64 | 2K5D | 99.25 | 92.40 | 75.50 | 24.30 | 0.20 | 0.00 | 2425 | 1.44 | 19-50,55-84,98-107 | 1.47 | 1.87 |
| 65 | 2KD1 | 98.31 | 90.36 | 83.60 | 16.30 | 0.00 | 0.10 | 2550 | 6.35 | 7-89 | 1.99 | 3.02 |
| 66 | 2LTL | 96.44 | 90.46 | 89.00 | 11.00 | 0.00 | 0.00 | 1667 | 1.07 | 19-35,39-41,46-110 | 2.37 | 2.85 |
| 67 | 2KVO | 98.17 | 90.00 | 79.20 | 20.80 | 0.00 | 0.00 | 2112 | 3.88 | 3-23,28-103 | 1.87 | 2.18 |
| 68 | 1T0Y | 97.27 | 85.04 | 76.10 | 23.80 | 0.10 | 0.00 | 2521 | 1.60 | 4-83 | 1.27 | 1.96 |
| 69 | 2KCD | 91.30 | 79.24 | 83.40 | 16.60 | 0.00 | 0.00 | 1421 | 10.14 | 3-108 | 3.13 | 3.59 |
| 70 | 2KRT | 97.01 | 82.72 | 81.80 | 18.20 | 0.00 | 0.00 | 1844 | 0.83 | 6-114 | 2.09 | 2.73 |
| 71 | 2LFI | 89.30 | 76.07 | 85.90 | 14.10 | 0.00 | 0.00 | 1661 | 0.80 | 2-104 | 2.42 | 2.89 |
| 72 | 2JQN | 97.81 | 90.68 | 88.50 | 11.50 | 0.00 | 0.00 | 1904 | 1.09 | 3-111 | 1.52 | 1.86 |
| 73 | 2L7Q | 96.11 | 82.62 | 83.80 | 16.20 | 0.00 | 0.00 | 2125 | 1.74 | 12-37,46-101,105-114 | 1.57 | 2.15 |
| 74 | 2KFP | 97.39 | 79.51 | 82.30 | 17.70 | 0.00 | 0.00 | 2650 | 8.46 | 3-115 | 2.23 | 2.89 |
| 75 | 1SE9 | 88.55 | 82.13 | 76.80 | 22.80 | 0.30 | 0.10 | 1507 | 1.85 | 17-84,94-101 | 2.10 | 2.59 |
| 76 | 2L3G | 98.04 | 93.19 | 83.90 | 16.10 | 0.00 | 0.00 | 2870 | 2.34 | 13-123 | 1.28 | 1.95 |
| 77 | 2L3B | 93.97 | 86.19 | 83.20 | 16.80 | 0.00 | 0.00 | 2185 | 0.81 | 14-38,45-113 | 1.05 | 1.63 |
| 78 | 2LRH | 94.91 | 77.45 | 90.60 | 9.40 | 0.00 | 0.00 | 3261 | 5.26 | 3-122 | 2.30 | 2.75 |
| 79 | 1VEE | 97.81 | 90.30 | 78.20 | 21.70 | 0.10 | 0.00 | 3627 | 5.14 | 6-123 | 1.03 | 1.41 |
| 80 | 2K1G | 97.22 | 92.18 | 79.00 | 20.90 | 0.00 | 0.00 | 3366 | 14.38 | 5-78,83-122 | 1.05 | 1.68 |
| 81 | 2KKZ | 96.50 | 87.83 | 82.40 | 16.80 | 0.80 | 0.00 | 2513 | 2.36 | 5-80,86-118 | 1.47 | 2.12 |
| 82 | 1VDY | 95.48 | 88.37 | 84.60 | 15.40 | 0.00 | 0.00 | 3833 | 9.13 | 9-102,113-128 | 0.95 | 1.43 |
| 83 | 2KKL | 94.06 | 80.60 | 71.40 | 28.40 | 0.10 | 0.00 | 2127 | 2.09 | 33-90,96-125 | 1.26 | 2.09 |
| 84 | 2N4B | 98.94 | 90.25 | 85.00 | 14.50 | 0.60 | 0.00 | 3659 | 10.49 | 2-26,40-54,66-134 | 1.14 | 1.95 |
| 85 | 2L8V | 93.82 | 76.58 | 85.00 | 15.00 | 0.00 | 0.00 | 1989 | 2.82 | 4-22,37-65,73-129 | 2.79 | 3.35 |
| 86 | 2LGH | 98.03 | 88.92 | 83.50 | 16.40 | 0.00 | 0.00 | 2831 | 3.18 | 2-109,113-135 | 2.43 | 2.70 |
| 87 | 2K1S | 98.83 | 92.02 | 86.50 | 13.50 | 0.00 | 0.00 | 2811 | 2.63 | 3-140 | 1.83 | 2.05 |
| 88 | 2M4F | 90.80 | 81.83 | 82.30 | 16.90 | 0.80 | 0.00 | 3427 | 6.07 | 23-46,51-57,63-94,103-114,120-129,136-148 | 1.11 | 1.75 |
| 89 | 2JXP | 97.08 | 89.62 | 86.50 | 13.50 | 0.00 | 0.00 | 2279 | 0.91 | 16-144 | 2.58 | 2.91 |
| 90 | 2L06 | 96.46 | 84.28 | 84.50 | 15.40 | 0.10 | 0.00 | 3374 | 13.42 | 15-38,45-141 | 1.57 | 2.05 |
| 91 | 2LAH | 93.23 | 85.67 | 84.90 | 15.10 | 0.00 | 0.00 | 3573 | 10.34 | 14-25,33-158 | 1.71 | 2.22 |
| 92 | 2LAK | 92.15 | 78.69 | 81.30 | 18.00 | 0.70 | 0.00 | 2180 | 4.21 | 10-37,68-77,93-139 | 1.54 | 1.82 |
| 93 | 2L82 | 97.87 | 81.05 | 85.40 | 14.40 | 0.20 | 0.00 | 3596 | 3.72 | 3-151 | 3.55 | 3.86 |
| 94 | 2M47 | 92.86 | 80.32 | 89.20 | 10.80 | 0.00 | 0.00 | 1930 | 1.46 | 5-25,40-56,66-156 | 4.72 | 5.42 |
| 95 | 2K3A | 96.79 | 84.40 | 72.30 | 27.60 | 0.10 | 0.00 | 1680 | 0.81 | 57-102,108-127,138-153 | 0.99 | 1.34 |
| 96 | 2M7U | 85.70 | 73.47 | 79.50 | 19.10 | 1.40 | 0.00 | 3243 | 15.64 | 12-151 | 2.14 | 2.96 |
| 97 | 2B3W | 93.39 | 80.09 | 81.70 | 16.70 | 0.90 | 0.80 | 4315 | 23.41 | 16-162 | 2.67 | 3.37 |
| 98 | KRAS4B | 98.00 | 76.81 | 89.00 | 11.00 | 0.00 | 0.00 | 4678 | 10.89 | 4-24,37-59,66-163 | 1.60 | 2.30 |
| 99 | 2G0Q | 93.99 | 84.33 | 79.30 | 20.30 | 0.00 | 0.40 | 3113 | 4.53 | 18-54,60-126 | 2.38 | 3.03 |
| 100 | 2LF2 | 97.56 | 85.86 | 86.10 | 13.80 | 0.10 | 0.00 | 3421 | 4.30 | 6-44,53-69,76-105,111-165 | 2.68 | 3.09 |

Residue ranges for RMSD calculation were determined by CYRANGE applied to the region between the first residue of the first secondary structure element and the last residue of the final secondary structure element of the reference PDB structure.

**Supplementary Table 5 Structure accuracy prediction.** Actual and predicted backbone RMSD between ARTINA and reference PDB structures. See Supplementary Fig. 2 for details.

| ID | Protein | RMSD to reference (Å) | Predicted RMSD to reference (Å) | Difference (Å) | ID | Protein | RMSD to reference (Å) | Predicted RMSD to reference (Å) | Difference (Å) |
|----|---------|------|------|-------|-----|---------|------|------|-------|
| 1 | 6SVC | 0.83 | 1.67 | 0.84 | 51 | 2KIF | 0.89 | 1.00 | 0.11 |
| 2 | 2JVD | 0.71 | 0.56 | -0.15 | 52 | 2KBN | 0.92 | 1.00 | 0.08 |
| 3 | 2K57 | 0.71 | 1.00 | 0.29 | 53 | 2MK2 | 1.56 | 1.78 | 0.22 |
| 4 | 6SOW | 1.16 | 2.44 | 1.28 | 54 | 2K50 | 1.00 | 1.00 | 0.00 |
| 5 | 2LX7 | 1.41 | 3.22 | 1.81 | 55 | 2KL5 | 2.58 | 2.11 | -0.47 |
| 6 | 2MA6 | 1.56 | 2.00 | 0.44 | 56 | 2LTA | 2.39 | 2.22 | -0.17 |
| 7 | 2JRM | 1.43 | 1.33 | -0.10 | 57 | 2KIW | 1.59 | 1.22 | -0.37 |
| 8 | 1YEZ | 0.73 | 0.78 | 0.05 | 58 | 2LVB | 1.56 | 1.11 | -0.45 |
| 9 | 2L9R | 0.59 | 0.67 | 0.08 | 59 | 2LND | 0.85 | 1.11 | 0.26 |
| 10 | 2K52 | 1.10 | 1.00 | -0.10 | 60 | 1WQU | 0.99 | 1.00 | 0.01 |
| 11 | 2KRS | 1.26 | 0.78 | -0.48 | 61 | 2KL6 | 0.87 | 1.33 | 0.46 |
| 12 | 2K53 | 0.88 | 0.78 | -0.10 | 62 | 6GT7 | 1.39 | 1.33 | -0.06 |
| 13 | 2JT1 | 0.94 | 0.67 | -0.27 | 63 | 2JN8 | 1.83 | 1.67 | -0.16 |
| 14 | 2JVO | 1.77 | 0.78 | -0.99 | 64 | 2K5D | 1.47 | 1.67 | 0.20 |
| 15 | 2ERR | 2.09 | 2.78 | 0.69 | 65 | 2KD1 | 1.99 | 2.78 | 0.79 |
| 16 | 2L1P | 2.13 | 1.33 | -0.80 | 66 | 2LTL | 2.37 | 3.89 | 1.52 |
| 17 | 2LN3 | 0.89 | 0.89 | -0.00 | 67 | 2KVO | 1.87 | 2.22 | 0.35 |
| 18 | 2HEQ | 0.60 | 0.89 | 0.29 | 68 | 1T0Y | 1.27 | 1.00 | -0.27 |
| 19 | 2KK8 | 1.25 | 1.56 | 0.31 | 69 | 2KCD | 3.13 | 3.00 | -0.13 |
| 20 | 2KD0 | 1.37 | 0.78 | -0.59 | 70 | 2KRT | 2.09 | 1.89 | -0.20 |
| 21 | 2LML | 1.53 | 1.11 | -0.42 | 71 | 2LFI | 2.42 | 3.11 | 0.69 |
| 22 | 2K3D | 1.44 | 2.00 | 0.56 | 72 | 2JQN | 1.52 | 1.67 | 0.15 |
| 23 | 2LK2 | 1.42 | 0.67 | -0.75 | 73 | 2L7Q | 1.57 | 1.22 | -0.35 |
| 24 | MH04 | 1.57 | 2.00 | 0.43 | 74 | 2KFP | 2.23 | 2.33 | 0.10 |
| 25 | 1PQX | 1.40 | 0.56 | -0.84 | 75 | 1SE9 | 2.10 | 2.00 | -0.10 |
| 26 | 2L33 | 0.79 | 0.67 | -0.12 | 76 | 2L3G | 1.28 | 2.00 | 0.72 |
| 27 | 2KZV | 2.62 | 1.89 | -0.73 | 77 | 2L3B | 1.05 | 1.33 | 0.28 |
| 28 | 2KCT | 0.77 | 1.33 | 0.56 | 78 | 2LRH | 2.30 | 3.00 | 0.70 |
| 29 | 2MDR | 1.72 | 1.67 | -0.05 | 79 | 1VEE | 1.03 | 0.56 | -0.47 |
| 30 | 2FB7 | 1.94 | 1.56 | -0.38 | 80 | 2K1G | 1.05 | 1.00 | -0.05 |
| 31 | 2MB0 | 1.11 | 1.00 | -0.11 | 81 | 2KKZ | 1.47 | 2.00 | 0.53 |
| 32 | 2L05 | 0.74 | 0.89 | 0.15 | 82 | 1VDY | 0.95 | 1.00 | 0.05 |
| 33 | 2KJR | 1.02 | 0.89 | -0.13 | 83 | 2KKL | 1.26 | 1.11 | -0.15 |
| 34 | 2M5O | 1.08 | 1.11 | 0.03 | 84 | 2N4B | 1.14 | 1.00 | -0.14 |
| 35 | MDM2 | 1.24 | 1.11 | -0.13 | 85 | 2L8V | 2.79 | 2.56 | -0.23 |
| 36 | 2LNA | 0.86 | 1.00 | 0.14 | 86 | 2LGH | 2.43 | 2.78 | 0.35 |
| 37 | 2LA6 | 0.81 | 0.78 | -0.03 | 87 | 2K1S | 1.83 | 2.00 | 0.17 |
| 38 | 6FIP | 2.05 | 3.00 | 0.95 | 88 | 2M4F | 1.11 | 1.33 | 0.22 |
| 39 | 2LEA | 1.45 | 0.89 | -0.56 | 89 | 2JXP | 2.58 | 3.33 | 0.75 |
| 40 | 2LL8 | 1.42 | 1.22 | -0.20 | 90 | 2L06 | 1.57 | 1.67 | 0.10 |
| 41 | 2KPN | 0.97 | 1.00 | 0.03 | 91 | 2LAH | 1.71 | 2.44 | 0.73 |
| 42 | 2K0M | 1.60 | 1.22 | -0.38 | 92 | 2LAK | 1.54 | 1.33 | -0.21 |
| 43 | 2K5V | 0.94 | 0.89 | -0.05 | 93 | 2L82 | 3.55 | 2.67 | -0.88 |
| 44 | 2MQL | 0.98 | 0.33 | -0.65 | 94 | 2M47 | 4.72 | 3.11 | -1.61 |
| 45 | 2K75 | 1.65 | 1.11 | -0.54 | 95 | 2K3A | 0.99 | 1.33 | 0.34 |
| 46 | 2LTM | 0.67 | 1.00 | 0.33 | 96 | 2M7U | 2.14 | 2.56 | 0.42 |
| 47 | 2KOB | 2.24 | 1.78 | -0.46 | 97 | 2B3W | 2.67 | 2.89 | 0.22 |
| 48 | 2KHD | 1.87 | 2.44 | 0.57 | 98 | KRAS4B | 1.60 | 1.22 | -0.38 |
| 49 | 2RN7 | 1.93 | 2.33 | 0.40 | 99 | 2G0Q | 2.38 | 2.44 | 0.06 |
| 50 | 2LXU | 1.19 | 1.11 | -0.08 | 100 | 2LF2 | 2.68 | 2.33 | -0.35 |

**Supplementary Table 6 ANSURR structure evaluation scores.** Correlation and RMSD scores given are the average of the corresponding scores calculated by ANSURR version 2.0.55 for the individual conformers of the structure calculated by ARTINA and the reference structure in the PDB, respectively.

| ID | Protein | ARTINA | | Reference | | ID | Protein | ARTINA | | Reference | |
|----|---------|--------|--------|-----------|--------|----|---------|--------|--------|-----------|--------|
| | | Correlation score (%) | RMSD score (%) | Correlation score (%) | RMSD score (%) | | | Correlation score (%) | RMSD score (%) | Correlation score (%) | RMSD score (%) |
| 1 | 6SVC | 18.6 | 34.6 | 21.4 | 12.2 | 51 | 2KIF | 26.4 | 58.4 | 53.6 | 85.9 |
| 2 | 2JVD | 72.0 | 83.5 | 94.5 | 74.2 | 52 | 2KBN | 44.6 | 20.8 | 54.0 | 15.3 |
| 3 | 2K57 | 16.3 | 49.3 | 18.3 | 51.0 | 53 | 2MK2 | 72.2 | 31.2 | 64.9 | 46.1 |
| 4 | 6SOW | 44.7 | 63.2 | 60.6 | 84.5 | 54 | 2K50 | 46.9 | 15.3 | 65.5 | 65.2 |
| 5 | 2LX7 | 28.7 | 4.5 | 35.1 | 19.3 | 55 | 2KL5 | 68.2 | 20.4 | 31.4 | 66.0 |
| 6 | 2MA6 | 21.8 | 2.4 | 61.9 | 70.7 | 56 | 2LTA | 21.3 | 45.9 | 38.0 | 92.7 |
| 7 | 2JRM | 88.7 | 90.6 | 75.8 | 63.5 | 57 | 2KIW | 71.4 | 78.2 | 94.2 | 38.9 |
| 8 | 1YEZ | 49.1 | 71.0 | 49.9 | 76.5 | 58 | 2LVB | 30.2 | 70.1 | 51.7 | 96.3 |
| 9 | 2L9R | 92.0 | 33.0 | 77.4 | 90.9 | 59 | 2LND | 43.0 | 84.7 | 55.3 | 96.7 |
| 10 | 2K52 | 4.1 | 7.4 | 23.5 | 81.4 | 60 | 1WQU | 29.7 | 36.7 | 7.1 | 38.3 |
| 11 | 2KRS | 43.5 | 49.7 | 19.7 | 6.4 | 61 | 2KL6 | 42.1 | 18.9 | 31.9 | 55.5 |
| 12 | 2K53 | 60.4 | 39.6 | 59.5 | 77.2 | 62 | 6GT7 | 14.1 | 76.4 | 27.9 | 67.8 |
| 13 | 2JT1 | 77.1 | 24.1 | 93.1 | 71.7 | 63 | 2JN8 | 67.1 | 47.5 | 90.5 | 47.1 |
| 14 | 2JVO | 1.4 | 55.1 | 10.3 | 88.6 | 64 | 2K5D | 80.2 | 27.0 | 77.9 | 42.9 |
| 15 | 2ERR | 29.4 | 34.9 | 25.8 | 86.7 | 65 | 2KD1 | 71.5 | 35.6 | 65.2 | 71.3 |
| 16 | 2L1P | 91.4 | 50.4 | 77.1 | 40.1 | 66 | 2LTL | 61.4 | 8.6 | 78.7 | 60.0 |
| 17 | 2LN3 | 64.7 | 89.6 | 56.9 | 85.4 | 67 | 2KVO | 27.3 | 40.1 | 68.5 | 22.2 |
| 18 | 2HEQ | 82.3 | 63.0 | 91.8 | 69.5 | 68 | 1T0Y | 95.3 | 42.3 | 56.9 | 64.6 |
| 19 | 2KK8 | 41.4 | 16.1 | 23.9 | 52.1 | 69 | 2KCD | 12.6 | 16.1 | 50.2 | 20.0 |
| 20 | 2KD0 | 59.1 | 43.1 | 17.1 | 40.5 | 70 | 2KRT | 28.0 | 29.6 | 23.1 | 62.7 |
| 21 | 2LML | 36.7 | 37.4 | 73.9 | 87.9 | 71 | 2LFI | 22.8 | 3.9 | 56.4 | 75.9 |
| 22 | 2K3D | 36.2 | 51.6 | 52.1 | 32.9 | 72 | 2JQN | 5.1 | 35.1 | 36.7 | 40.7 |
| 23 | 2LK2 | 96.7 | 36.7 | 97.7 | 44.1 | 73 | 2L7Q | 83.5 | 9.7 | 60.5 | 51.7 |
| 24 | MH04 | 51.3 | 24.4 | 18.1 | 25.5 | 74 | 2KFP | 43.5 | 34.5 | 47.2 | 12.9 |
| 25 | 1PQX | 33.9 | 25.2 | 72.7 | 13.2 | 75 | 1SE9 | 78.1 | 11.5 | 82.0 | 44.6 |
| 26 | 2L33 | 80.0 | 46.3 | 89.5 | 64.8 | 76 | 2L3G | 68.7 | 32.2 | 74.8 | 76.9 |
| 27 | 2KZV | 89.8 | 45.8 | 64.2 | 11.8 | 77 | 2L3B | 49.2 | 12.6 | 68.2 | 59.6 |
| 28 | 2KCT | 85.0 | 48.8 | 88.3 | 57.9 | 78 | 2LRH | 10.3 | 88.2 | 25.1 | 95.1 |
| 29 | 2MDR | 38.6 | 57.4 | 36.0 | 36.1 | 79 | 1VEE | 0.9 | 37.6 | 1.1 | 50.5 |
| 30 | 2FB7 | 73.8 | 10.2 | 44.8 | 18.4 | 80 | 2K1G | 35.5 | 64.0 | 77.0 | 73.8 |
| 31 | 2MB0 | 49.4 | 52.1 | 43.1 | 56.9 | 81 | 2KKZ | 74.2 | 23.9 | 71.3 | 91.1 |
| 32 | 2L05 | 68.9 | 45.1 | 54.9 | 62.5 | 82 | 1VDY | 3.7 | 26.9 | 5.7 | 50.8 |
| 33 | 2KJR | 49.1 | 32.9 | 31.9 | 7.3 | 83 | 2KKL | 55.1 | 29.9 | 46.8 | 6.5 |
| 34 | 2M5O | 71.0 | 42.8 | 92.5 | 91.5 | 84 | 2N4B | 22.1 | 23.3 | 48.0 | 62.4 |
| 35 | MDM2 | 64.3 | 26.0 | 65.1 | 17.3 | 85 | 2L8V | 31.6 | 31.9 | 23.7 | 39.4 |
| 36 | 2LNA | 73.3 | 43.8 | 69.9 | 64.9 | 86 | 2LGH | 48.4 | 36.5 | 65.8 | 85.1 |
| 37 | 2LA6 | 45.6 | 31.4 | 76.6 | 79.2 | 87 | 2K1S | 57.1 | 48.9 | 58.2 | 77.1 |
| 38 | 6FIP | 54.9 | 33.2 | 57.3 | 44.0 | 88 | 2M4F | 81.5 | 26.5 | 76.9 | 46.5 |
| 39 | 2LEA | 39.1 | 16.1 | 30.8 | 34.1 | 89 | 2JXP | 48.1 | 12.7 | 52.1 | 46.6 |
| 40 | 2LL8 | 14.4 | 77.9 | 45.3 | 90.4 | 90 | 2L06 | 49.8 | 57.0 | 52.9 | 49.7 |
| 41 | 2KPN | 92.8 | 26.5 | 65.6 | 33.3 | 91 | 2LAH | 24.8 | 63.1 | 19.8 | 92.2 |
| 42 | 2K0M | 24.0 | 45.8 | 24.9 | 72.2 | 92 | 2LAK | 59.0 | 11.9 | 47.3 | 6.1 |
| 43 | 2K5V | 33.7 | 7.6 | 33.2 | 51.4 | 93 | 2L82 | 24.4 | 91.0 | 35.8 | 98.5 |
| 44 | 2MQL | 40.8 | 36.4 | 61.9 | 21.0 | 94 | 2M47 | 13.8 | 8.3 | 36.6 | 60.7 |
| 45 | 2K75 | 15.2 | 9.4 | 28.7 | 20.0 | 95 | 2K3A | 81.0 | 28.0 | 74.7 | 61.0 |
| 46 | 2LTM | 43.5 | 56.1 | 20.7 | 67.1 | 96 | 2M7U | 58.8 | 33.2 | 37.1 | 13.6 |
| 47 | 2KOB | 63.3 | 32.6 | 80.9 | 75.2 | 97 | 2B3W | 73.2 | 24.4 | 55.9 | 55.5 |
| 48 | 2KHD | 79.5 | 19.3 | 77.2 | 23.4 | 98 | KRAS4B | 3.8 | 43.9 | 2.8 | 40.8 |
| 49 | 2RN7 | 91.1 | 24.1 | 88.6 | 41.7 | 99 | 2G0Q | 87.6 | 27.6 | 34.2 | 68.4 |
| 50 | 2LXU | 79.4 | 42.2 | 68.9 | 77.7 | 100 | 2LF2 | 31.8 | 29.4 | 28.6 | 35.2 |

**Supplementary Table 7 RPF structure evaluation scores.**

| ID | Protein | Recall | Precision | F-measure | DP-score | ID | Protein | Recall | Precision | F-measure | DP-score |
|---|---|---|---|---|---|---|---|---|---|---|---|
| 1 | 6SVC | 0.91 | 0.87 | 0.89 | 0.70 | 51 | 2KIF | 0.88 | 0.90 | 0.89 | 0.76 |
| 2 | 2JVD | 0.93 | 0.93 | 0.93 | 0.77 | 52 | 2KBN | 0.91 | 0.88 | 0.89 | 0.72 |
| 3 | 2K57 | 0.94 | 0.86 | 0.90 | 0.75 | 53 | 2MK2 | 0.87 | 0.88 | 0.87 | 0.70 |
| 4 | 6SOW | 0.82 | 0.84 | 0.83 | 0.64 | 54 | 2K50 | 0.90 | 0.84 | 0.87 | 0.76 |
| 5 | 2LX7 | 0.86 | 0.84 | 0.85 | 0.52 | 55 | 2KL5 | 0.83 | 0.79 | 0.81 | 0.63 |
| 6 | 2MA6 | 0.83 | 0.83 | 0.83 | 0.55 | 56 | 2LTA | 0.86 | 0.79 | 0.82 | 0.70 |
| 7 | 2JRM | 0.86 | 0.88 | 0.87 | 0.67 | 57 | 2KIW | 0.89 | 0.85 | 0.87 | 0.61 |
| 8 | 1YEZ | 0.85 | 0.82 | 0.83 | 0.65 | 58 | 2LVB | 0.92 | 0.85 | 0.89 | 0.70 |
| 9 | 2L9R | 0.93 | 0.86 | 0.89 | 0.83 | 59 | 2LND | 0.81 | 0.85 | 0.83 | 0.66 |
| 10 | 2K52 | 0.91 | 0.82 | 0.86 | 0.75 | 60 | 1WQU | 0.87 | 0.92 | 0.89 | 0.80 |
| 11 | 2KRS | 0.93 | 0.90 | 0.92 | 0.78 | 61 | 2KL6 | 0.94 | 0.84 | 0.89 | 0.80 |
| 12 | 2K53 | 0.89 | 0.89 | 0.89 | 0.66 | 62 | 6GT7 | 0.90 | 0.84 | 0.87 | 0.76 |
| 13 | 2JT1 | 0.92 | 0.88 | 0.90 | 0.72 | 63 | 2JN8 | 0.88 | 0.84 | 0.86 | 0.65 |
| 14 | 2JVO | 0.81 | 0.90 | 0.85 | 0.70 | 64 | 2K5D | 0.87 | 0.92 | 0.89 | 0.74 |
| 15 | 2ERR | 0.88 | 0.89 | 0.88 | 0.69 | 65 | 2KD1 | 0.87 | 0.86 | 0.87 | 0.65 |
| 16 | 2L1P | 0.81 | 0.88 | 0.85 | 0.68 | 66 | 2LTL | 0.90 | 0.85 | 0.87 | 0.63 |
| 17 | 2LN3 | 0.92 | 0.88 | 0.90 | 0.74 | 67 | 2KVO | 0.85 | 0.81 | 0.83 | 0.61 |
| 18 | 2HEQ | 0.73 | 0.69 | 0.71 | 0.55 | 68 | 1T0Y | 0.87 | 0.85 | 0.86 | 0.76 |
| 19 | 2KK8 | 0.87 | 0.88 | 0.88 | 0.64 | 69 | 2KCD | 0.90 | 0.79 | 0.84 | 0.53 |
| 20 | 2KD0 | 0.83 | 0.91 | 0.87 | 0.67 | 70 | 2KRT | 0.91 | 0.71 | 0.80 | 0.80 |
| 21 | 2LML | 0.94 | 0.89 | 0.91 | 0.68 | 71 | 2LFI | 0.78 | 0.83 | 0.81 | 0.53 |
| 22 | 2K3D | 0.85 | 0.90 | 0.87 | 0.68 | 72 | 2JQN | 0.94 | 0.84 | 0.88 | 0.77 |
| 23 | 2LK2 | 0.90 | 0.88 | 0.89 | 0.65 | 73 | 2L7Q | 0.88 | 0.82 | 0.85 | 0.69 |
| 24 | MH04 | 0.85 | 0.88 | 0.86 | 0.60 | 74 | 2KFP | 0.80 | 0.87 | 0.83 | 0.59 |
| 25 | 1PQX | 0.80 | 0.86 | 0.83 | 0.65 | 75 | 1SE9 | 0.82 | 0.78 | 0.80 | 0.59 |
| 26 | 2L33 | 0.88 | 0.88 | 0.88 | 0.74 | 76 | 2L3G | 0.92 | 0.86 | 0.89 | 0.77 |
| 27 | 2KZV | 0.85 | 0.87 | 0.86 | 0.65 | 77 | 2L3B | 0.81 | 0.83 | 0.82 | 0.67 |
| 28 | 2KCT | 0.85 | 0.88 | 0.87 | 0.75 | 78 | 2LRH | 0.80 | 0.86 | 0.82 | 0.63 |
| 29 | 2MDR | 0.80 | 0.87 | 0.83 | 0.66 | 79 | 1VEE | 0.89 | 0.91 | 0.90 | 0.76 |
| 30 | 2FB7 | 0.78 | 0.81 | 0.79 | 0.57 | 80 | 2K1G | 0.87 | 0.88 | 0.88 | 0.80 |
| 31 | 2MB0 | 0.90 | 0.90 | 0.90 | 0.83 | 81 | 2KKZ | 0.90 | 0.82 | 0.86 | 0.68 |
| 32 | 2L05 | 0.85 | 0.88 | 0.86 | 0.72 | 82 | 1VDY | 0.89 | 0.92 | 0.90 | 0.82 |
| 33 | 2KJR | 0.89 | 0.90 | 0.89 | 0.74 | 83 | 2KKL | 0.84 | 0.88 | 0.86 | 0.65 |
| 34 | 2M5O | 0.84 | 0.88 | 0.86 | 0.68 | 84 | 2N4B | 0.80 | 0.90 | 0.85 | 0.65 |
| 35 | MDM2 | 0.93 | 0.87 | 0.90 | 0.65 | 85 | 2L8V | 0.86 | 0.77 | 0.81 | 0.62 |
| 36 | 2LNA | 0.85 | 0.90 | 0.87 | 0.73 | 86 | 2LGH | 0.87 | 0.90 | 0.89 | 0.70 |
| 37 | 2LA6 | 0.89 | 0.87 | 0.88 | 0.72 | 87 | 2K1S | 0.80 | 0.84 | 0.82 | 0.61 |
| 38 | 6FIP | 0.67 | 0.79 | 0.73 | 0.51 | 88 | 2M4F | 0.83 | 0.88 | 0.85 | 0.66 |
| 39 | 2LEA | 0.79 | 0.88 | 0.83 | 0.68 | 89 | 2JXP | 0.90 | 0.82 | 0.86 | 0.73 |
| 40 | 2LL8 | 0.88 | 0.90 | 0.89 | 0.72 | 90 | 2L06 | 0.89 | 0.87 | 0.88 | 0.72 |
| 41 | 2KPN | 0.89 | 0.87 | 0.88 | 0.78 | 91 | 2LAH | 0.90 | 0.79 | 0.84 | 0.80 |
| 42 | 2K0M | 0.94 | 0.85 | 0.89 | 0.73 | 92 | 2LAK | 0.90 | 0.84 | 0.87 | 0.72 |
| 43 | 2K5V | 0.94 | 0.92 | 0.93 | 0.81 | 93 | 2L82 | 0.82 | 0.83 | 0.82 | 0.70 |
| 44 | 2MQL | 0.80 | 0.89 | 0.84 | 0.68 | 94 | 2M47 | 0.89 | 0.81 | 0.85 | 0.69 |
| 45 | 2K75 | 0.91 | 0.85 | 0.88 | 0.69 | 95 | 2K3A | 0.88 | 0.77 | 0.82 | 0.70 |
| 46 | 2LTM | 0.91 | 0.87 | 0.89 | 0.79 | 96 | 2M7U | 0.80 | 0.71 | 0.75 | 0.72 |
| 47 | 2KOB | 0.90 | 0.85 | 0.87 | 0.65 | 97 | 2B3W | 0.81 | 0.85 | 0.83 | 0.68 |
| 48 | 2KHD | 0.92 | 0.85 | 0.88 | 0.64 | 98 | KRAS4B | 0.92 | 0.88 | 0.90 | 0.83 |
| 49 | 2RN7 | 0.92 | 0.67 | 0.78 | 0.63 | 99 | 2G0Q | 0.86 | 0.80 | 0.83 | 0.72 |
| 50 | 2LXU | 0.92 | 0.88 | 0.90 | 0.73 | 100 | 2LF2 | 0.84 | 0.88 | 0.86 | 0.68 |

Calculated by the RPF web server at https://montelionelab.chem.rpi.edu/rpf/ running ASDP version 2.3.

**Supplementary Table 8 Consensus structure bundles.** NMR structure bundles calculated by ARTINA and by using consensus distance restraints with the *multnoeassign* command of CYANA. RMSD values are the average of the 20 backbone RMSDs between the individual conformers and their mean coordinates for the residue given in Supplementary Table 4.

| ID | Protein | ARTINA | | Consensus | | ID | Protein | ARTINA | | Consensus | |
|---|---|---|---|---|---|---|---|---|---|---|---|
| | | Target function value (Å²) | RMSD to mean (Å) | Target function value (Å²) | RMSD to mean (Å) | | | Target function value (Å²) | RMSD to mean (Å) | Target function value (Å²) | RMSD to mean (Å) |
| 1 | 6SVC | 1.80 | 0.04 | 0.61 | 0.12 | 51 | 2KIF | 3.64 | 0.14 | 0.68 | 0.41 |
| 2 | 2JVD | 0.30 | 0.14 | 0.03 | 0.33 | 52 | 2KBN | 0.45 | 0.22 | 1.58 | 0.42 |
| 3 | 2K57 | 0.35 | 0.15 | 0.29 | 0.25 | 53 | 2MK2 | 2.62 | 0.39 | 0.21 | 0.71 |
| 4 | 6SOW | 2.32 | 0.10 | 0.45 | 0.40 | 54 | 2K50 | 0.86 | 0.25 | 1.04 | 0.40 |
| 5 | 2LX7 | 0.32 | 0.80 | 0.32 | 0.95 | 55 | 2KL5 | 5.90 | 0.44 | 0.26 | 1.12 |
| 6 | 2MA6 | 3.25 | 0.14 | 1.95 | 0.41 | 56 | 2LTA | 0.73 | 0.46 | 1.96 | 0.63 |
| 7 | 2JRM | 0.50 | 0.13 | 0.20 | 0.37 | 57 | 2KIW | 1.10 | 0.28 | 0.37 | 0.63 |
| 8 | 1YEZ | 0.58 | 0.14 | 0.37 | 0.35 | 58 | 2LVB | 0.86 | 0.50 | 0.41 | 0.52 |
| 9 | 2L9R | 0.37 | 0.21 | 0.36 | 0.30 | 59 | 2LND | 1.29 | 0.15 | 1.12 | 0.62 |
| 10 | 2K52 | 2.68 | 0.17 | 0.84 | 0.48 | 60 | 1WQU | 9.51 | 0.14 | 0.30 | 0.30 |
| 11 | 2KRS | 0.35 | 0.18 | 0.21 | 0.40 | 61 | 2KL6 | 1.99 | 0.17 | 2.90 | 0.56 |
| 12 | 2K53 | 0.60 | 0.38 | 0.42 | 0.33 | 62 | 6GT7 | 4.21 | 0.13 | 0.63 | 0.36 |
| 13 | 2JT1 | 0.11 | 0.29 | 0.10 | 0.35 | 63 | 2JN8 | 1.12 | 0.39 | 1.73 | 0.72 |
| 14 | 2JVO | 10.44 | 0.05 | 3.66 | 0.16 | 64 | 2K5D | 1.44 | 0.19 | 0.38 | 0.49 |
| 15 | 2ERR | 13.61 | 0.46 | 8.86 | 0.62 | 65 | 2KD1 | 6.35 | 0.26 | 0.46 | 0.63 |
| 16 | 2L1P | 2.10 | 0.08 | 0.27 | 0.25 | 66 | 2LTL | 1.07 | 0.77 | 1.99 | 2.42 |
| 17 | 2LN3 | 1.15 | 0.14 | 0.09 | 0.27 | 67 | 2KVO | 3.88 | 0.22 | 0.66 | 0.75 |
| 18 | 2HEQ | 3.20 | 0.03 | 0.82 | 0.30 | 68 | 1T0Y | 1.60 | 0.15 | 1.73 | 0.46 |
| 19 | 2KK8 | 2.24 | 0.29 | 0.82 | 0.78 | 69 | 2KCD | 10.14 | 0.57 | 0.24 | 1.50 |
| 20 | 2KD0 | 4.43 | 0.07 | 1.63 | 0.24 | 70 | 2KRT | 0.83 | 0.58 | 5.44 | 0.90 |
| 21 | 2LML | 0.67 | 0.48 | 0.57 | 0.69 | 71 | 2LFI | 0.80 | 0.73 | 0.30 | 2.02 |
| 22 | 2K3D | 2.00 | 0.12 | 0.94 | 0.44 | 72 | 2JQN | 1.09 | 0.58 | 0.43 | 0.72 |
| 23 | 2LK2 | 0.94 | 0.36 | 0.45 | 0.43 | 73 | 2L7Q | 1.74 | 0.23 | 1.02 | 0.55 |
| 24 | MH04 | 0.86 | 0.46 | 0.54 | 0.81 | 74 | 2KFP | 8.46 | 0.23 | 0.26 | 0.57 |
| 25 | 1PQX | 10.18 | 0.13 | 8.54 | 0.29 | 75 | 1SE9 | 1.85 | 0.48 | 4.15 | 1.05 |
| 26 | 2L33 | 1.37 | 0.16 | 0.16 | 0.30 | 76 | 2L3G | 2.34 | 0.21 | 0.70 | 0.63 |
| 27 | 2KZV | 0.41 | 0.46 | 0.30 | 0.81 | 77 | 2L3B | 0.81 | 0.30 | 0.76 | 0.55 |
| 28 | 2KCT | 4.45 | 0.11 | 2.67 | 0.21 | 78 | 2LRH | 5.26 | 0.18 | 0.32 | 0.80 |
| 29 | 2MDR | 3.06 | 0.12 | 0.86 | 0.42 | 79 | 1VEE | 5.14 | 0.13 | 0.76 | 0.26 |
| 30 | 2FB7 | 1.51 | 0.29 | 0.37 | 0.69 | 80 | 2K1G | 14.38 | 0.12 | 1.90 | 0.29 |
| 31 | 2MB0 | 9.63 | 0.09 | 4.22 | 0.25 | 81 | 2KKZ | 2.36 | 0.35 | 3.50 | 0.86 |
| 32 | 2L05 | 2.07 | 0.17 | 0.53 | 0.53 | 82 | 1VDY | 9.13 | 0.13 | 0.69 | 0.47 |
| 33 | 2KJR | 1.04 | 0.15 | 0.39 | 0.45 | 83 | 2KKL | 2.09 | 0.42 | 4.31 | 0.89 |
| 34 | 2M5O | 7.18 | 0.07 | 3.54 | 0.24 | 84 | 2N4B | 10.49 | 0.10 | 0.44 | 0.38 |
| 35 | MDM2 | 1.29 | 0.72 | 0.63 | 1.07 | 85 | 2L8V | 2.82 | 0.80 | 5.82 | 1.27 |
| 36 | 2LNA | 4.06 | 0.14 | 1.06 | 0.61 | 86 | 2LGH | 3.18 | 0.56 | 1.06 | 0.83 |
| 37 | 2LA6 | 6.62 | 0.11 | 4.89 | 0.19 | 87 | 2K1S | 2.63 | 0.26 | 1.37 | 0.72 |
| 38 | 6FIP | 17.06 | 0.13 | 6.79 | 1.39 | 88 | 2M4F | 6.07 | 0.18 | 0.59 | 0.54 |
| 39 | 2LEA | 7.98 | 0.09 | 1.13 | 0.35 | 89 | 2JXP | 0.91 | 1.28 | 0.86 | 4.17 |
| 40 | 2LL8 | 2.64 | 0.24 | 1.08 | 0.44 | 90 | 2L06 | 13.42 | 0.25 | 0.31 | 0.51 |
| 41 | 2KPN | 2.33 | 0.14 | 0.72 | 0.56 | 91 | 2LAH | 10.34 | 0.35 | 5.11 | 0.89 |
| 42 | 2K0M | 0.69 | 0.43 | 0.53 | 0.57 | 92 | 2LAK | 4.21 | 0.38 | 6.29 | 0.81 |
| 43 | 2K5V | 1.10 | 0.22 | 0.48 | 0.40 | 93 | 2L82 | 3.72 | 0.55 | 2.02 | 0.86 |
| 44 | 2MQL | 22.17 | 0.09 | 5.11 | 0.18 | 94 | 2M47 | 1.46 | 1.59 | 0.56 | 2.39 |
| 45 | 2K75 | 0.77 | 0.41 | 0.21 | 0.65 | 95 | 2K3A | 0.81 | 0.25 | 0.78 | 0.45 |
| 46 | 2LTM | 2.68 | 0.07 | 0.96 | 0.25 | 96 | 2M7U | 15.64 | 0.25 | 0.42 | 0.93 |
| 47 | 2KOB | 2.66 | 0.33 | 1.16 | 0.55 | 97 | 2B3W | 23.41 | 0.33 | 7.67 | 0.89 |
| 48 | 2KHD | 0.04 | 0.48 | 0.01 | 0.73 | 98 | KRAS4B | 10.89 | 0.22 | 7.11 | 0.65 |
| 49 | 2RN7 | 1.80 | 1.02 | 0.61 | 0.12 | 99 | 2G0Q | 4.53 | 0.55 | 1.86 | 1.25 |
| 50 | 2LXU | 0.30 | 0.27 | 0.03 | 0.33 | 100 | 2LF2 | 4.30 | 0.38 | 1.13 | 0.82 |

**Supplementary Table 9 Results of restrained energy refinement.** CYANA structure bundles calculated by ARTINA were energy-refined in explicit water using OPALp. Backbone RMSDs to the reference PDB structure before and after energy refinement show a standard deviation of 0.07 Å, maximal deviation of 0.16 Å, and linear correlation coefficient of 0.998.

| | | CYANA | | OPALp | | | | ARTINA | | OPALp | |
|---|---|---|---|---|---|---|---|---|---|---|---|
| ID | Protein | AMBER energy (kcal/mol) | RMSD to reference (Å) | AMBER energy (kcal/mol) | RMSD to reference (Å) | ID | Protein | AMBER energy (kcal/mol) | RMSD to reference (Å) | AMBER energy (kcal/mol) | RMSD to reference (Å) |
| 1 | 6SVC | -664 | 0.83 | -974 | 0.91 | 51 | 2KIF | -2906 | 0.89 | -3905 | 0.81 |
| 2 | 2JVD | -1590 | 0.71 | -2044 | 0.59 | 52 | 2KBN | -3061 | 0.92 | -3974 | 0.87 |
| 3 | 2K57 | -1547 | 0.71 | -2270 | 0.61 | 53 | 2MK2 | -2936 | 1.56 | -4013 | 1.49 |
| 4 | 6SOW | -1892 | 1.16 | -2369 | 1.09 | 54 | 2K50 | -3051 | 1.00 | -3976 | 0.85 |
| 5 | 2LX7 | -1981 | 1.41 | -2372 | 1.37 | 55 | 2KL5 | -3039 | 2.58 | -4149 | 2.56 |
| 6 | 2MA6 | -1200 | 1.56 | -1740 | 1.40 | 56 | 2LTA | -3214 | 2.39 | -4298 | 2.36 |
| 7 | 2JRM | -1733 | 1.43 | -2309 | 1.38 | 57 | 2KIW | -2779 | 1.59 | -3733 | 1.56 |
| 8 | 1YEZ | -1961 | 0.73 | -2463 | 0.61 | 58 | 2LVB | -3533 | 1.56 | -4492 | 1.52 |
| 9 | 2L9R | -1319 | 0.59 | -1877 | 0.54 | 59 | 2LND | -3512 | 0.85 | -4359 | 0.81 |
| 10 | 2K52 | -2597 | 1.10 | -3242 | 1.06 | 60 | 1WQU | -3058 | 0.99 | -4019 | 0.89 |
| 11 | 2KRS | -2228 | 1.26 | -2809 | 1.19 | 61 | 2KL6 | -3287 | 0.87 | -4413 | 0.78 |
| 12 | 2K53 | -2055 | 0.88 | -2753 | 0.87 | 62 | 6GT7 | -3421 | 1.39 | -4474 | 1.34 |
| 13 | 2JT1 | -2064 | 0.94 | -2792 | 0.87 | 63 | 2JN8 | -3285 | 1.83 | -4290 | 1.77 |
| 14 | 2JVO | -1912 | 1.77 | -2573 | 1.77 | 64 | 2K5D | -3500 | 1.47 | -4544 | 1.43 |
| 15 | 2ERR | -2788 | 2.09 | -3397 | 2.04 | 65 | 2KD1 | -2298 | 1.99 | -3492 | 1.99 |
| 16 | 2L1P | -2259 | 2.13 | -2932 | 2.14 | 66 | 2LTL | -3266 | 2.37 | -4322 | 2.38 |
| 17 | 2LN3 | -2501 | 0.89 | -3284 | 0.82 | 67 | 2KVO | -3111 | 1.87 | -4156 | 1.76 |
| 18 | 2HEQ | -1952 | 0.60 | -2780 | 0.57 | 68 | 1T0Y | -3261 | 1.27 | -4412 | 1.23 |
| 19 | 2KK8 | -2369 | 1.25 | -3099 | 1.20 | 69 | 2KCD | -2950 | 3.13 | -4099 | 3.14 |
| 20 | 2KD0 | -1865 | 1.37 | -2553 | 1.29 | 70 | 2KRT | -4018 | 2.09 | -5052 | 2.04 |
| 21 | 2LML | -2331 | 1.53 | -3095 | 1.44 | 71 | 2LFI | -3451 | 2.42 | -4437 | 2.45 |
| 22 | 2K3D | -2347 | 1.44 | -3094 | 1.43 | 72 | 2JQN | -3560 | 1.52 | -4547 | 1.48 |
| 23 | 2LK2 | -2288 | 1.42 | -3056 | 1.37 | 73 | 2L7Q | -3272 | 1.57 | -4451 | 1.45 |
| 24 | MH04 | -2942 | 1.57 | -3645 | 1.56 | 74 | 2KFP | -3419 | 2.23 | -4485 | 2.15 |
| 25 | 1PQX | -2398 | 1.40 | -3198 | 1.31 | 75 | 1SE9 | -2800 | 2.10 | -4006 | 2.07 |
| 26 | 2L3G | -2613 | 0.79 | -3353 | 0.80 | 76 | 2L3G | -4106 | 1.28 | -5120 | 1.17 |
| 27 | 2KZV | -2599 | 2.62 | -3276 | 2.60 | 77 | 2L3B | -3838 | 1.05 | -4904 | 1.01 |
| 28 | 2KCT | -2484 | 0.77 | -3225 | 0.73 | 78 | 2LRH | -3853 | 2.30 | -5364 | 2.26 |
| 29 | 2MDR | -2770 | 1.72 | -3540 | 1.74 | 79 | 1VEE | -3642 | 1.03 | -4911 | 0.93 |
| 30 | 2FB7 | -2359 | 1.94 | -3153 | 1.89 | 80 | 2K1G | -4044 | 1.05 | -5109 | 1.02 |
| 31 | 2MB0 | -2934 | 1.11 | -3905 | 1.03 | 81 | 2KKZ | -3854 | 1.47 | -5136 | 1.38 |
| 32 | 2L05 | -2134 | 0.74 | -3053 | 0.69 | 82 | 1VDY | -4235 | 0.95 | -5516 | 0.87 |
| 33 | 2KJR | -2797 | 1.02 | -3557 | 1.03 | 83 | 2KKL | -3245 | 1.26 | -4606 | 1.23 |
| 34 | 2M5O | -1612 | 1.08 | -2526 | 1.04 | 84 | 2N4B | -4485 | 1.14 | -5526 | 1.07 |
| 35 | MDM2 | -2857 | 1.24 | -3644 | 1.24 | 85 | 2L8V | -4036 | 2.79 | -5337 | 2.81 |
| 36 | 2LNA | -3227 | 0.86 | -4005 | 0.76 | 86 | 2LGH | -3831 | 2.43 | -5001 | 2.39 |
| 37 | 2LA6 | -2771 | 0.81 | -3645 | 0.78 | 87 | 2K1S | -5004 | 1.83 | -6232 | 1.74 |
| 38 | 6FIP | -2719 | 2.05 | -3637 | 2.04 | 88 | 2M4F | -4357 | 1.11 | -5942 | 1.06 |
| 39 | 2LEA | -2738 | 1.45 | -3599 | 1.38 | 89 | 2JXP | -4596 | 2.58 | -5895 | 2.55 |
| 40 | 2LL8 | -2611 | 1.42 | -3403 | 1.34 | 90 | 2L06 | -4389 | 1.57 | -5848 | 1.47 |
| 41 | 2KPN | -2678 | 0.97 | -3623 | 0.87 | 91 | 2LAH | -4976 | 1.71 | -6375 | 1.67 |
| 42 | 2K0M | -3002 | 1.60 | -3776 | 1.52 | 92 | 2LAK | -4201 | 1.54 | -5388 | 1.43 |
| 43 | 2K5V | -2818 | 0.94 | -3728 | 0.84 | 93 | 2L82 | -5405 | 3.55 | -7021 | 3.55 |
| 44 | 2MQL | -1780 | 0.98 | -2762 | 0.93 | 94 | 2M47 | -3907 | 4.72 | -5279 | 4.72 |
| 45 | 2K75 | -3142 | 1.65 | -4036 | 1.57 | 95 | 2K3A | -4339 | 0.99 | -5567 | 0.92 |
| 46 | 2LTM | -3030 | 0.67 | -3870 | 0.57 | 96 | 2M7U | -4846 | 2.14 | -6344 | 2.12 |
| 47 | 2KOB | -2974 | 2.24 | -3982 | 2.19 | 97 | 2B3W | -4335 | 2.67 | -5976 | 2.68 |
| 48 | 2KHD | -2886 | 1.87 | -3686 | 1.80 | 98 | KRAS4B | -5147 | 1.60 | -6814 | 1.64 |
| 49 | 2RN7 | -3214 | 1.93 | -4240 | 1.92 | 99 | 2G0Q | -3999 | 2.38 | -5697 | 2.36 |
| 50 | 2LXU | -3344 | 1.19 | -4281 | 1.08 | 100 | 2LF2 | -5149 | 2.68 | -6737 | 2.68 |

**Supplementary Table 10 Chemical shift assignment accuracy of protein core residues.**
Chemical shift assignment accuracy (%) is reported for core residues of the given amino acid types. The penultimate column (ARO) presents the accuracy calculated for His, Phe, Tyr, and Trp shifts. Core residues are those with a solvent-accessible surface area (SASA) in the protein of up to 20% of the corresponding SASA of the isolated residue. SASAs were calculated with Biopython 1.7.9.

| Protein | ALA | ARG | ASN | ASP | CYS | GLU | GLN | GLY | HIS | ILE | LEU | LYS | MET | PHE | PRO | SER | THR | TRP | TYR | VAL | ARO | RMSD |
|---|---|---|---|---|---|---|---|---|---|---|---|---|---|---|---|---|---|---|---|---|---|---|
| 6SVC | 100.0 | - | 80.0 | - | - | - | - | 100.0 | - | - | 69.2 | - | - | - | 95.2 | 100.0 | - | - | 100.0 | 92.3 | 95.6 | 0.83 |
| 2JVD | 100.0 | - | - | - | - | 97.0 | 93.3 | - | - | 100.0 | 100.0 | 94.1 | - | - | - | - | - | - | - | - | - | 0.71 |
| 2K57 | 100.0 | - | 100.0 | 100.0 | - | - | 100.0 | 100.0 | 100.0 | - | 100.0 | 100.0 | - | - | 87.5 | 100.0 | 100.0 | - | 85.7 | 100.0 | 86.7 | 0.71 |
| 6SOW | 100.0 | 100.0 | 91.7 | - | - | - | - | 93.3 | - | - | 95.2 | 88.6 | - | - | 78.0 | 92.3 | 100.0 | - | - | - | 78.0 | 1.16 |
| 2LX7 | 85.7 | - | - | 100.0 | 93.9 | - | 100.0 | - | 100.0 | 100.0 | 100.0 | - | - | 71.0 | 91.7 | - | 94.7 | 95.0 | - | 100.0 | 80.4 | 1.41 |
| 2MA6 | - | - | - | 100.0 | 100.0 | - | - | 100.0 | 92.3 | 100.0 | 100.0 | - | - | 87.5 | - | 88.2 | - | 80.0 | - | 100.0 | 86.2 | 1.56 |
| 2JRM | 100.0 | - | - | 100.0 | 71.4 | - | - | 90.0 | - | - | 100.0 | - | - | - | - | 100.0 | - | 95.0 | 100.0 | - | 96.3 | 1.43 |
| 1YEZ | 85.7 | - | - | 100.0 | - | - | - | 100.0 | - | 98.8 | 92.9 | - | - | 87.5 | 100.0 | 100.0 | 100.0 | - | 100.0 | 97.7 | 91.3 | 0.73 |
| 2L9R | 100.0 | 88.1 | 100.0 | - | - | 100.0 | 97.6 | - | - | 100.0 | 97.1 | 94.1 | 100.0 | 84.4 | - | - | 100.0 | 85.0 | - | 95.5 | 84.6 | 0.59 |
| 2K52 | 100.0 | - | 95.5 | 95.8 | - | 100.0 | 100.0 | 91.7 | - | 88.0 | 97.6 | - | 84.6 | 93.3 | 100.0 | - | 100.0 | - | 100.0 | 100.0 | 96.6 | 1.10 |
| 2KRS | 100.0 | 100.0 | - | - | - | 100.0 | - | 97.6 | - | 100.0 | 100.0 | - | - | - | 100.0 | 100.0 | - | 100.0 | 100.0 | 97.7 | 100.0 | 1.26 |
| 2K53 | - | - | 95.0 | 100.0 | 100.0 | 95.0 | - | 100.0 | - | 100.0 | 100.0 | 100.0 | 94.4 | 100.0 | - | 100.0 | 100.0 | - | 100.0 | 100.0 | 100.0 | 0.88 |
| 2JT1 | 85.7 | 93.8 | - | 100.0 | - | - | 86.7 | 100.0 | 80.0 | 90.0 | 95.2 | - | - | - | 91.7 | - | 77.8 | 95.0 | 100.0 | 83.3 | 93.2 | 0.94 |
| 2JVO | 87.5 | 84.6 | - | - | - | 90.0 | - | 100.0 | 90.9 | 84.6 | 94.9 | - | 50.0 | 70.5 | 27.3 | 95.2 | 100.0 | - | - | 94.1 | 73.0 | 1.77 |
| 2ERR | 100.0 | 81.8 | - | 71.4 | - | 90.0 | - | 88.0 | 83.3 | 93.8 | 94.9 | - | 100.0 | 84.2 | 12.5 | 100.0 | 100.0 | - | - | 92.0 | 84.0 | 2.09 |
| 2L1P | 92.9 | - | - | 100.0 | - | 92.9 | 50.0 | - | 92.9 | 87.1 | 82.4 | - | - | 93.8 | - | 91.7 | 100.0 | 97.5 | 92.9 | 100.0 | 95.7 | 2.13 |
| 2LN3 | 100.0 | 93.8 | - | 100.0 | - | 100.0 | - | 100.0 | - | 100.0 | 93.9 | 92.2 | 100.0 | 81.3 | - | 82.4 | 88.9 | - | - | 100.0 | 81.3 | 0.89 |
| 2HEQ | 95.2 | - | 100.0 | 100.0 | 100.0 | 93.9 | - | 83.3 | 83.3 | 100.0 | 87.8 | 88.2 | - | 75.0 | 100.0 | 100.0 | 100.0 | - | 74.1 | 93.9 | 76.4 | 0.60 |
| 2KK8 | - | - | 100.0 | 100.0 | 100.0 | 81.8 | 90.5 | - | 95.2 | 92.8 | 100.0 | 90.9 | - | 84.6 | - | 100.0 | 92.0 | - | 100.0 | 96.1 | 92.6 | 1.25 |
| 2KD0 | - | - | - | 100.0 | - | 100.0 | 100.0 | - | - | 100.0 | 90.4 | 100.0 | - | 100.0 | 95.8 | 95.8 | 91.9 | - | - | 98.2 | 100.0 | 1.37 |
| 2LML | 100.0 | - | 100.0 | 87.5 | - | 100.0 | - | 100.0 | 85.7 | 100.0 | 100.0 | 100.0 | - | 83.3 | - | 100.0 | 76.0 | 60.0 | 100.0 | 100.0 | 81.3 | 1.53 |
| 2K3D | 100.0 | - | 82.9 | 100.0 | - | - | 71.4 | 100.0 | - | 94.2 | 100.0 | 100.0 | 100.0 | 87.1 | - | 97.0 | 96.4 | - | 97.6 | 93.9 | 91.1 | 1.44 |
| 2LK2 | 92.9 | 62.5 | - | 100.0 | 90.9 | - | 83.3 | 100.0 | 100.0 | 94.5 | 100.0 | 90.9 | - | 100.0 | 100.0 | 90.6 | 90.0 | 91.9 | - | 100.0 | 94.5 | 1.42 |
| MH04 | 100.0 | - | 100.0 | 100.0 | - | 93.9 | 78.6 | 100.0 | - | 98.6 | 100.0 | - | 88.5 | 83.3 | 94.4 | - | 77.8 | 100.0 | 100.0 | 96.1 | 90.8 | 1.57 |
| 1PQX | 100.0 | - | 100.0 | 100.0 | - | - | 100.0 | 100.0 | 80.0 | 96.4 | 97.6 | 94.1 | 84.6 | 68.8 | - | 91.7 | 94.4 | 95.0 | 100.0 | 96.6 | 82.9 | 1.40 |
| 2L33 | 100.0 | - | 95.0 | - | - | 100.0 | - | 100.0 | - | - | 95.4 | 100.0 | 79.2 | 95.3 | 100.0 | - | 100.0 | - | - | 100.0 | 96.4 | 0.79 |
| 2KZV | 97.1 | - | - | 100.0 | - | 100.0 | - | 91.7 | 88.9 | 92.9 | 85.4 | - | 88.5 | 73.9 | 91.7 | 88.2 | 84.2 | - | 92.9 | 97.7 | 79.7 | 2.62 |
| 2KCT | 100.0 | 87.5 | - | 100.0 | - | 100.0 | - | 97.2 | - | 100.0 | 94.0 | - | - | 100.0 | 100.0 | - | 95.6 | - | 100.0 | 98.7 | 100.0 | 0.77 |
| 2MDR | 96.3 | - | - | 100.0 | - | 100.0 | 53.8 | 96.0 | - | 100.0 | 92.3 | 100.0 | - | 79.5 | 86.4 | 84.6 | - | - | 92.3 | 100.0 | 85.5 | 1.72 |
| 2FB7 | 100.0 | - | - | 100.0 | - | - | - | 91.7 | - | 90.6 | 90.0 | 92.6 | - | 70.0 | - | 93.8 | 94.3 | - | 100.0 | 100.0 | 85.0 | 1.94 |
| 2MB0 | 97.2 | - | 100.0 | 89.5 | - | 100.0 | 100.0 | 96.0 | - | 97.4 | 100.0 | 100.0 | 91.7 | 80.8 | - | 100.0 | 100.0 | 96.3 | 100.0 | 100.0 | 83.3 | 1.11 |
| 2L05 | 100.0 | 93.8 | - | 100.0 | 100.0 | 90.9 | 100.0 | - | 100.0 | 83.3 | 100.0 | 97.4 | - | 81.3 | 100.0 | 100.0 | 96.3 | 100.0 | - | 98.9 | 91.9 | 0.74 |
| 2KJR | - | - | 91.7 | 87.5 | - | 100.0 | 93.3 | 100.0 | 71.4 | 100.0 | 92.9 | 94.1 | 100.0 | 75.0 | - | 100.0 | 94.7 | - | 92.9 | 100.0 | 79.5 | 1.02 |
| 2M5O | - | - | 83.3 | 100.0 | - | 100.0 | 86.7 | - | 90.5 | 100.0 | 76.9 | 81.3 | - | 87.5 | 100.0 | - | - | - | 92.9 | 97.7 | 86.7 | 1.08 |
| MDM2 | 100.0 | - | 100.0 | 83.3 | 0.0 | - | 100.0 | 100.0 | 70.0 | 91.5 | 97.8 | - | 100.0 | 64.0 | 100.0 | 100.0 | 95.8 | - | 63.1 | 100.0 | 64.0 | 1.24 |
| 2LNA | - | 87.5 | 91.7 | 100.0 | - | 100.0 | 93.3 | 100.0 | - | 100.0 | 94.6 | - | - | 72.5 | - | - | 100.0 | 86.8 | 92.9 | 100.0 | 80.1 | 0.86 |
| 2LA6 | 100.0 | - | 73.8 | 100.0 | - | 90.9 | - | 95.8 | - | 99.0 | 100.0 | - | - | - | 89.5 | 100.0 | 95.8 | 100.0 | - | 100.0 | 90.8 | 0.81 |
| 6FIP | 100.0 | - | - | 93.8 | - | 100.0 | - | 100.0 | - | 95.9 | 85.7 | - | 84.6 | 76.6 | 91.7 | 87.5 | 100.0 | 60.0 | 69.2 | 94.5 | 72.2 | 2.05 |
| 2LEA | 100.0 | 63.6 | 90.0 | 96.4 | - | 100.0 | - | 93.1 | - | 92.3 | 93.2 | 68.8 | 51.2 | 96.4 | 88.9 | 90.5 | 97.4 | - | - | 98.6 | 96.4 | 1.45 |
| 2LL8 | 100.0 | 87.5 | 91.7 | 100.0 | 100.0 | 100.0 | - | - | 85.7 | 100.0 | 99.0 | - | - | 88.9 | 100.0 | 93.8 | 89.3 | 70.0 | 71.4 | 97.7 | 82.9 | 1.42 |
| 2KPN | 96.4 | - | - | 100.0 | - | - | 93.3 | 100.0 | 91.7 | 100.0 | 97.6 | - | - | 100.0 | 100.0 | - | 97.8 | - | 93.1 | 99.0 | 94.7 | 0.97 |
| 2K0M | 100.0 | 100.0 | 100.0 | 100.0 | 87.5 | 100.0 | 100.0 | 100.0 | 90.0 | 96.4 | 98.6 | 94.1 | 96.2 | 75.0 | - | 20.0 | - | 63.2 | 92.3 | 100.0 | 77.9 | 1.60 |
| 2K5V | 97.1 | - | 91.7 | 100.0 | - | 100.0 | 80.0 | 100.0 | - | 96.4 | 100.0 | 94.1 | - | 87.5 | - | 100.0 | 100.0 | - | - | 100.0 | 91.3 | 0.94 |
| 2MQL | 91.7 | - | 90.0 | 100.0 | 71.4 | 90.0 | 30.8 | 90.0 | 54.5 | 79.0 | 80.0 | - | 85.7 | 53.6 | 63.6 | 76.2 | - | - | 70.8 | 89.0 | 60.3 | 0.98 |
| 2K75 | 100.0 | - | 91.7 | 100.0 | - | - | 86.7 | 95.8 | - | 96.0 | 85.4 | - | - | 75.0 | - | 97.2 | 100.0 | - | 76.9 | 99.0 | 75.9 | 1.65 |
| 2LTM | 100.0 | 100.0 | 100.0 | 83.3 | 100.0 | - | 100.0 | 100.0 | - | 100.0 | 91.8 | 98.0 | 69.6 | 86.3 | 95.8 | 100.0 | 100.0 | 85.0 | 100.0 | 98.2 | 87.6 | 0.67 |
| 2KOB | 100.0 | 85.7 | 90.9 | 100.0 | 87.5 | - | 70.0 | 33.3 | 37.5 | 99.2 | 76.5 | 88.2 | - | 59.1 | 91.7 | 60.0 | 88.9 | 95.0 | 64.3 | 100.0 | 65.8 | 2.24 |
| 2KHD | 100.0 | 75.0 | - | - | - | 100.0 | - | 100.0 | - | 100.0 | 85.7 | 91.2 | - | 77.1 | - | - | 100.0 | - | 100.0 | 100.0 | 85.5 | 1.87 |
| 2RN7 | 100.0 | - | - | - | 100.0 | - | 100.0 | - | - | 100.0 | 64.3 | - | 92.3 | - | - | 93.3 | 100.0 | 75.0 | - | 100.0 | 75.0 | 1.93 |
| 2LXU | 100.0 | - | 83.3 | 100.0 | 100.0 | 98.2 | 93.3 | 95.7 | - | 97.6 | 98.2 | 98.0 | 96.2 | 81.3 | 100.0 | 100.0 | 100.0 | - | - | 99.0 | 81.3 | 1.19 |

**Supplementary Table 10 continued**

| Protein | ALA | ARG | ASN | ASP | CYS | GLU | GLN | GLY | HIS | ILE | LEU | LYS | MET | PHE | PRO | SER | THR | TRP | TYR | VAL | ARO | RMSD |
|---|---|---|---|---|---|---|---|---|---|---|---|---|---|---|---|---|---|---|---|---|---|---|
| 2KIF | 100.0 | 96.4 | 90.9 | 100.0 | - | 90.0 | 92.9 | 100.0 | - | 91.3 | 95.2 | 90.6 | - | 75.6 | 100.0 | 90.5 | 100.0 | 97.4 | 92.3 | 100.0 | 87.7 | 0.89 |
| 2KBN | 95.2 | - | 97.0 | 100.0 | - | 90.9 | 78.2 | 93.3 | - | 97.9 | 95.9 | 80.0 | - | 100.0 | 83.3 | 100.0 | 100.0 | 90.0 | 100.0 | 96.1 | 96.0 | 0.92 |
| 2MK2 | 100.0 | 93.8 | - | 100.0 | 93.8 | 100.0 | - | 94.4 | 71.4 | 96.4 | 95.7 | - | - | 94.1 | 100.0 | 94.1 | 100.0 | 100.0 | 92.9 | 100.0 | 91.8 | 1.56 |
| 2K50 | 100.0 | 85.7 | 100.0 | 96.8 | - | 100.0 | - | 86.7 | 85.7 | 96.4 | 93.9 | 100.0 | 76.9 | 100.0 | 79.2 | 88.9 | 94.4 | 90.0 | - | 100.0 | 95.1 | 1.00 |
| 2KL5 | 88.9 | 46.2 | 21.4 | 92.9 | 8.3 | 95.0 | 44.4 | 90.0 | - | 76.9 | 71.9 | 81.3 | - | 83.0 | - | - | 16.7 | - | 80.0 | 100.0 | 81.7 | 2.58 |
| 2LTA | 92.9 | - | - | 87.5 | - | 100.0 | 100.0 | - | - | 97.4 | 91.4 | 87.1 | - | 100.0 | - | 68.8 | 100.0 | 70.0 | 100.0 | 95.5 | 88.0 | 2.39 |
| 2KIW | 95.9 | 92.9 | 83.3 | 93.8 | - | - | - | - | - | 95.2 | 85.7 | - | - | 66.1 | 100.0 | 95.2 | 55.6 | 73.7 | 100.0 | 98.5 | 78.9 | 1.59 |
| 2LVB | 100.0 | - | 100.0 | 100.0 | - | 90.9 | 100.0 | 61.5 | - | 96.9 | 92.9 | 78.8 | 61.5 | 89.6 | 83.3 | 84.4 | 100.0 | - | 92.6 | 97.0 | 90.7 | 1.56 |
| 2LND | 94.3 | - | 95.2 | 100.0 | - | 77.3 | 89.3 | 63.2 | - | 98.0 | 95.2 | - | 69.2 | 87.5 | 100.0 | 94.6 | 94.4 | - | 97.4 | 97.0 | 92.0 | 0.85 |
| 1WQU | 100.0 | 87.5 | - | 100.0 | - | 95.5 | 96.4 | 92.9 | 100.0 | 98.2 | 79.5 | - | - | 77.1 | 95.8 | 93.9 | 100.0 | 95.0 | 100.0 | 97.7 | 91.6 | 0.99 |
| 2KL6 | 100.0 | 75.0 | 93.5 | 100.0 | - | 100.0 | - | 100.0 | 100.0 | 100.0 | 83.3 | 94.1 | - | 96.9 | 95.8 | 100.0 | 93.1 | 76.9 | 95.3 | 99.4 | 90.5 | 0.87 |
| 6GT7 | 100.0 | - | 50.0 | 95.7 | 100.0 | 85.0 | - | 80.0 | 100.0 | 94.5 | 86.0 | 66.7 | 90.9 | 66.7 | 81.3 | 91.7 | 88.5 | 97.2 | 97.4 | 89.9 | 93.0 | 1.39 |
| 2JN8 | 97.6 | 93.8 | 100.0 | 75.0 | - | 100.0 | 100.0 | 100.0 | 55.6 | 100.0 | 97.1 | 94.1 | 38.5 | 87.5 | 100.0 | 75.0 | 100.0 | 97.5 | 100.0 | 86.4 | 87.5 | 1.83 |
| 2K5D | 100.0 | 75.0 | - | 100.0 | - | 100.0 | 93.3 | - | - | 100.0 | 100.0 | 88.2 | - | 84.4 | - | 96.9 | 100.0 | 95.5 | 96.6 | 98.2 | 91.4 | 1.47 |
| 2KD1 | 100.0 | 93.3 | 81.8 | - | - | 90.9 | - | 93.3 | 90.9 | 96.8 | 95.8 | 100.0 | 84.6 | 75.0 | - | 84.4 | 100.0 | 85.0 | 79.2 | 97.7 | 82.0 | 1.99 |
| 2LTL | 91.7 | - | 95.0 | 74.3 | 100.0 | - | 90.4 | 100.0 | - | 99.4 | 93.1 | 96.0 | 95.8 | 75.6 | - | 92.9 | 90.6 | 64.7 | - | 100.0 | 72.6 | 2.37 |
| 2KVO | 100.0 | - | 100.0 | 87.5 | - | 100.0 | 71.4 | 96.7 | - | 100.0 | 88.1 | 88.2 | 89.2 | 75.7 | 97.2 | 97.5 | 94.7 | 90.0 | 100.0 | 98.5 | 83.9 | 1.87 |
| 1T0Y | 100.0 | - | 100.0 | 100.0 | - | 90.0 | 100.0 | 100.0 | 100.0 | 97.6 | 89.1 | 94.1 | 100.0 | 92.9 | - | 92.3 | 100.0 | - | 97.6 | 96.8 | 97.1 | 1.27 |
| 2KCD | 100.0 | - | 73.9 | 100.0 | - | 100.0 | 18.2 | 94.1 | 75.0 | 99.0 | 81.9 | 50.0 | 100.0 | 81.0 | 79.2 | 100.0 | 100.0 | 40.0 | 67.5 | 100.0 | 68.2 | 3.13 |
| 2KRT | 85.7 | - | 85.7 | 87.5 | - | 81.8 | 78.6 | 100.0 | - | 93.6 | 77.9 | 90.0 | 100.0 | 90.9 | 25.0 | 100.0 | 100.0 | - | 100.0 | 100.0 | 93.5 | 2.09 |
| 2LFI | 95.2 | - | 91.7 | 81.3 | - | 90.5 | 38.0 | 91.3 | 60.0 | 84.3 | 90.2 | - | - | 58.5 | - | 81.3 | 96.3 | 85.3 | 69.2 | 98.5 | 68.7 | 2.42 |
| 2JQN | 96.4 | - | - | 100.0 | - | 100.0 | 76.7 | 95.5 | 90.0 | 100.0 | 87.9 | 94.1 | - | 82.5 | 91.7 | 95.8 | 100.0 | 100.0 | 100.0 | 100.0 | 90.4 | 1.52 |
| 2L7Q | 100.0 | 50.0 | 83.3 | 91.7 | - | 95.5 | 70.5 | 100.0 | - | 94.6 | 90.2 | - | 100.0 | 85.2 | 91.7 | 91.4 | 94.7 | 60.0 | 88.8 | 100.0 | 84.3 | 1.57 |
| 2KFP | 91.9 | - | 91.7 | 100.0 | - | - | - | 96.7 | 100.0 | 98.4 | 85.5 | 81.8 | 95.8 | 87.5 | 52.2 | 100.0 | 79.3 | 70.3 | 90.9 | 96.9 | 84.8 | 2.23 |
| 1SE9 | 100.0 | 100.0 | - | 100.0 | - | 100.0 | 100.0 | 100.0 | - | 82.1 | 94.3 | 74.4 | 90.9 | 75.0 | 75.0 | 93.3 | 100.0 | 100.0 | 100.0 | 92.4 | 87.9 | 2.10 |
| 2L3G | 100.0 | - | 95.5 | 93.8 | 92.3 | 93.9 | 100.0 | 96.7 | - | 92.7 | 95.6 | - | - | 89.7 | 100.0 | 96.9 | 97.3 | 90.0 | 100.0 | 96.6 | 91.3 | 1.28 |
| 2L3B | 100.0 | 82.4 | - | 100.0 | 100.0 | 90.9 | 75.0 | 100.0 | - | 95.2 | 94.7 | - | 100.0 | 89.6 | 83.3 | 79.2 | 93.3 | - | 93.5 | 100.0 | 91.6 | 1.05 |
| 2LRH | 85.7 | - | 93.9 | 91.7 | - | 87.7 | 96.4 | - | - | 96.4 | 90.6 | 89.7 | 89.7 | 64.1 | - | 88.9 | 90.0 | - | 100.0 | 88.3 | 78.3 | 2.30 |
| 1VEE | 100.0 | 77.8 | 100.0 | 100.0 | - | 100.0 | 95.8 | - | 100.0 | 100.0 | 95.7 | 94.1 | - | 88.7 | 89.2 | 89.6 | 96.4 | 100.0 | 98.2 | 100.0 | 94.9 | 1.03 |
| 2K1G | 100.0 | 89.3 | 91.7 | 93.8 | 87.5 | 100.0 | 83.6 | 98.1 | 79.2 | 100.0 | 100.0 | - | 84.6 | 84.4 | 100.0 | 97.9 | 100.0 | 100.0 | 96.2 | 100.0 | 90.2 | 1.05 |
| 2KKZ | 95.2 | 44.4 | 94.3 | 100.0 | 100.0 | 87.9 | - | 90.0 | 76.9 | 98.0 | 92.8 | 88.2 | 77.8 | 83.3 | 79.2 | 91.3 | 100.0 | 65.0 | - | 100.0 | 77.8 | 1.47 |
| 1VDY | 100.0 | 89.5 | 100.0 | 100.0 | 100.0 | 90.9 | 100.0 | 66.7 | - | 97.3 | 84.9 | 92.6 | 92.3 | 88.6 | 100.0 | 89.5 | 89.7 | - | 96.4 | 100.0 | 91.7 | 0.95 |
| 2KKL | 89.3 | - | 91.7 | 84.4 | - | 45.5 | 50.0 | 91.2 | 28.6 | 96.4 | 91.2 | - | - | 84.2 | - | 90.9 | 89.3 | - | - | 93.2 | 75.6 | 1.26 |
| 2N4B | 100.0 | 88.2 | 87.5 | 100.0 | - | 97.0 | 93.3 | 100.0 | 80.0 | 100.0 | 100.0 | - | 100.0 | 82.3 | 100.0 | 100.0 | 95.7 | 80.0 | 90.5 | 97.0 | 83.0 | 1.14 |
| 2L8V | 95.1 | 71.4 | 41.7 | 93.8 | - | 86.4 | 50.0 | 100.0 | 80.0 | 91.4 | 76.1 | 70.6 | - | 50.0 | 91.7 | 78.8 | 94.7 | - | 82.0 | 97.2 | 67.9 | 2.79 |
| 2LGH | 98.4 | 88.2 | 55.9 | 100.0 | 100.0 | 100.0 | 92.0 | 100.0 | 91.7 | 97.6 | 96.3 | 100.0 | 100.0 | 82.5 | 100.0 | 100.0 | 100.0 | 79.1 | 100.0 | 98.9 | 82.8 | 2.43 |
| 2K1S | 100.0 | 93.8 | 95.2 | 87.5 | - | 100.0 | 100.0 | 95.7 | - | 97.1 | 92.9 | - | 55.3 | 100.0 | 100.0 | 95.7 | 95.5 | - | 78.3 | 99.2 | 87.2 | 1.83 |
| 2M4F | 100.0 | - | 100.0 | 100.0 | - | 92.1 | - | 100.0 | - | 100.0 | 97.3 | 97.1 | 92.3 | 70.1 | - | 87.5 | 100.0 | 75.0 | 72.7 | 100.0 | 71.8 | 1.11 |
| 2JXP | 100.0 | 37.5 | - | 100.0 | - | 94.5 | 90.6 | 100.0 | - | 98.2 | 89.6 | 100.0 | 80.0 | 95.1 | - | 93.8 | 100.0 | - | 96.4 | 99.1 | 95.7 | 2.58 |
| 2L06 | 98.6 | 43.8 | 91.7 | 100.0 | - | 93.2 | 91.7 | 100.0 | 80.0 | 98.9 | 93.4 | 64.7 | 84.6 | 81.4 | 72.2 | 84.8 | 100.0 | - | 76.2 | 100.0 | 79.2 | 1.57 |
| 2LAH | 97.2 | 92.3 | 91.7 | 100.0 | 87.5 | 90.0 | 84.6 | 80.0 | 78.6 | 93.6 | 94.5 | 100.0 | 75.0 | 74.3 | 90.9 | 81.8 | 93.8 | 94.3 | 92.3 | 96.7 | 84.2 | 1.71 |
| 2LAK | 100.0 | 75.0 | 45.5 | 93.3 | 100.0 | 90.5 | 81.8 | 90.5 | 78.6 | 96.4 | 96.7 | - | 92.3 | 37.5 | 88.9 | 95.1 | 94.7 | 66.7 | 67.9 | 96.5 | 63.9 | 1.54 |
| 2L82 | 100.0 | 74.1 | 100.0 | 95.9 | - | 76.0 | 98.1 | 100.0 | - | 96.9 | 92.3 | - | - | 77.5 | 100.0 | 78.3 | 95.8 | 89.5 | 97.4 | 87.4 | 83.8 | 3.55 |
| 2M47 | 97.6 | - | 100.0 | 87.5 | 100.0 | 54.5 | 76.9 | 92.0 | - | 81.1 | 78.3 | 100.0 | 76.9 | 83.3 | 78.9 | 89.6 | 92.9 | 62.9 | 92.7 | 96.6 | 75.7 | 4.72 |
| 2K3A | 100.0 | - | 97.0 | - | 87.5 | 95.5 | 85.7 | 90.0 | 57.1 | 98.2 | 100.0 | - | 100.0 | - | 83.3 | 94.6 | 100.0 | 91.7 | 93.5 | 100.0 | 87.3 | 0.99 |
| 2M7U | 89.3 | 61.1 | 75.0 | 73.9 | 50.0 | 84.2 | 85.7 | 96.2 | 68.2 | 88.3 | 75.8 | - | 83.3 | 89.5 | 63.6 | 65.2 | 84.8 | 83.8 | 91.7 | 87.2 | 85.0 | 2.14 |
| 2B3W | 92.8 | 91.1 | 73.9 | 95.5 | - | 75.8 | 75.9 | 82.1 | 80.8 | 100.0 | 84.4 | 79.4 | 72.2 | 77.2 | 95.8 | 87.5 | 81.4 | 71.8 | 53.8 | 96.0 | 71.6 | 2.67 |
| KRAS4B | 94.4 | 77.8 | 80.0 | 97.1 | 95.2 | 84.6 | 78.4 | 94.3 | 100.0 | 79.1 | 84.0 | 63.2 | 90.0 | 81.4 | 68.4 | 64.3 | 86.2 | - | 92.6 | 95.7 | 87.7 | 1.60 |
| 2G0Q | 100.0 | 87.0 | 100.0 | 92.3 | 75.0 | 69.7 | 85.7 | 95.2 | 100.0 | 90.0 | 82.5 | 92.3 | 83.9 | 87.5 | 78.6 | 90.9 | 94.6 | 65.0 | 82.9 | 94.1 | 85.0 | 2.38 |
| 2LF2 | 100.0 | 87.5 | 83.3 | 100.0 | 87.5 | 86.4 | 93.3 | 95.1 | 100.0 | 99.3 | 89.7 | - | 87.5 | 89.7 | 100.0 | 92.0 | 91.8 | 61.2 | 96.2 | 100.0 | 82.4 | 2.68 |
| Mean | 97.7 | 81.8 | 89.2 | 95.7 | 88.4 | 92.8 | 85.4 | 93.6 | 81.3 | 96.1 | 91.6 | 90.3 | 86.6 | 82.2 | 88.3 | 91.0 | 94.2 | 83.6 | 91.4 | 97.5 | 85.2 | 1.55 |
| Median | 100.0 | 87.5 | 93.5 | 100.0 | 100.0 | 95.0 | 92.4 | 96.7 | 83.3 | 97.6 | 93.1 | 94.1 | 90.0 | 83.3 | 94.8 | 93.8 | 96.4 | 88.2 | 96.2 | 98.6 | 85.5 | 1.44 |

**Supplementary Table 11 Quantitative analysis of sources of backbone RMSD in automatically determined structures.** Each row presents backbone the RMSD to reference [Å] of an automatically determined protein structure. The quantity was calculated for 7 different residue ranges. The broadest one (**REF**) has been determined by CYRANGE[27] and corresponds to the values reported in Supplementary Table 4 and Fig. 4. Remaining residue ranges cover the same fragments of the sequence as REF with the exclusion of: (**A**) the first secondary structure element, (**B**) the last secondary structure element, (**C**) α-helices, (**D**) β-sheets, (**E**) α-helices and β-sheets, and (**F**) flexible loops. The last column (**G**) presents difference between REF and the minimum value of A–F. Proteins that benefit by more than 0.5 Å RMSD decrease from the exclusion of one group of residues are highlighted in green. If a residue range contains less than 10 residues, its RMSD is not calculated.

| ID | Protein | REF | A | B | C | D | E | F | G |
|----|---------|-----|-----|-----|-----|-----|-----|-----|-----|
| 1 | 6SVC | 0.83 | 0.84 | 0.78 | 0.83 | - | - | 0.54 | 0.29 |
| 2 | 2JVD | 0.71 | 0.58 | 0.68 | - | 0.71 | - | 0.69 | 0.13 |
| 3 | 2K57 | 0.71 | 0.74 | 0.69 | 0.71 | 0.66 | 0.68 | 0.60 | 0.11 |
| 4 | 6SOW | 1.16 | 0.57 | 1.20 | - | 1.16 | - | 1.20 | 0.59 |
| 5 | 2LX7 | 1.41 | 1.43 | 1.43 | 1.41 | 1.59 | 1.59 | 0.70 | 0.71 |
| 6 | 2MA6 | 1.56 | 1.56 | 1.51 | 1.54 | 1.44 | 1.37 | 1.38 | 0.19 |
| 7 | 2JRM | 1.43 | 1.46 | 1.58 | 1.87 | 1.43 | - | 1.04 | 0.39 |
| 8 | 1YEZ | 0.73 | 0.70 | 0.74 | 0.73 | 0.86 | 0.86 | 0.63 | 0.10 |
| 9 | 2L9R | 0.59 | 0.56 | 0.60 | - | 0.59 | - | 0.6 | 0.03 |
| 10 | 2K52 | 1.10 | 1.16 | 1.13 | 1.10 | 1.35 | 1.37 | 0.81 | 0.29 |
| 11 | 2KRS | 1.26 | 1.28 | 1.22 | 1.26 | 1.43 | 1.40 | 1.16 | 0.10 |
| 12 | 2K53 | 0.88 | 0.88 | 0.92 | 0.80 | 0.88 | - | 0.85 | 0.08 |
| 13 | 2JT1 | 0.94 | 0.86 | 0.95 | 0.74 | 0.98 | - | 0.84 | 0.20 |
| 14 | 2JVO | 1.77 | 1.76 | 1.73 | 1.83 | 1.63 | 1.73 | 1.67 | 0.14 |
| 15 | 2ERR | 2.09 | 2.13 | 2.14 | 2.27 | 1.97 | 2.44 | 1.89 | 0.20 |
| 16 | 2L1P | 2.13 | 2.25 | 2.07 | 2.96 | 2.13 | 2.96 | 1.5 | 0.63 |
| 17 | 2LN3 | 0.89 | 0.87 | 0.91 | 0.79 | 0.92 | 0.84 | 0.85 | 0.10 |
| 18 | 2HEQ | 0.60 | 0.62 | 0.61 | 0.60 | 0.76 | 0.63 | 0.43 | 0.17 |
| 19 | 2KK8 | 1.25 | 1.23 | 1.26 | 1.28 | 1.16 | 1.39 | 1.08 | 0.17 |
| 20 | 2KD0 | 1.37 | 1.42 | 1.40 | 1.15 | 1.57 | 1.37 | 1.30 | 0.22 |
| 21 | 2LML | 1.53 | 1.46 | 1.50 | 1.39 | 1.53 | 1.39 | 1.57 | 0.14 |
| 22 | 2K3D | 1.44 | 1.28 | 1.44 | 1.32 | 1.54 | 1.49 | 1.33 | 0.16 |
| 23 | 2LK2 | 1.42 | 1.49 | 1.36 | 1.21 | 1.42 | 1.21 | 1.13 | 0.29 |
| 24 | MH04 | 1.57 | 1.57 | 1.58 | 1.58 | 1.85 | 1.97 | 0.97 | 0.60 |
| 25 | 1PQX | 1.40 | 1.44 | 1.39 | 1.24 | 1.61 | 1.62 | 1.19 | 0.21 |
| 26 | 2L33 | 0.79 | 0.72 | 0.77 | 0.70 | 0.90 | - | 0.70 | 0.09 |
| 27 | 2KZV | 2.62 | 2.82 | 2.70 | 3.07 | 2.84 | 3.33 | 1.98 | 0.64 |
| 28 | 2KCT | 0.77 | 0.71 | 0.78 | 0.77 | 0.79 | 0.79 | 0.65 | 0.12 |
| 29 | 2MDR | 1.72 | 1.73 | 1.70 | 1.77 | 1.79 | 1.96 | 1.13 | 0.59 |
| 30 | 2FB7 | 1.94 | 2.02 | 2.01 | 1.94 | 2.5 | 2.5 | 1.41 | 0.53 |
| 31 | 2MB0 | 1.11 | 1.14 | 1.11 | 1.04 | 1.18 | 1.06 | 1.08 | 0.07 |
| 32 | 2L05 | 0.74 | 0.76 | 0.73 | 0.75 | 0.66 | 0.69 | 0.73 | 0.08 |
| 33 | 2KJR | 1.02 | 1.05 | 1.02 | 1.06 | 1.08 | 1.26 | 0.83 | 0.19 |
| 34 | 2M5O | 1.08 | 1.12 | 1.08 | 1.02 | 1.13 | 1.26 | 0.93 | 0.15 |
| 35 | MDM2 | 1.24 | 1.22 | 1.19 | 1.17 | 1.26 | 1.20 | 1.08 | 0.16 |
| 36 | 2LNA | 0.86 | 0.86 | 0.84 | 0.85 | 0.92 | 1.00 | 0.77 | 0.09 |
| 37 | 2LA6 | 0.81 | 0.82 | 0.81 | 0.81 | 0.87 | 0.91 | 0.56 | 0.25 |
| 38 | 6FIP | 2.05 | 2.07 | 2.08 | 1.98 | 2.37 | 2.29 | 1.72 | 0.33 |
| 39 | 2LEA | 1.45 | 1.36 | 1.47 | 1.51 | 1.36 | 1.42 | 1.34 | 0.11 |
| 40 | 2LL8 | 1.42 | 1.42 | 1.50 | 1.47 | 1.42 | 1.47 | 1.32 | 0.10 |

**Supplementary Table 11 continued**

| ID | Protein | REF | A | B | C | D | E | F | G |
|----|---------|-----|------|------|------|------|------|------|------|
| 41 | 2KPN | 0.97 | 0.98 | 1.04 | 0.83 | 1.19 | 0.93 | 0.93 | 0.14 |
| 42 | 2K0M | 1.60 | 1.62 | 1.57 | 1.38 | 1.82 | 1.45 | 1.52 | 0.22 |
| 43 | 2K5V | 0.94 | 0.95 | 0.85 | 0.94 | 1.18 | 0.93 | 0.92 | 0.09 |
| 44 | 2MQL | 0.98 | 1.01 | 0.98 | 0.96 | 1.10 | 1.14 | 0.82 | 0.16 |
| 45 | 2K75 | 1.65 | 1.65 | 1.64 | 1.65 | 1.73 | 1.76 | 1.47 | 0.18 |
| 46 | 2LTM | 0.67 | 0.68 | 0.65 | 0.66 | 0.72 | 0.69 | 0.62 | 0.05 |
| 47 | 2KOB | 2.24 | 2.20 | 2.28 | 2.08 | 2.24 | 2.23 | 2.20 | 0.16 |
| 48 | 2KHD | 1.87 | 1.88 | 1.13 | 1.11 | 1.99 | 0.83 | 1.78 | 1.04 |
| 49 | 2RN7 | 1.93 | 1.12 | 2.12 | - | 1.93 | - | 1.81 | 0.81 |
| 50 | 2LXU | 1.19 | 1.21 | 1.24 | 1.29 | 1.22 | 1.45 | 0.96 | 0.23 |
| 51 | 2KIF | 0.89 | 0.90 | 0.90 | 0.98 | 0.92 | 1.02 | 0.79 | 0.10 |
| 52 | 2KBN | 0.92 | 0.88 | 0.92 | 0.92 | 1.00 | 0.92 | 0.91 | 0.04 |
| 53 | 2MK2 | 1.56 | 1.58 | 1.57 | 1.57 | 1.80 | 1.91 | 1.11 | 0.45 |
| 54 | 2K50 | 1.00 | 1.00 | 0.91 | 1.00 | 0.90 | 0.88 | 1.01 | 0.12 |
| 55 | 2KL5 | 2.58 | 2.63 | 2.69 | 2.09 | 3.25 | 2.93 | 2.15 | 0.49 |
| 56 | 2LTA | 2.39 | 2.47 | 2.32 | 2.18 | 2.65 | 2.88 | 1.92 | 0.47 |
| 57 | 2KIW | 1.59 | 1.63 | 1.56 | 1.69 | 1.59 | 1.86 | 1.47 | 0.12 |
| 58 | 2LVB | 1.56 | 1.57 | 1.65 | 1.83 | 1.56 | 2.36 | 1.20 | 0.36 |
| 59 | 2LND | 0.85 | 0.86 | 0.82 | 0.66 | 0.84 | 0.69 | 0.87 | 0.19 |
| 60 | 1WQU | 0.99 | 0.99 | 0.99 | 0.96 | 1.04 | 1.04 | 0.89 | 0.10 |
| 61 | 2KL6 | 0.87 | 0.89 | 0.88 | 0.87 | 1.09 | 1.09 | 0.63 | 0.24 |
| 62 | 6GT7 | 1.39 | 1.36 | 1.29 | 1.16 | 1.39 | 1.16 | 1.40 | 0.23 |
| 63 | 2JN8 | 1.83 | 1.74 | 1.87 | 2.17 | 1.83 | 2.22 | 1.61 | 0.22 |
| 64 | 2K5D | 1.47 | 1.49 | 1.54 | 1.47 | 1.77 | 1.77 | 0.96 | 0.51 |
| 65 | 2KD1 | 1.99 | 1.97 | 1.89 | 1.94 | 1.99 | 2.04 | 1.91 | 0.10 |
| 66 | 2LTL | 2.37 | 2.32 | 2.10 | 2.33 | 2.40 | 2.72 | 2.09 | 0.28 |
| 67 | 2KVO | 1.87 | 1.89 | 2.00 | 2.01 | 2.11 | 2.64 | 1.33 | 0.54 |
| 68 | 1T0Y | 1.27 | 1.29 | 1.30 | 1.31 | 1.43 | 1.71 | 0.88 | 0.39 |
| 69 | 2KCD | 3.13 | 2.98 | 2.40 | 2.40 | 3.61 | 2.75 | 3.22 | 0.73 |
| 70 | 2KRT | 2.09 | 1.92 | 2.00 | 1.94 | 2.41 | 2.18 | 1.90 | 0.19 |
| 71 | 2LFI | 2.42 | 2.34 | 2.40 | 2.42 | 2.64 | 2.59 | 2.28 | 0.14 |
| 72 | 2JQN | 1.52 | 1.55 | 1.51 | 1.50 | 1.65 | 1.57 | 1.37 | 0.15 |
| 73 | 2L7Q | 1.57 | 1.35 | 1.58 | 1.57 | 1.61 | 1.60 | 1.53 | 0.22 |
| 74 | 2KFP | 2.23 | 2.29 | 2.24 | 2.21 | 2.44 | 2.56 | 1.71 | 0.52 |
| 75 | 1SE9 | 2.10 | 2.19 | 2.20 | 2.19 | 2.48 | 2.93 | 1.10 | 1.00 |
| 76 | 2L3G | 1.28 | 1.32 | 1.33 | 1.72 | 1.28 | 1.76 | 0.84 | 0.44 |
| 77 | 2L3B | 1.05 | 0.97 | 1.02 | 1.05 | 0.96 | 0.96 | 1.07 | 0.09 |
| 78 | 2LRH | 2.3 | 2.35 | 1.45 | 1.50 | 2.57 | 1.85 | 2.25 | 0.85 |
| 79 | 1VEE | 1.03 | 1.03 | 1.03 | 1.02 | 1.09 | 1.10 | 0.89 | 0.14 |
| 80 | 2K1G | 1.05 | 1.05 | 1.08 | 1.06 | 1.12 | 1.24 | 0.92 | 0.13 |
| 81 | 2KKZ | 1.47 | 1.43 | 1.46 | 1.51 | 1.50 | 1.72 | 1.32 | 0.15 |
| 82 | 1VDY | 0.95 | 0.88 | 1.01 | 1.12 | 0.95 | 1.20 | 0.83 | 0.12 |
| 83 | 2KKL | 1.26 | 1.29 | 1.28 | 1.26 | 1.50 | 1.50 | 0.85 | 0.41 |
| 84 | 2N4B | 1.14 | 1.17 | 1.13 | 1.17 | 1.22 | 1.40 | 0.94 | 0.20 |
| 85 | 2L8V | 2.79 | 2.80 | 2.79 | 2.79 | 2.79 | 2.80 | 2.54 | 0.25 |
| 86 | 2LGH | 2.43 | 2.49 | 2.58 | 2.67 | 2.83 | 3.57 | 1.67 | 0.76 |
| 87 | 2K1S | 1.83 | 1.76 | 1.87 | 1.81 | 2.00 | 2.29 | 1.56 | 0.27 |
| 88 | 2M4F | 1.11 | 1.14 | 1.13 | 1.16 | 1.48 | 1.93 | 0.83 | 0.28 |
| 89 | 2JXP | 2.58 | 2.60 | 2.46 | 2.22 | 3.11 | 2.63 | 2.51 | 0.36 |
| 90 | 2L06 | 1.57 | 1.58 | 1.58 | 1.74 | 1.57 | 1.79 | 1.43 | 0.14 |
| 91 | 2LAH | 1.71 | 1.75 | 1.64 | 1.90 | 1.71 | 1.95 | 1.63 | 0.08 |
| 92 | 2LAK | 1.54 | 1.54 | 1.61 | 1.33 | 1.34 | 1.52 | 1.53 | 0.21 |
| 93 | 2L82 | 3.55 | 3.61 | 1.04 | 1.08 | 3.97 | 1.24 | 3.83 | 2.51 |
| 94 | 2M47 | 4.72 | 4.88 | 4.79 | 3.77 | 4.8 | 4.33 | 4.71 | 0.95 |
| 95 | 2K3A | 0.99 | 0.97 | 1.00 | 1.02 | 1.05 | 1.14 | 0.89 | 0.10 |
| 96 | 2M7U | 2.14 | 2.19 | 2.18 | 1.96 | 2.31 | 2.27 | 1.92 | 0.22 |
| 97 | 2B3W | 2.67 | 2.62 | 2.60 | 2.91 | 2.59 | 2.77 | 2.50 | 0.17 |
| 98 | KRAS4B | 1.60 | 1.61 | 1.61 | 1.57 | 1.67 | 1.58 | 1.56 | 0.04 |
| 99 | 2G0Q | 2.38 | 2.44 | 2.48 | 2.50 | 3.08 | 3.67 | 1.14 | 1.24 |
| 100 | 2LF2 | 2.68 | 2.71 | 2.24 | 2.37 | 3.03 | 3.00 | 2.51 | 0.44 |

**Supplementary Table 12 Quality metrics of 26 protein structures determined without and with 4D CC-NOESY spectra.**

| Protein | Backbone RMSD to reference [Å] | | Heavy-atom RMSD to reference [Å] | | Backbone chemical shift assignment accuracy [%] | | Side-chain chemical shift assignment accuracy [%] | |
|---|---|---|---|---|---|---|---|---|
| | without CCNOESY | with CCNOESY | without CCNOESY | with CCNOESY | without CCNOESY | with CCNOESY | without CCNOESY | with CCNOESY |
| 2B3W | 2.25 | 2.67 | 3.05 | 3.37 | 95.28 | 93.39 | 81.41 | 80.09 |
| 2HEQ | 0.50 | 0.60 | 1.27 | 1.48 | 95.30 | 94.48 | 84.03 | 84.63 |
| 2K1S | 1.49 | 1.83 | 1.82 | 2.05 | 98.39 | 98.83 | 92.14 | 92.02 |
| 2K3D | 1.54 | 1.44 | 2.33 | 2.13 | 98.96 | 98.96 | 88.89 | 89.76 |
| 2K75 | 1.65 | 1.65 | 2.03 | 1.98 | 98.36 | 98.77 | 88.36 | 88.36 |
| 2KBN | 1.09 | 0.92 | 1.76 | 1.46 | 98.18 | 97.98 | 89.49 | 88.49 |
| 2KCD | 1.77 | 3.13 | 2.48 | 3.59 | 97.22 | 91.30 | 86.86 | 79.24 |
| 2KFP | 2.03 | 2.23 | 2.65 | 2.89 | 97.02 | 97.39 | 81.96 | 79.51 |
| 2KIW | 2.78 | 1.59 | 3.23 | 2.08 | 94.96 | 93.99 | 79.50 | 80.94 |
| 2KKL | 1.80 | 1.26 | 2.45 | 2.09 | 94.06 | 94.06 | 81.67 | 80.60 |
| 2KOB | 2.27 | 2.24 | 2.77 | 2.93 | 91.18 | 88.18 | 85.05 | 82.29 |
| 2KRS | 1.02 | 1.26 | 1.35 | 1.50 | 98.47 | 98.47 | 96.40 | 96.18 |
| 2KVO | 2.24 | 1.87 | 2.58 | 2.18 | 98.35 | 98.17 | 90.27 | 90.00 |
| 2KZV | 2.04 | 2.62 | 2.68 | 3.13 | 96.58 | 95.84 | 83.45 | 81.67 |
| 2L3B | 1.23 | 1.05 | 1.74 | 1.63 | 93.79 | 93.97 | 86.31 | 86.19 |
| 2L8V | 7.49 | 2.79 | 7.87 | 3.35 | 93.63 | 93.82 | 76.32 | 76.58 |
| 2LAK | 1.49 | 1.54 | 1.82 | 1.82 | 92.59 | 92.15 | 80.73 | 78.69 |
| 2LF2 | 3.79 | 2.68 | 4.10 | 3.09 | 97.80 | 97.56 | 86.58 | 85.86 |
| 2LGH | 1.54 | 2.43 | 1.99 | 2.70 | 97.43 | 98.03 | 88.44 | 88.92 |
| 2LK2 | 0.90 | 1.42 | 1.66 | 1.98 | 96.81 | 96.08 | 87.44 | 86.08 |
| 2LML | 1.15 | 1.53 | 1.54 | 1.90 | 97.44 | 97.44 | 91.53 | 91.71 |
| 2LNA | 0.89 | 0.86 | 1.51 | 1.45 | 98.14 | 99.30 | 91.57 | 90.78 |
| 2LX7 | 1.42 | 1.41 | 1.93 | 2.02 | 97.81 | 98.54 | 88.59 | 87.84 |
| 2LXU | 1.64 | 1.19 | 2.33 | 1.81 | 99.20 | 99.40 | 92.77 | 91.33 |
| 2MA6 | 1.52 | 1.56 | 2.09 | 2.09 | 98.59 | 98.94 | 89.36 | 89.36 |
| 2N4B | 1.26 | 1.14 | 2.04 | 1.95 | 98.94 | 98.94 | 89.81 | 90.25 |
| **Median** | **1.54** | **1.55** | **2.07** | **2.07** | **97.43** | **97.50** | **87.90** | **87.01** |
| **Mean** | **1.88** | **1.73** | **2.43** | **2.26** | **96.71** | **96.31** | **86.88** | **86.05** |

Rows presenting proteins with more than 1 Å difference in backbone RMSD to reference are highlighted in green (if better without 4D CC-NOESY) or red (if better with 4D CC-NOESY).

**Supplementary Table 13 Accuracy of automated chemical shift assignment using all input spectra.** Chemical shift assignment accuracy (%) is reported for the 45 proteins in the automated chemical shift assignment experiment (see Results section).

| Protein | All shifts | Backbone | Side-chain | ALA | ARG | ASN | ASP | CYS | GLU | GLN | GLY | HIS | ILE | LEU | LYS | MET | PHE | PRO | SER | THR | TRP | TYR | VAL |
|---|---|---|---|---|---|---|---|---|---|---|---|---|---|---|---|---|---|---|---|---|---|---|---|
| 2JVD | 95.8 | 99.6 | 93.2 | 100.0 | 100.0 | 91.7 | - | - | 98.7 | 91.1 | 100.0 | - | 100.0 | 100.0 | 90.8 | 100.0 | 75.0 | - | 95.8 | 100.0 | - | 92.9 | 100.0 |
| 2K57 | 95.9 | 99.6 | 93.2 | 100.0 | 95.8 | 100.0 | 95.8 | - | 100.0 | 98.2 | 100.0 | - | 100.0 | 100.0 | 95.3 | - | 65.6 | 100.0 | 100.0 | 97.8 | - | 78.6 | 100.0 |
| 2LX7 | 93.1 | 98.2 | 89.6 | 100.0 | 93.1 | - | 100.0 | 100.0 | 96.4 | 90.0 | 93.2 | 57.1 | 100.0 | 99.1 | 100.0 | 80.0 | 74.5 | 97.2 | 100.0 | 89.5 | 85.0 | 100.0 | 100.0 |
| 1YEZ | 94.7 | 99.7 | 91.0 | 100.0 | 93.9 | - | 100.0 | - | 93.9 | 85.7 | 100.0 | - | 100.0 | 85.7 | 90.0 | 0.0 | 81.7 | 97.2 | 100.0 | 100.0 | - | 100.0 | 97.7 |
| 2K52 | 92.5 | 95.8 | 90.3 | 100.0 | 87.0 | 97.7 | 95.0 | - | 97.7 | 100.0 | 96.7 | - | 91.0 | 87.4 | 91.6 | 91.7 | 82.6 | 91.7 | 100.0 | 100.0 | - | 88.1 | 100.0 |
| 2JT1 | 89.2 | 92.8 | 86.7 | 100.0 | 90.0 | 87.5 | 100.0 | - | 92.5 | 86.7 | 100.0 | 80.0 | 92.9 | 85.7 | 80.9 | 100.0 | - | 87.5 | 75.9 | 94.4 | 95.0 | 92.9 | 84.1 |
| 2HEQ | 87.5 | 95.0 | 82.0 | 95.0 | 96.9 | 97.0 | 96.8 | 100.0 | 95.5 | - | 85.4 | 61.3 | 100.0 | 91.7 | 92.9 | - | 50.0 | 95.0 | 87.5 | 100.0 | 55.0 | 71.1 | 92.7 |
| 2KD0 | 95.2 | 98.6 | 92.7 | 100.0 | 92.9 | 100.0 | 100.0 | 100.0 | 100.0 | 97.1 | 100.0 | 83.3 | 100.0 | 97.8 | 90.6 | 87.5 | 71.9 | 95.8 | 95.8 | 94.5 | - | - | 100.0 |
| 2LK2 | 90.8 | 97.5 | 86.1 | 100.0 | 83.1 | 91.7 | 96.8 | 100.0 | 87.0 | 92.2 | 83.3 | 91.3 | 98.2 | 89.7 | 85.0 | 96.0 | 90.5 | 97.9 | 89.3 | 96.4 | 86.5 | 95.1 | 100.0 |
| MH04 | 92.9 | 98.8 | 88.7 | 98.6 | 86.4 | 100.0 | 100.0 | - | 92.4 | 84.5 | 100.0 | - | 97.1 | 97.1 | 95.2 | 80.8 | 85.7 | 96.7 | 90.0 | 80.0 | 80.0 | 96.4 | 97.7 |
| 2KZV | 87.2 | 95.4 | 81.3 | 97.6 | 77.1 | 100.0 | 100.0 | - | 95.5 | 90.5 | 91.7 | 67.4 | 92.9 | 86.3 | 86.3 | 78.6 | 76.1 | 91.3 | 89.3 | 89.5 | 71.1 | 92.9 | 100.0 |
| 2L05 | 94.8 | 99.5 | 91.7 | 100.0 | 91.7 | 87.0 | 100.0 | 100.0 | 98.5 | 93.3 | 100.0 | 83.3 | 100.0 | 100.0 | 94.1 | 93.9 | 87.5 | 83.9 | 100.0 | 97.2 | 77.5 | 100.0 | 100.0 |
| 6FIP | 88.6 | 98.5 | 81.8 | 100.0 | 78.1 | 100.0 | 90.9 | - | 100.0 | 96.4 | 100.0 | 0.0 | 96.9 | 92.9 | 86.3 | 84.6 | 72.4 | 95.8 | 93.8 | 100.0 | 70.0 | 61.5 | 100.0 |
| 2KPN | 92.2 | 96.7 | 88.6 | 97.1 | 81.3 | 63.6 | 97.2 | - | 96.0 | 86.7 | 92.3 | 90.9 | 100.0 | 95.3 | 82.4 | 52.4 | 81.3 | 100.0 | 90.6 | 98.3 | - | 89.7 | 98.7 |
| 2K0M | 91.7 | 95.3 | 89.1 | 94.6 | 96.4 | 100.0 | 97.9 | 68.8 | 97.3 | 100.0 | 97.6 | 72.7 | 97.1 | 94.0 | 95.7 | 92.3 | 63.0 | 95.7 | 89.2 | 100.0 | 57.9 | 87.5 | 98.7 |
| 2K5V | 94.2 | 98.9 | 90.8 | 97.1 | 90.6 | 90.0 | 100.0 | - | 91.9 | 86.7 | 98.3 | - | 92.0 | 93.7 | 93.6 | 79.2 | 85.4 | 100.0 | 97.2 | 100.0 | 100.0 | 100.0 | 100.0 |
| 2KOB | 81.8 | 87.4 | 77.8 | 100.0 | 72.5 | 85.3 | 85.4 | 87.5 | 83.1 | 73.3 | 66.7 | 25.0 | 100.0 | 84.2 | 76.2 | 0.0 | 56.7 | 91.7 | 71.3 | 88.9 | 95.0 | 71.4 | 100.0 |
| 2KHD | 90.6 | 96.9 | 85.6 | 100.0 | 73.7 | 83.3 | 100.0 | 100.0 | 86.7 | 92.0 | 97.1 | 66.7 | 98.5 | 90.5 | 90.2 | - | 66.2 | 100.0 | 96.7 | 100.0 | - | 100.0 | 99.0 |
| 2MK2 | 94.0 | 99.2 | 90.0 | 100.0 | 87.7 | 100.0 | 100.0 | 100.0 | 98.2 | 94.0 | 97.9 | 76.2 | 96.4 | 95.1 | 100.0 | 81.8 | 95.7 | 97.9 | 90.4 | 100.0 | 80.0 | 87.5 | 97.9 |
| 2K50 | 92.3 | 98.0 | 88.3 | 100.0 | 94.3 | 100.0 | 100.0 | - | 96.6 | - | 96.7 | 80.6 | 98.2 | 91.1 | 95.1 | 63.4 | 87.1 | 65.0 | 93.9 | 91.7 | 90.0 | - | 97.7 |
| 2KIW | 85.5 | 96.3 | 78.2 | 95.2 | 92.8 | 87.5 | 93.8 | - | 92.7 | 90.0 | 100.0 | 67.7 | 98.4 | 83.3 | 69.6 | 68.0 | 69.4 | 75.0 | 92.5 | 82.2 | 68.4 | 94.0 | 94.9 |
| 1WQU | 90.5 | 98.6 | 84.9 | 100.0 | 93.4 | 100.0 | 100.0 | - | 95.5 | 87.8 | 95.0 | 86.1 | 98.6 | 80.5 | 77.9 | - | 75.0 | 96.9 | 95.2 | 100.0 | 97.5 | 95.2 | 97.9 |
| 2KL6 | 94.1 | 98.1 | 91.3 | 100.0 | 84.4 | 92.4 | 100.0 | - | 93.0 | 86.7 | 97.9 | 83.3 | 96.4 | 92.9 | 97.5 | 85.7 | 72.9 | 100.0 | 100.0 | 97.8 | 89.7 | 87.7 | 98.5 |
| 6GT7 | 85.1 | 96.7 | 78.1 | 100.0 | 90.6 | 58.8 | 95.5 | 100.0 | 76.6 | 100.0 | 86.7 | 90.0 | 94.2 | 88.7 | 77.0 | 100.0 | 76.2 | 79.2 | 94.4 | 83.3 | 64.6 | 88.2 | 94.9 |
| 2K5D | 94.8 | 98.5 | 92.1 | 100.0 | 90.4 | 89.6 | 96.6 | - | 100.0 | 93.3 | 97.2 | - | 100.0 | 94.4 | 91.6 | 96.4 | 81.3 | 100.0 | 95.8 | 98.4 | 60.0 | 96.5 | 98.0 |
| 2KVO | 91.3 | 96.7 | 87.3 | 100.0 | 77.0 | 93.6 | 92.5 | - | 96.1 | 90.5 | 93.3 | 75.0 | 100.0 | 89.8 | 90.4 | 94.0 | 75.5 | 79.2 | 96.9 | 96.9 | 90.0 | 90.5 | 98.5 |
| 2KCD | 81.2 | 86.7 | 77.6 | 100.0 | 95.2 | 75.9 | 100.0 | 50.0 | 77.0 | 81.2 | 91.3 | 67.5 | 99.2 | 72.7 | 83.5 | 92.0 | 60.7 | 83.3 | 96.9 | 94.9 | 55.0 | 68.0 | 100.0 |
| 2L7Q | 86.8 | 94.5 | 81.5 | 100.0 | 64.1 | 82.3 | 96.3 | - | 93.2 | 69.2 | 97.9 | 100.0 | 96.4 | 84.5 | 89.7 | 68.0 | 85.5 | 88.3 | 93.0 | 97.3 | 92.5 | 82.9 | 94.5 |
| 2KFP | 86.2 | 96.8 | 79.6 | 97.3 | 68.2 | 91.2 | 100.0 | - | 89.6 | 90.5 | 94.4 | 81.0 | 98.4 | 90.1 | 75.1 | 95.8 | 82.1 | 76.8 | 93.8 | 86.2 | 86.0 | 74.3 | 96.9 |
| 2L3B | 88.8 | 94.0 | 85.0 | 90.5 | 79.3 | 61.1 | 90.9 | 50.0 | 81.8 | 80.6 | 97.6 | 86.7 | 94.6 | 93.3 | 93.1 | 78.6 | 90.0 | 88.1 | 90.3 | 97.8 | - | 90.0 | 98.5 |
| 1VEE | 92.9 | 98.0 | 89.1 | 100.0 | 93.5 | 100.0 | 100.0 | - | 98.5 | 100.0 | 93.4 | - | 100.0 | 98.0 | 84.6 | - | 76.7 | 96.5 | 85.8 | 95.3 | 97.5 | 83.9 | 100.0 |
| 2K1G | 89.7 | 95.4 | 85.6 | 100.0 | 92.4 | 87.0 | 97.5 | 25.0 | 90.9 | 86.4 | 92.2 | 70.8 | 97.6 | 85.7 | 95.8 | 63.3 | 72.7 | 100.0 | 91.3 | 98.0 | 82.5 | 90.0 | 100.0 |
| 2KKZ | 89.7 | 95.2 | 86.0 | 91.5 | 86.9 | 84.3 | 100.0 | 100.0 | 90.9 | 88.9 | 90.3 | 88.2 | 97.6 | 84.0 | 83.0 | 93.7 | 100.0 | 80.0 | 100.0 | 100.0 | 100.0 | 100.0 | 100.0 |
| 1VDY | 90.2 | 95.5 | 86.4 | 100.0 | 91.9 | 84.6 | 100.0 | - | 78.5 | 100.0 | 71.7 | 95.2 | 99.1 | 84.3 | 90.1 | 84.6 | 78.9 | 96.6 | 87.3 | 93.1 | 100.0 | 92.9 | 100.0 |
| 2KKL | 82.3 | 90.2 | 76.3 | 95.2 | 66.4 | 83.3 | 76.1 | - | 84.4 | 78.0 | 90.8 | 44.9 | 97.6 | 81.8 | 96.1 | - | 66.1 | 90.1 | 86.3 | 89.4 | - | 78.6 | 95.1 |
| 2L8V | 82.2 | 93.1 | 74.6 | 97.9 | 77.8 | 83.3 | 98.4 | - | 83.6 | 70.0 | 96.7 | 88.9 | 92.9 | 76.8 | 74.6 | - | 65.3 | 89.4 | 72.3 | 97.2 | 89.5 | 72.1 | 84.7 |
| 2LGH | 90.5 | 96.2 | 86.1 | 98.2 | 79.3 | 58.8 | 100.0 | 100.0 | 98.3 | 93.3 | 97.2 | 74.1 | 85.5 | 100.0 | 96.0 | 94.1 | 80.9 | 100.0 | 98.4 | 97.8 | 73.8 | 78.6 | 100.0 |
| 2M4F | 85.1 | 91.1 | 80.4 | 96.4 | 100.0 | 90.7 | 93.8 | - | 87.2 | 46.2 | 97.6 | 44.4 | 93.9 | 98.2 | 78.9 | 76.9 | 70.3 | - | 76.8 | 94.4 | 70.0 | 65.4 | 99.1 |
| 2JXP | 89.5 | 93.9 | 86.1 | 97.6 | 86.7 | 100.0 | 100.0 | - | 72.7 | 89.9 | 82.9 | 64.3 | 99.6 | 77.8 | 87.5 | 86.1 | 92.5 | 100.0 | - | 88.1 | 99.2 |
| 2LAK | 84.2 | 91.7 | 78.6 | 93.7 | 79.0 | 45.5 | 100.0 | 100.0 | 82.2 | 92.0 | 91.5 | 72.9 | 100.0 | 90.2 | 94.1 | 77.8 | 65.4 | 81.2 | 92.4 | 85.5 | 63.2 | 62.7 | 93.6 |
| 2M47 | 84.6 | 91.6 | 79.0 | 98.2 | 83.0 | 93.8 | 74.1 | 100.0 | 84.7 | 79.5 | 91.2 | 66.7 | 85.6 | 89.2 | 96.1 | 53.8 | 86.7 | 76.1 | 80.3 | 92.9 | 58.1 | 95.8 | 85.8 |
| 2K3A | 86.8 | 94.2 | 80.0 | 89.5 | 87.5 | 84.8 | 96.9 | 87.5 | 87.9 | 69.2 | 83.0 | 63.6 | 88.1 | 90.8 | 63.3 | 76.9 | 87.5 | 91.7 | 89.5 | 88.6 | 90.0 | 91.7 | 98.3 |
| 2B3W | 82.2 | 89.6 | 77.4 | 96.2 | 79.4 | 80.7 | 94.5 | - | 83.1 | 82.6 | 94.1 | 61.4 | 98.8 | 88.6 | 81.9 | 74.2 | 81.1 | 94.3 | 83.0 | 57.7 | 65.4 | 52.6 | 96.5 |
| KRAS4B | 85.2 | 97.7 | 77.5 | 96.3 | 76.2 | 85.0 | 95.2 | 85.7 | 88.5 | 90.4 | 78.4 | 81.6 | 80.6 | 78.7 | 80.0 | 74.3 | 70.7 | 81.5 | 85.6 | - | 89.8 | 96.0 |
| 2LF2 | 89.1 | 96.6 | 83.5 | 96.9 | 71.4 | 90.3 | 98.9 | 87.5 | 96.8 | 93.3 | 95.4 | 81.0 | 98.4 | 90.6 | 77.2 | 85.0 | 85.8 | 96.7 | 93.0 | 95.4 | 69.3 | 88.5 | 100.0 |
| **Median** | **89.7** | **96.6** | **86.0** | **100.0** | **87.0** | **90.0** | **98.7** | **100.0** | **92.7** | **90.0** | **96.4** | **72.9** | **98.2** | **90.3** | **90.1** | **80.4** | **76.5** | **94.3** | **92.5** | **96.9** | **80.0** | **89.7** | **98.7** |

**Supplementary Table 14 Accuracy of automated chemical shift assignment using all except NOESY-type input spectra.** Chemical shift assignment accuracy (%) is reported for the 45 proteins in the automated chemical shift assignment experiment (see Results section).

| Protein | All shifts | Backbone | Side-chain | ALA | ARG | ASN | ASP | CYS | GLU | GLN | GLY | HIS | ILE | LEU | LYS | MET | PHE | PRO | SER | THR | TRP | TYR | VAL |
|---|---|---|---|---|---|---|---|---|---|---|---|---|---|---|---|---|---|---|---|---|---|---|---|
| 2JVD | 91.4 | 100.0 | 85.4 | 100.0 | 100.0 | 70.8 | - | - | 100.0 | 80.0 | 100.0 | - | 100.0 | 100.0 | 83.2 | 100.0 | 66.7 | - | 95.8 | 77.8 | - | 80.0 | 100.0 |
| 2K57 | 92.5 | 98.9 | 88.0 | 100.0 | 89.6 | 84.2 | 95.8 | - | 100.0 | 87.7 | 100.0 | - | 100.0 | 100.0 | 95.3 | - | 53.1 | 100.0 | 100.0 | 97.8 | - | 60.7 | 100.0 |
| 2LX7 | 85.8 | 93.8 | 80.4 | 100.0 | 93.1 | - | 100.0 | 87.5 | 83.6 | 65.0 | 84.7 | 85.7 | 100.0 | 99.1 | 100.0 | 100.0 | 57.4 | 94.4 | 100.0 | 89.5 | 55.0 | 85.2 | 100.0 |
| 1YEZ | 92.3 | 98.7 | 87.5 | 100.0 | 92.7 | - | 100.0 | - | 87.8 | 78.6 | 100.0 | - | 100.0 | 85.7 | 90.0 | 100.0 | 69.0 | 97.2 | 95.8 | 100.0 | - | 100.0 | 97.7 |
| 2K52 | 85.4 | 94.1 | 79.7 | 85.7 | 84.8 | 77.3 | 85.0 | - | 96.6 | 67.9 | 86.7 | - | 99.1 | 69.5 | 96.4 | 83.3 | 63.0 | 79.2 | 100.0 | 100.0 | - | 61.9 | 100.0 |
| 2JT1 | 75.6 | 99.1 | 58.6 | 100.0 | 61.7 | 75.0 | 100.0 | - | 79.2 | 65.0 | 100.0 | 80.0 | 73.8 | 70.5 | 63.2 | 90.9 | - | 79.2 | 79.3 | 88.9 | 50.0 | 80.0 | 78.4 |
| 2HEQ | 75.1 | 87.0 | 66.2 | 87.5 | 75.0 | 63.6 | 87.1 | 100.0 | 71.2 | - | 64.6 | 35.5 | 92.9 | 82.3 | 76.5 | - | 66.7 | 86.7 | 87.5 | 88.6 | 40.0 | 76.8 | 92.7 |
| 2KD0 | 92.9 | 97.5 | 89.5 | 100.0 | 92.9 | 91.3 | 100.0 | - | 91.4 | 90.5 | 66.7 | 100.0 | 92.8 | 91.2 | 95.8 | 94.4 | 94.4 | - | - | - | - | - | 100.0 |
| 2LK2 | 84.0 | 96.1 | 75.6 | 100.0 | 74.2 | 68.8 | 93.5 | 100.0 | 83.3 | 77.8 | 88.9 | 84.8 | 98.2 | 86.8 | 80.0 | 72.0 | 85.7 | 97.9 | 92.9 | 96.4 | 67.6 | 75.6 | 100.0 |
| MH04 | 87.3 | 98.6 | 79.1 | 94.3 | 80.0 | 84.8 | 100.0 | - | 97.0 | 86.9 | 100.0 | - | 98.6 | 82.9 | 94.0 | 76.9 | 63.3 | 88.3 | 93.3 | 88.0 | 45.0 | 80.0 | 98.9 |
| 2KZV | 77.0 | 91.0 | 66.9 | 97.6 | 63.5 | 75.8 | 89.1 | - | 90.9 | 52.4 | 83.3 | 69.6 | 89.3 | 66.1 | 79.4 | 92.9 | 60.9 | 95.7 | 81.3 | 78.9 | 57.9 | 78.6 | 93.2 |
| 2L05 | 81.8 | 95.5 | 73.1 | 100.0 | 82.3 | 60.9 | 100.0 | 100.0 | 90.9 | 53.3 | 86.7 | 66.7 | 97.1 | 87.3 | 77.3 | 87.9 | 75.0 | 46.2 | 91.7 | 100.0 | 55.0 | 78.6 | 100.0 |
| 6FIP | 85.7 | 99.4 | 76.2 | 100.0 | 75.0 | 100.0 | 89.1 | - | 87.3 | 91.1 | 100.0 | 11.1 | 91.8 | 92.9 | 82.9 | 85.9 | 81.5 | 94.4 | 100.0 | 100.0 | 35.0 | 88.9 | 95.5 |
| 2KPN | 87.9 | 96.3 | 81.3 | 97.1 | 87.5 | 86.4 | 95.8 | - | 87.9 | 66.7 | 94.9 | 68.2 | 100.0 | 89.3 | 75.9 | 61.9 | 58.3 | 89.6 | 87.5 | 96.6 | - | 76.2 | 97.4 |
| 2K0M | 82.5 | 91.1 | 76.3 | 95.9 | 77.5 | 86.4 | 97.9 | 56.3 | 85.1 | 82.1 | 95.2 | 50.0 | 95.7 | 81.0 | 92.2 | 84.6 | 47.1 | 95.7 | 89.2 | 100.0 | 15.8 | 78.6 | 96.0 |
| 2K5V | 89.5 | 98.5 | 83.1 | 97.1 | 81.3 | 75.0 | 100.0 | - | 88.9 | 66.7 | 93.3 | - | 91.1 | 97.6 | 90.9 | 87.5 | 64.6 | 100.0 | 97.2 | 100.0 | 60.0 | 73.2 | 100.0 |
| 2KOB | 74.5 | 85.6 | 66.3 | 100.0 | 81.3 | 50.5 | 93.8 | 12.5 | 80.0 | 73.3 | 66.7 | 31.3 | 87.8 | 82.0 | 75.5 | 0.0 | 54.5 | 70.8 | 72.5 | 96.3 | 30.0 | 52.5 | 100.0 |
| 2KHD | 83.0 | 92.5 | 75.4 | 100.0 | 71.9 | 69.4 | 100.0 | 93.8 | 80.6 | 80.0 | 88.6 | 62.5 | 58.2 | 88.9 | 89.2 | - | 54.9 | 100.0 | 93.3 | 100.0 | - | 69.0 | 96.0 |
| 2MK2 | 85.0 | 98.7 | 74.5 | 100.0 | 76.2 | 72.7 | 97.5 | 75.0 | 90.9 | 73.5 | 93.8 | 69.0 | 89.3 | 86.9 | 100.0 | 72.7 | 74.5 | 97.9 | 78.8 | 85.3 | 70.0 | 75.0 | 94.8 |
| 2K50 | 85.6 | 95.6 | 78.4 | 100.0 | 87.9 | 75.8 | 89.1 | - | 88.8 | - | 93.3 | 58.1 | 97.3 | 93.8 | 93.7 | 48.8 | 69.6 | 48.3 | 93.9 | 91.7 | 40.0 | - | 100.0 |
| 2KIW | 78.3 | 95.2 | 67.0 | 95.2 | 90.4 | 66.7 | 93.8 | - | 87.3 | 70.0 | 100.0 | 51.6 | 92.1 | 85.7 | 62.9 | 64.0 | 61.3 | 77.8 | 83.6 | 77.8 | 73.7 | 76.2 | 89.9 |
| 1WQU | 83.4 | 97.9 | 73.6 | 100.0 | 92.1 | 81.8 | 100.0 | - | 87.5 | 72.4 | 90.0 | 81.9 | 97.1 | 78.8 | 64.7 | - | 50.0 | 94.8 | 94.0 | 100.0 | 80.0 | 54.8 | 91.8 |
| 2KL6 | 85.9 | 95.5 | 79.2 | 100.0 | 53.1 | 73.7 | 90.6 | - | 85.2 | 66.7 | 97.9 | 83.3 | 98.2 | 94.6 | 86.6 | 95.2 | 45.8 | 98.3 | 100.0 | 95.6 | 46.2 | 67.9 | 98.0 |
| 6GT7 | 61.0 | 95.4 | 40.0 | 90.9 | 59.4 | 52.9 | 81.8 | 76.9 | 58.9 | 50.0 | 60.0 | 60.0 | 62.5 | 59.4 | 51.5 | 63.6 | 57.1 | 58.3 | 83.3 | 72.2 | 62.5 | 52.9 | 65.3 |
| 2K5D | 88.0 | 96.4 | 81.8 | 100.0 | 86.5 | 68.8 | 90.9 | - | 96.6 | 66.7 | 88.9 | - | 100.0 | 86.5 | 88.2 | 64.3 | 68.8 | 97.2 | 95.8 | 100.0 | 75.0 | 63.2 | 98.0 |
| 2KVO | 84.4 | 95.2 | 76.5 | 100.0 | 62.1 | 76.6 | 92.5 | - | 91.5 | 69.0 | 91.7 | 50.0 | 100.0 | 89.8 | 87.5 | 68.0 | 64.3 | 77.1 | 93.8 | 96.9 | 70.0 | 57.1 | 98.5 |
| 2KCD | 76.3 | 85.6 | 70.1 | 100.0 | 81.0 | 60.2 | 88.9 | 62.5 | 86.2 | 61.4 | 91.3 | 65.0 | 99.2 | 67.4 | 80.0 | 100.0 | 50.0 | 80.0 | 93.8 | 94.9 | 60.0 | 61.6 | 100.0 |
| 2L7Q | 61.1 | 72.9 | 52.7 | 95.2 | 34.4 | 49.4 | 86.3 | - | 66.7 | 32.4 | 91.7 | 50.0 | 81.0 | 64.3 | 62.9 | 68.0 | 43.1 | 86.7 | 83.7 | 78.4 | 5.0 | 25.9 | 96.4 |
| 2KFP | 75.3 | 91.1 | 65.4 | 94.6 | 56.5 | 59.6 | 90.0 | - | 90.9 | 64.9 | 97.2 | 81.0 | 98.4 | 91.4 | 61.7 | 87.5 | 53.6 | 67.4 | 87.5 | 82.8 | 43.9 | 50.0 | 95.3 |
| 2L3B | 68.7 | 83.1 | 57.8 | 90.5 | 51.2 | 50.0 | 78.4 | 25.0 | 69.1 | 46.8 | 92.9 | 53.3 | 94.6 | 59.3 | 70.3 | 71.4 | 68.6 | 78.6 | 83.3 | 97.8 | - | 53.0 | 83.1 |
| 1VEE | 84.4 | 96.4 | 75.3 | 100.0 | 87.0 | 77.3 | 98.6 | - | 93.9 | 85.7 | 86.8 | - | 100.0 | 93.4 | 72.8 | - | 68.5 | 91.8 | 80.0 | 90.6 | 55.0 | 58.9 | 100.0 |
| 2K1G | 82.6 | 94.6 | 74.1 | 100.0 | 86.4 | 72.5 | 100.0 | 0.0 | 83.1 | 72.7 | 79.2 | 70.8 | 97.6 | 85.7 | 92.4 | 67.3 | 61.0 | 100.0 | 82.6 | 92.0 | 40.0 | 73.8 | 100.0 |
| 2KKZ | 71.2 | 83.5 | 62.8 | 83.0 | 76.2 | 80.0 | 78.6 | 100.0 | 79.2 | 76.2 | 30.8 | 81.1 | 61.0 | 71.6 | 52.9 | 59.6 | 50.0 | 82.5 | 79.2 | 70.0 | - | 85.7 | 97.4 |
| 1VDY | 81.4 | 91.8 | 73.8 | 100.0 | 65.4 | 53.8 | 100.0 | 100.0 | 76.9 | 82.1 | 63.0 | 76.2 | 98.2 | 87.9 | 78.7 | 84.6 | 47.4 | 91.5 | 80.0 | 93.1 | 100.0 | 82.1 | 99.1 |
| 2KKL | 57.8 | 69.4 | 49.0 | 86.7 | 32.9 | 57.1 | 67.0 | - | 72.5 | 49.2 | 80.3 | 30.6 | 50.0 | 48.3 | 88.2 | - | 34.7 | 64.8 | 64.4 | 51.1 | - | 100.0 | 64.0 |
| 2L8V | 73.5 | 89.5 | 61.9 | 93.8 | 65.7 | 70.0 | 85.9 | - | 84.9 | 50.0 | 86.7 | 72.2 | 78.6 | 74.0 | 74.6 | - | 65.4 | 74.5 | 58.5 | 97.2 | 50.0 | 59.3 | 82.3 |
| 2LGH | 82.0 | 92.6 | 73.7 | 84.7 | 74.1 | 50.0 | 93.1 | 100.0 | 89.2 | 75.2 | 93.1 | 67.2 | 84.3 | 98.1 | 94.0 | 85.3 | 61.8 | 100.0 | 96.9 | 95.5 | 47.6 | 92.9 | 99.1 |
| 2M4F | 84.5 | 94.2 | 76.9 | 96.4 | 93.8 | 72.9 | 93.8 | - | 82.0 | 69.2 | 96.4 | 61.1 | 96.6 | 95.5 | 76.8 | 76.9 | 69.2 | - | 89.3 | 97.8 | 80.0 | 62.8 | 96.4 |
| 2JXP | 79.3 | 86.7 | 73.4 | 96.0 | 78.8 | 63.6 | 80.0 | - | 54.0 | 66.1 | 74.3 | 75.0 | 91.6 | 79.4 | 84.3 | 55.6 | 68.8 | 86.1 | 78.8 | 100.0 | - | 71.4 | 95.9 |
| 2LAK | 78.0 | 89.1 | 69.7 | 90.1 | 69.8 | 18.2 | 97.3 | 100.0 | 77.8 | 84.0 | 89.6 | 55.7 | 100.0 | 93.8 | 94.1 | 61.1 | 53.3 | 77.9 | 93.9 | 89.1 | 40.4 | 62.2 | 76.4 |
| 2M47 | 76.4 | 86.7 | 68.3 | 98.2 | 74.1 | 77.1 | 74.1 | 79.2 | 77.3 | 66.5 | 70.6 | 88.3 | 66.7 | 88.3 | 94.2 | 69.2 | 62.7 | 75.7 | 73.2 | 86.6 | 37.9 | 75.0 | 89.4 |
| 2K3A | 82.8 | 94.5 | 72.1 | 91.0 | 87.5 | 73.3 | 100.0 | 87.5 | 83.8 | 73.1 | 83.0 | 70.5 | 84.5 | 93.9 | 95.9 | 84.6 | 6.3 | 91.7 | 82.4 | 86.2 | 75.0 | 67.8 | 98.3 |
| 2B3W | 76.6 | 92.3 | 66.3 | 97.1 | 68.9 | 84.9 | 90.9 | - | 87.7 | 75.7 | 92.2 | 61.4 | 100.0 | 85.0 | 77.8 | 71.0 | 57.9 | 54.3 | 87.2 | 73.1 | 48.7 | 48.7 | 96.5 |
| KRAS4B | 66.9 | 85.7 | 55.4 | 92.6 | 52.3 | 55.0 | 76.2 | 95.2 | 73.8 | 54.0 | 96.4 | 70.3 | 59.2 | 77.1 | 55.1 | 80.0 | 60.8 | 31.7 | 57.4 | 73.3 | - | 48.1 | 89.3 |
| 2LF2 | 78.7 | 89.7 | 70.5 | 95.9 | 64.8 | 66.7 | 70.5 | 25.0 | 92.5 | 70.0 | 92.3 | 57.1 | 95.6 | 86.2 | 72.4 | 70.0 | 70.8 | 58.3 | 82.5 | 92.7 | 49.1 | 71.2 | 97.4 |
| **Median** | **82.5** | **94.2** | **73.7** | **97.6** | **76.2** | **70.0** | **93.3** | **87.5** | **86.2** | **70.0** | **91.7** | **65.0** | **96.6** | **86.5** | **82.9** | **76.9** | **61.6** | **86.7** | **87.5** | **93.1** | **52.5** | **71.4** | **97.4** |

**Supplementary Movie 1** Visualization of 100 automatically determined protein structures (blue) along with their PDB depositions (orange). The displayed proteins are sorted by their RMSD to PDB reference in ascending order.

**Supplementary Movie 2** Visualization of automated visual spectrum analysis. The movie presents sequentially all planes from a challenging $^{13}$C-resolved [$^1$H,$^1$H]-NOESY spectrum of the 20 kDa protein 2B3W. All planes were annotated automatically by ARTINA within 5 minutes and without any human involvement in the process. Model confidence about a signal being a true peak is presented by color-coding, according to the legend in the upper right corner of the movie.

**Supplementary Movie 3** Visualization of the ability of automated visual spectrum analysis to handle strong background artefacts. The movie presents sequentially all planes from a noisy $^{13}$C-resolved [$^1$H,$^1$H]-NOESY spectrum of the 7 kDa protein 6SOW with only 20% $^{13}$C labeling. All planes were annotated automatically by ARTINA within 5 min and without any human involvement in the process. The model confidence about a signal being a true peak is presented by color-coding, according to the legend in the upper right corner of the movie.

**Supplementary Movie 4** Visualization of automated visual spectrum analysis. The movie presents sequentially all planes from an HCCH-TOCSY spectrum, which has been acquired on the 16 kDa protein 2LGH. All planes have been annotated automatically by ARTINA without any human involvement in the process. The spectrum analysis time was under 5 minutes. The model confidence about a signal being a true peak is presented by color-coding according to the legend in the upper right corner of the movie.

**Supplementary Movie 5** Video tutorial presenting all steps of automated protein structure determination with ARTINA as available on the nmrtist.org website.